\documentclass[sn-mathphys,Numbered]{sn-jnl}

\usepackage{graphicx}%
\usepackage{subcaption}%
\usepackage{multirow}%
\usepackage{geometry}%
\usepackage{amsmath,amssymb,amsfonts}%
\usepackage{amsthm}%
\usepackage{mathrsfs}%
\usepackage[title]{appendix}%
\usepackage{xcolor}%
\usepackage{textcomp}%
\usepackage{manyfoot}%
\usepackage{booktabs}%
\usepackage{algorithm}%
\usepackage{algorithmicx}%
\usepackage{algpseudocode}%
\usepackage{listings}%
\usepackage{fixltx2e}%
\usepackage{xcolor}%
\usepackage{soul}%
\usepackage{placeins}
\usepackage{tabularx,afterpage}
\usepackage{xtab}
\usepackage{longtable}
\usepackage{soul}
\usepackage{tablefootnote}

\theoremstyle{thmstyleone}%
\theoremstyle{thmstyletwo}%

\theoremstyle{thmstylethree}%

\begin{document}

\title[Article Title]{Universal Numerical Simulation Model for Laser Material Processing}


\author*[1]{\fnm{Andreas} \sur{Otto}}\email{andreas.otto@tuwien.ac.at}

\author[1]{\fnm{Michele} \sur{Buttazzoni}}\email{michele.buttazzoni@tuwien.ac.at}

\author[1]{\fnm{Carlos} \sur{Durán}}\email{carlos.duran@tuwien.ac.at}

\author[1]{\fnm{Tobias} \sur{Florian}}\email{tobias.florian@tuwien.ac.at}

\author[1]{\fnm{Constantin} \sur{Zenz}}\email{constantin.zenz@tuwien.ac.at}

\affil[1]{ \orgname{TU Wien}, \orgdiv{Institute of Production Engineering and Photonic Technologies}, \orgaddress{\street{Getreidemarkt 9}, \city{Vienna}, \postcode{1060}, \country{Austria}}}


\abstract{High power lasers are used for a variety of manufacturing processes on time- and length scales that cover many orders of magnitude and on different types of materials. The variety of manufacturing processes that can be achieved through laser-material-interaction is a result of numerous underlying coupled, nonlinear physical phenomena. Simulation models can be effectively used to gain process understanding, help to explain experimentally observed phenomena and to optimize existing and design new processes. However, the inherent complexity and variety of the involved phenomena makes modeling a challenging task.

Within this chapter, a universal model is presented that is capable of accurately simulating a broad spectrum of different processes, covering applications such as welding, additive manufacturing, cutting, ablation, drilling and surface structuring, encompassing both continuous wave lasers and ultrashort pulsed lasers, and their interaction with different materials.

Starting from the fundamental principles of conservation of mass, momentum and energy, a continuum mechanical framework is presented that captures the main physical phenomena at play in this coupled multiphysical problem. A numerical solution to the derived mathematical model using the Finite Volume Method is provided and its validation is demonstrated on
a series of benchmark problems, which consist of laser material processing scenarios where detailed experimental results are available for comparison. Examples for macroscopic problems, featuring continuous wave laser sources, include keyhole drilling and collapse under stationary illumination and melt pool dynamics and pore formation in copper welding. In the realm of microscopic processes, copper ablation with femtosecond-long pulses is simulated.

Finally, a variety of real-world applications are presented, where phenomena and defects encountered in industrial applications are explained and optimized with the help of the simulation model. Examples include deep penetration welding of steel using dynamic beam shaping, multilayer additive manufacturing via powder bed fusion, the ultrashort pulsed drilling of micro vias in dielectric materials and many more.}


\keywords{multiphysical simulation, laser material processing, welding, ablation, powder bed fusion, computational fluid dynamics, process modeling}



\maketitle

\section{Introduction and State of the Art}\label{sec:Introduction}
\subsection{Motivation and Introduction}\label{subsec:Motivation}
Physics-based simulations are a powerful tool to gain understanding of physical phenomena and their interactions as they offer the possibility to observe quantities that are difficult or even impossible to capture experimentally at relevant time- and length scales~\cite{Otto2011}. Even with recent advances in in situ synchrotron X-ray observation~\cite{Lu2025}, temporal acquisition rates are limited, and direct visualization of melt pool or gas flow velocity patterns or pressure distributions remain virtually impossible. Additionally, models allow for simulative experiments under idealized conditions or isolating the influence of individual parameters~\cite{Otto2011}. They enable prediction process outcomes before setting up costly and time-consuming experimental campaigns, allowing for virtual planning of production processes and the use of digital twins. However, in contrast to many other manufacturing processes, laser-based material processing poses a highly coupled problem of nonlinear physical phenomena (a true \textit{multiphysical} problem), making it especially difficult to derive simulation models that possess both strong predictive capabilities and allow for reasonable computational times and numerical stability.

\begin{figure}[h]
\centering
\includegraphics[width=0.99\textwidth]{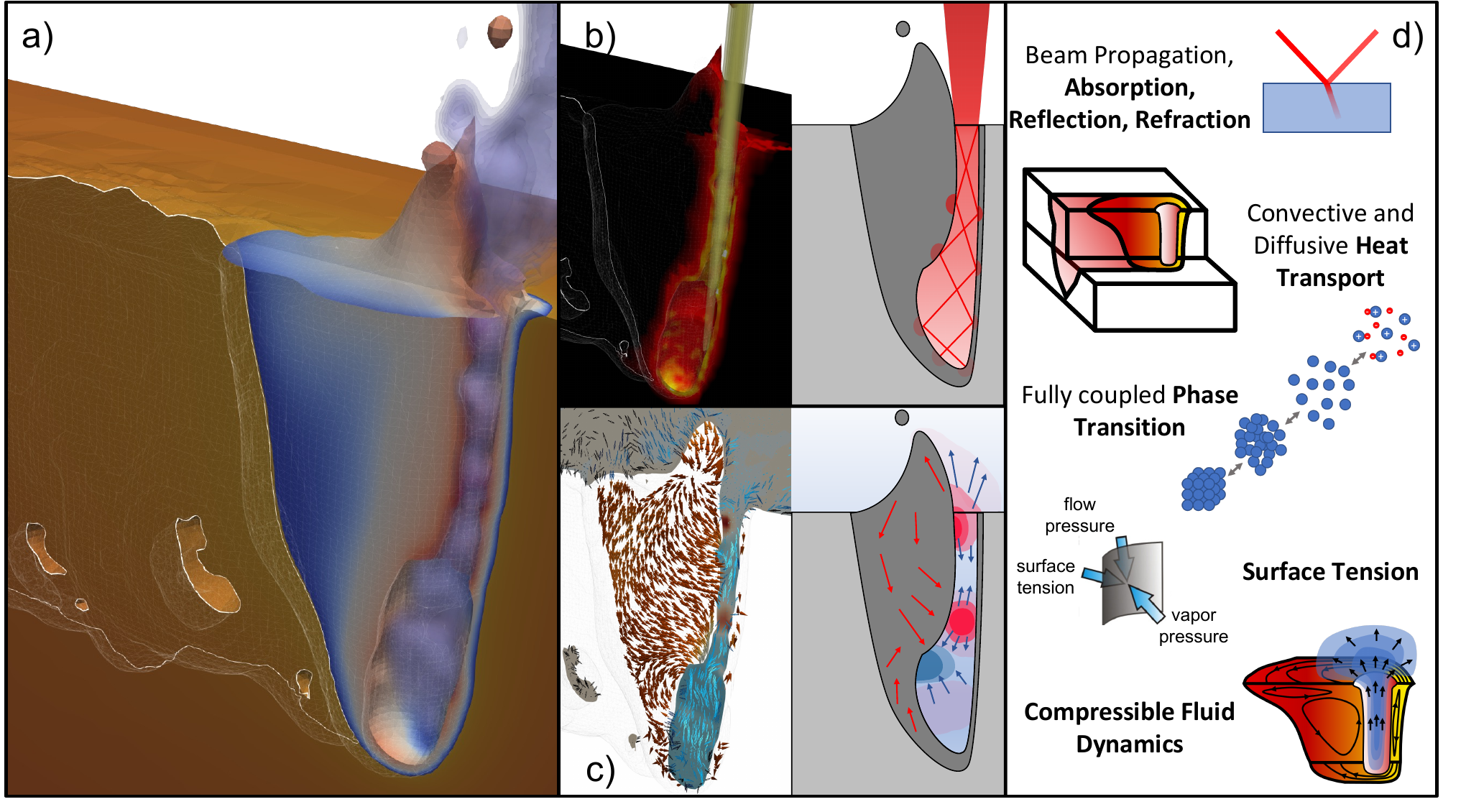}
\caption{Schematic overview of physical phenomena recorded from a copper welding simulation. a) Temperature in condensed matter, purple contours depict metal vapor - absence of metal vapor indicates presence of ambient gas. b) Propagating laser intensity up to first impact and absorbed power after ray tracing (simulation and schematic sketch). c) Velocity vectors in liquid and gaseous phases and pressure distribution in gaseous phases. Red and blue color denote pressure levels above and  below 1atm, respectively (simulation and schematic sketch). d) Non-comprehensive collection of key physical phenomena at play}\label{fig:overview}
\end{figure}

Furthermore, the laser beam is a very versatile tool with a wide range of applications, where a change in processing parameters can lead to a completely different outcome, for example either a joining process or a subtractive process (e.g., in the transition from welding to ablation cutting~\cite{Otto2018a}). Hence, a useful simulation model should be \textit{universal}, in the sense of being able to predict the very type of process that will emerge from a given set of initial conditions, boundary conditions, material and processing parameters. Therefore, no a priori assumptions on the type of resulting process should be needed. Figure~\ref{fig:overview} provides a schematic overview of the various physical phenomena at play in a typical scenario of laser-material-processing.

\begin{figure}[h]
\centering
\includegraphics[width=0.99\textwidth]{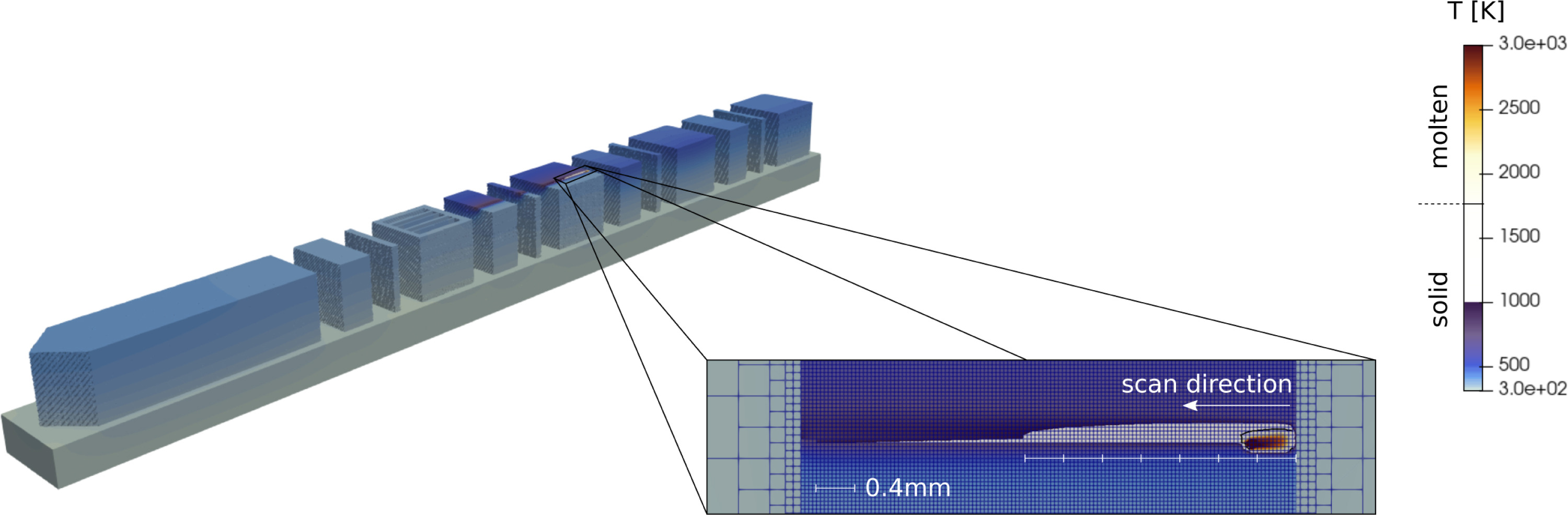}
\caption{Scan-track resolved simulation of an entire PBF-LB/M part build process by Proell et al.~\cite{Proell2024a}. In this macro-scale model, only heat conduction is included (Reprinted from \cite{Proell2024a}, Addit. Manuf., Vol \textbf{79}, Proell et al., A highly efficient computational approach for fast scan-resolved simulations of metal additive manufacturing processes on the scale of real parts, 103921, Copyright (2024), with permission from Elsevier)}\label{fig:intro:Proell2024}
\end{figure}

\begin{figure}[h]
\centering
\includegraphics[width=0.8\textwidth]{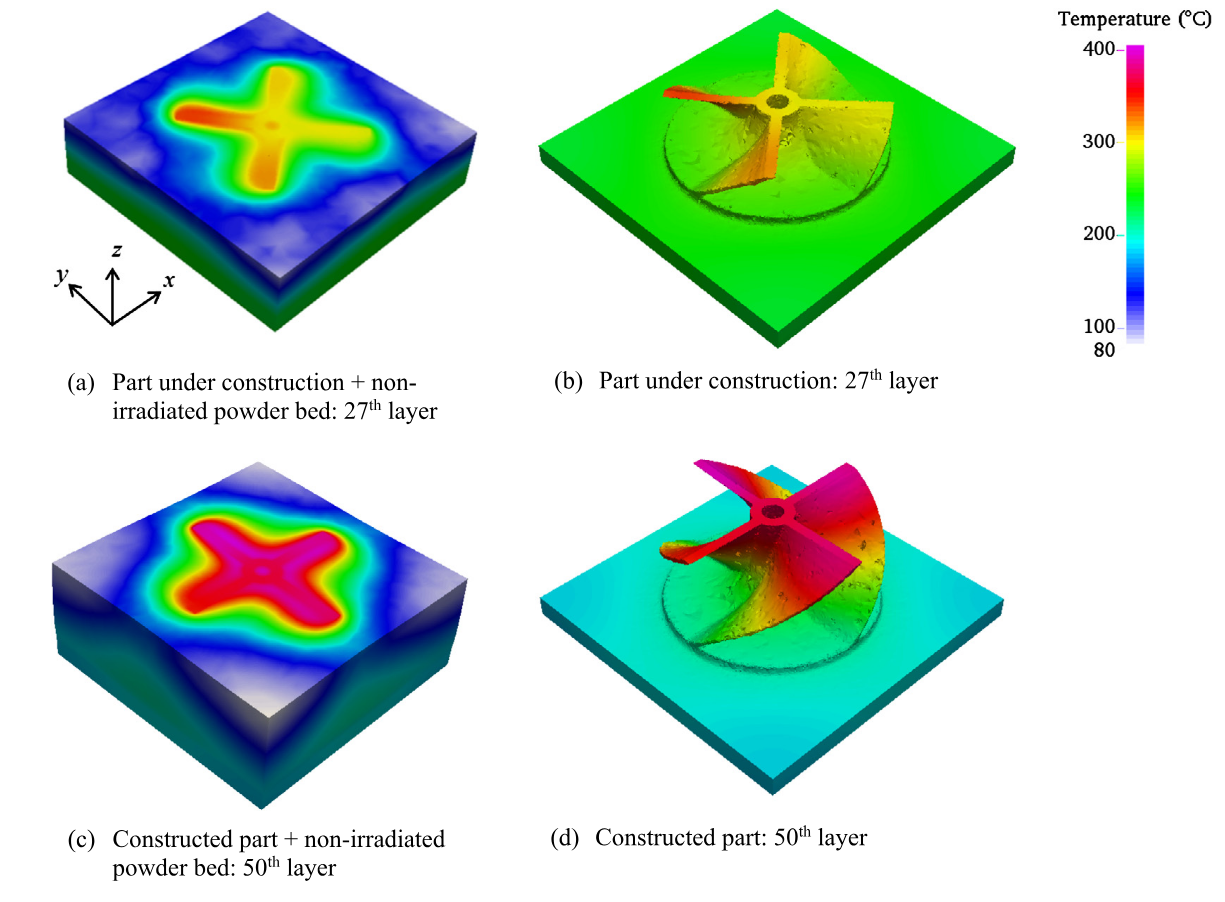}
\caption{Part-scale simulation of residual stress distribution during PBF-LB/M process by Zhang et al.~\cite{Zhang2018}. Here, each build layer is decomposed into 4 -- 20 fractions, omitting scan-track resolution, to allow for the simulation of the entire part (Reprinted from \cite{Zhang2018}, C.R. Mec., Vol \textbf{346}(11), Zhang et al., Numerical modelling of fluid and solid thermomechanics in additive manufacturing by powder-bed fusion: Continuum and level set formulation applied to track- and part-scale simulations, 1055--1071, Copyright (2018) under Creative Commons BY-NC-ND 4.0 license)}\label{fig:intro:Zhang2018}
\end{figure}

\begin{figure}[h]
\centering
\includegraphics[width=0.99\textwidth]{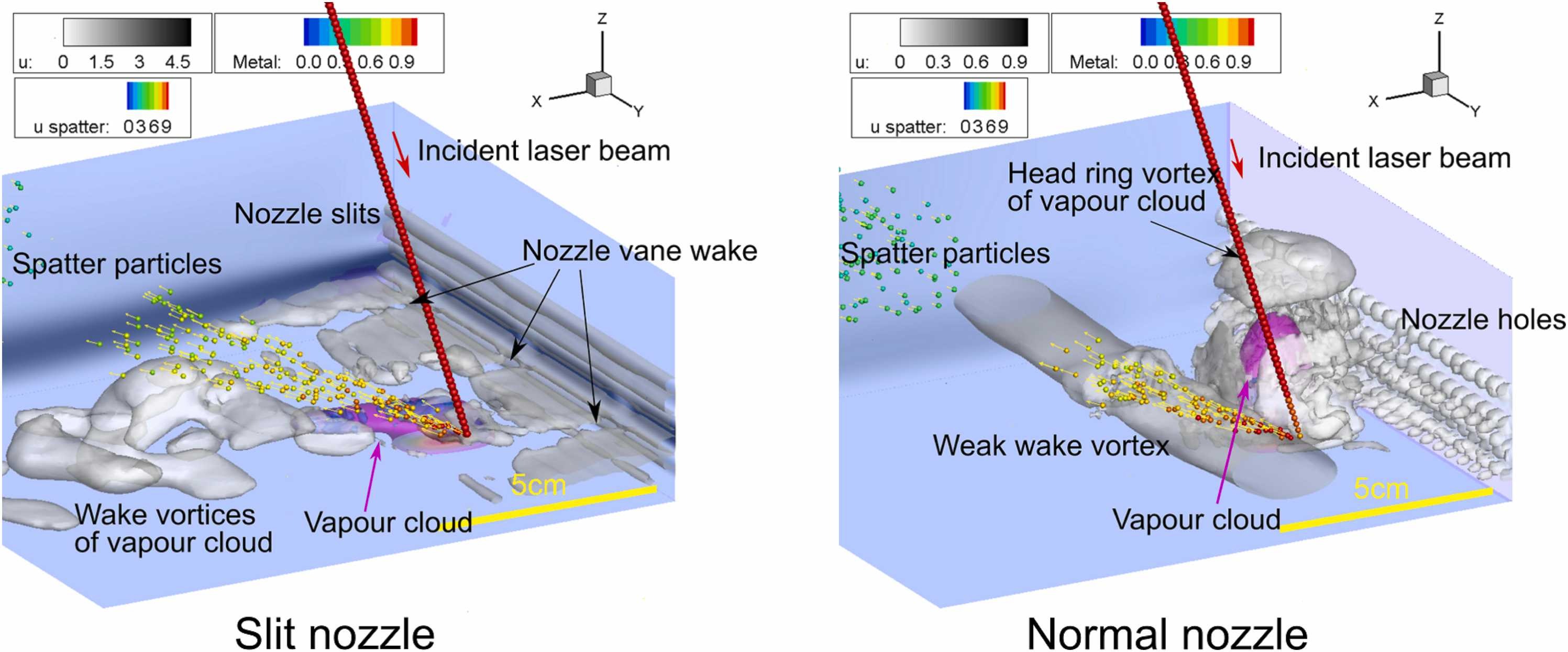}
\caption{Simulation of vapor cloud and spatter dynamics by Shinjo et al.~\cite{Shinjo2024} comparing the influence of different types of shielding gas nozzles (Reprinted from \cite{Shinjo2024}, Addit. Manuf., Vol \textbf{80}, Shinjo et al., In-process monitoring and direct simulation of argon shielding gas and vapour dynamics to control laser-matter interaction in laser powder bed fusion additive manufacturing, 103953, Copyright (2024) under Creative Commons BY 4.0 license)}\label{fig:intro:Shinjo2024}
\end{figure}

The simulation model presented hereafter focuses on the meso-scale of the process, starting from the basic principles of conservation of mass, momentum and energy, and predicting the resulting coupled thermo-fluidmechanical phenomena, at the scale of approximately the melt pool. Depending on the envisaged scope, other simulation approaches focus on larger scales (macro-scale), aiming to predict, for example, thermal histories, solid mechanical deformations and residual stresses at the scale of the entire manufactured part~\cite{Proell2024a,Proell2024b,Denlinger2016,Denlinger2017}, or at very small scales well below the fluid mechanical length scales, focusing for example on the interatomic forces during laser irradiation in Molecular Dynamics (MD) simulations~\cite{Zhigilei1997, Ivanov2003, Povarnitsyn2015, Wu2014} or on the microstructure evolution in the solidifying material. In Figure~\ref{fig:intro:Proell2024}, a recent example of a part-scale simulation of PBF-LB/M is shown, where only heat conduction is accounted for, to allow simulation of all 312 layers of the entire printed part just under 24 hours on 1248 CPU cores~\cite{Proell2024a}. Figure~\ref{fig:intro:Zhang2018} shows another example of a macro-scale model, where solid body mechanics are included in addition to heat conduction, which is made possible by decomposing the printing process into only a few fractions per layer, omitting scan-track resolution~\cite{Zhang2018}.
There also exist high-fidelity models for specific sub-domains of the problem, such as the turbulent shielding gas flow and its interaction with powder particles, but neglecting coupling with liquid material dynamics~\cite{Shinjo2024}, cf. Figure~\ref{fig:intro:Shinjo2024} for an example. While such models allow for a detailed analysis of certain aspects of the process, they lack multiphysical coupling and do not allow predictions of the general process behavior at the meso-scale. 
Often, different models, e.g., a meso-scale model and either a solid-mechanical or microstructural model, are coupled to obtain a multi-scale model (e.g.,~\cite{Zhang2018,Zenz2024b,Bayat2024,Mosbah2025}).

\subsection{State of the Art}\label{subsec:SOA}
\subsubsection{Continuous Wave Laser Processing}
Within the meso-scale introduced above, considering continuous wave laser processing, most simulation frameworks are developed to simulate either the process of laser beam welding (LBW) or additive manufacturing (AM) via laser-based powder bed fusion (PBF-LB/M). From a physics perspective those processes are identical, except for the additional powder material to be welded onto a substrate in AM and the joining of multiple parts in LBW. The similarities between these processes are further highlighted if one takes into account that a large percentage of fundamental investigations of AM (both in experiment and simulation) are so-called bare plate experiments, i.e., lacking powder, while investigations in welding mostly feature so-called bead-on-plate welds. Both of the aforementioned configurations are simply different names for the same process.

\begin{figure}[h]
\centering
\includegraphics[width=0.99\textwidth]{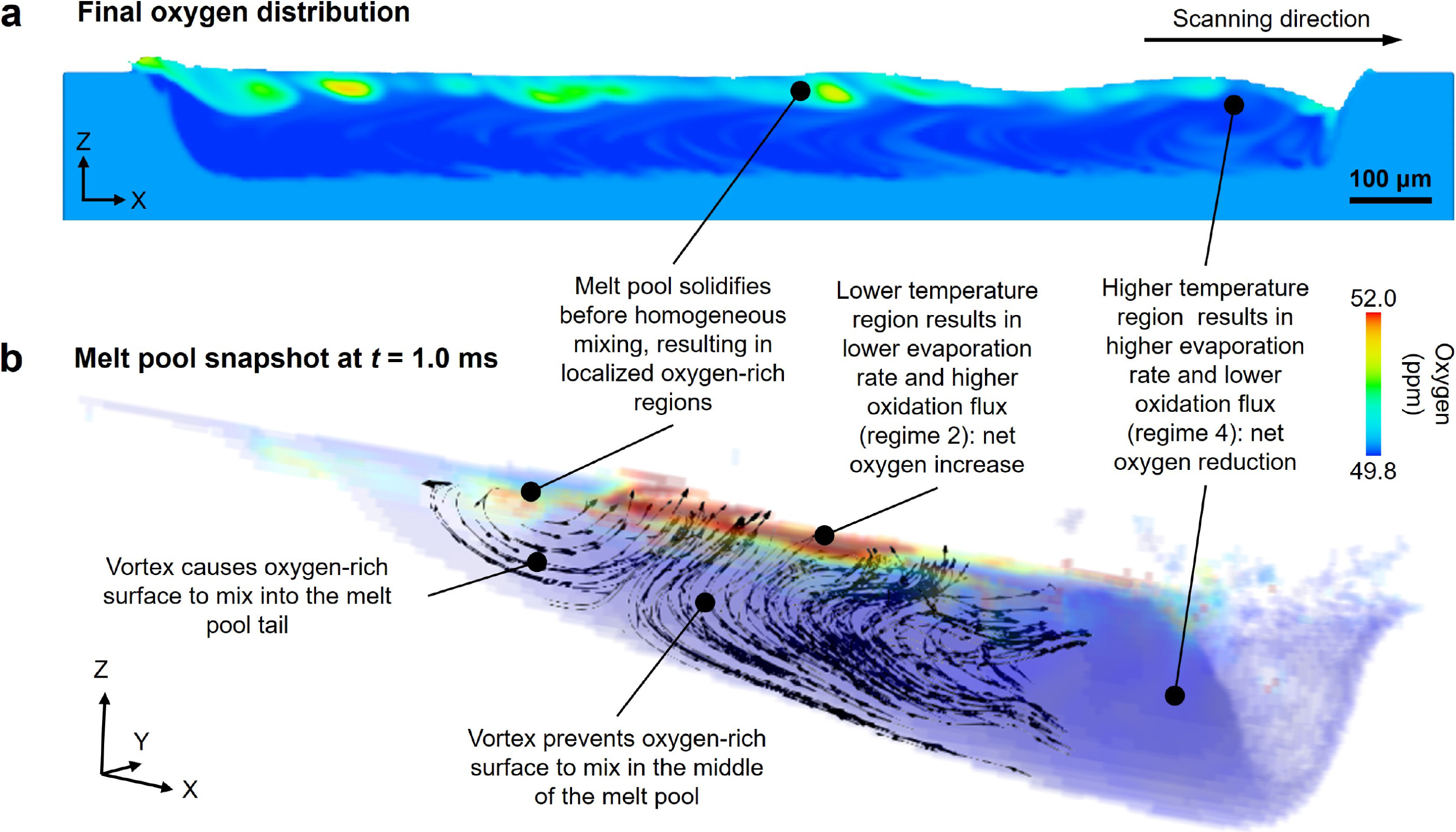}
\caption{Simulation of melt pool dynamics including the evolution of oxygen within the melt pool via chemical reactions and advection by Chia et al.~\cite{Chia2024} (Reprinted from \cite{Chia2024}, Acta Mater., Vol \textbf{275}, Chia et al., Unveiling gas–liquid metal reactions in metal additive manufacturing: High-fidelity modeling validated with experiments, 120029, Copyright (2024), with permission from Elsevier)}\label{fig:intro:Chia2024}
\end{figure}

\begin{figure}[h]
\centering
\includegraphics[width=0.8\textwidth]{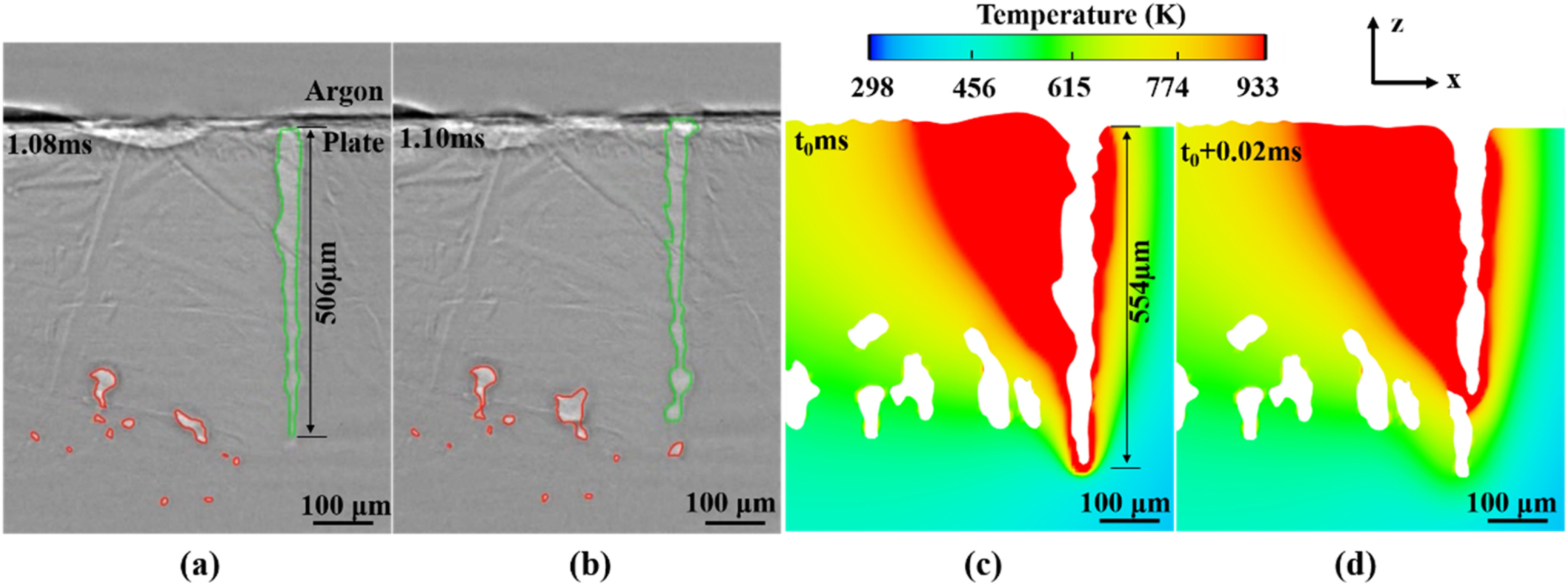}
\caption{Simulation~\cite{Guo2023} of pore formation in PBF-LB/M of aluminum (c, d) compared to the corresponding experiment of~\cite{Huang2022} (a, b) (Reprinted from \cite{Guo2023}, Int. J. Mach. Tools Manuf., Vol \textbf{184}, Guo et al., Understanding keyhole induced-porosities in laser powder bed fusion of aluminum and elimination strategy, 103977, Copyright (2023), with permission from Elsevier; images (a, b) originally from~\cite{Huang2022}, Copyright (2022) under Creative Commons BY 4.0 license)}\label{fig:intro:Guo2023}
\end{figure}

In LBW and AM, available simulation models can broadly be divided into those that include gaseous phases (termed CGP (\textbf{C}ondensed and \textbf{G}aseous \textbf{P}hases) models hereafter) and those that only consider condensed matter (termed CP (\textbf{C}ondensed \textbf{P}hases) models hereafter) and account for liquid-vapor interactions via phenomenological models. Naturally, models exist that bridge the gap between these two categories, exhibiting varying levels of complexity and incorporating highly process-specific features. Examples include magnetohydrodynamics~\cite{Flint2021mhd, Yang2024}, selective element evaporation modeling~\cite{Flint2022volumeDilation}, and the inclusion of chemical reactions~ \cite{Chia2024} (cf. Figure~\ref{fig:intro:Chia2024} for an illustrative example).
One of the most widely used CP models is arguably the commercial software \textit{FLOW-3D\textsuperscript{\textregistered}}, which is utilized both in industrial and academic research and only includes condensed matter in the modeling domain, employing phenomenological models at the domain boundary (the liquid-vapor interface) to account for effects such as the recoil of evaporating material. Due to the lack of coupling between liquid and vapor phases, the predictive capabilities of such CP models are lower, especially when entering process regimes where evaporation- and condensation-induced effects become dominant. However, this drawback comes at the advantage of relatively high computational speed and numerical stability, which (in combination with the user-friendliness of a commercial software interface) leads to a large user base. Also, for many applications, such as simulation of AM close to the conduction mode regime (i.e., where evaporation only plays a minor role), or when properly calibrating a CP model to a certain, confined process regime, these approaches prove useful and efficient. An example of such model calibration is laid out in~\cite{Jabar2024}, where a model coefficient in the phenomenological recoil pressure equation was fitted to reproduce experimentally obtained results (where notably a different value for the coefficient was needed for different laser intensity distributions). A recent example of CP model simulations used to investigate spiking pore formation in AM of aluminum \cite{Guo2023} is given in Figure~\ref{fig:intro:Guo2023}, where the obtained results are compared to in-situ x-ray images.

Furthermore, efforts have been made to enhance the degree of accuracy and coupling of CP models without actually including the gaseous phase in the modeling domain. An example is the work of Schmidt et al.~\cite{Schmidt2024}, who aimed at incorporating effects of local shielding gas supply by adding a dynamic pressure term locally in the keyhole region within a CP model simulation framework to analyze the effect of shielding gas flow on resulting process instabilities in high speed welding of steel. Another example is the work of Wang et al.~\cite{Wang2020}, who derived a new evaporation model to replace the usually employed model of Anisimov~\cite{Anisimov1968} that is used in virtually all CP models, leading to improved accuracy in continuous wave laser drilling simulations under both atmospheric conditions and reduced pressure.

Some published CGP models that do include gaseous phases in their modeling domain still rely on a phenomenological recoil pressure, mainly for increased computational speed that comes at the cost of reduced predictive capabilities, an example being the open source software \texttt{beamWeldFoam}~\cite{Flint2022beamWeldFoam}. Another notable CGP modeling approach is that of Yu and Zhao~\cite{Yu2022,Yu2023}, where (while still explicitly calculating a recoil pressure to be added to the momentum equations) the evaporation rate is not calculated following Anisimov's model, but adopted from~\cite{Wang2020} in a CGP model framework. Consequently, this model is able to quantitatively predict various dynamic keyhole events in additive manufacturing scenarios.

When aiming to predict highly dynamic events, such as the formation of process instabilities, across large parameter spaces (going towards a universal model), higher fidelity approaches are necessary that incorporate the physical coupling between liquid and vapor phases relevant at the surface irradiated by the laser. This necessitates fully resolving state transitions across the liquid-vapor interface, omitting the need for phenomenological models which explicitly calculate a recoil pressure. While various evaporation-interface-resolved multiphase models exist~\cite{Schlottke2008,Palmore2019,Scapin2020}, applying such methods in the framework of laser-material-interaction has, to date, only been achieved within the volume dilation model of Flint et al.~\cite{Flint2020} and the Mass-of-Fluid (MoF) model~\cite{Zenz2024a}. During laser-induced evaporation, the created vapor is not able to freely expand (even in a vacuum at very short time scales), leading to a local increase in pressure on the vapor side of the interface~\cite{Gusarov2002}. As most simulation models of laser material processing rely on the incompressibility assumption, interface-resolved evaporation would lead to unphysically high acceleration of the vapor phase instead of an increase in pressure. Flint et. al overcome this problem by locally (at the evaporating or condensing interface) allowing for a deviation from the incompressibility constraint, through adding a volume dilation source term to the continuity equation~\cite{Flint2020,Flint2022volumeDilation}. In the present MoF model~\cite{Zenz2024a} this problem is overcome by treating all phases as fully compressible and including the gaseous phase in the calculation. In addition to fully resolving state transitions and hence omitting phenomenological recoil pressure models, another degree of liquid-vapor coupling can be incorporated by making evaporation- and condensation rates dependent on the local pressure distribution. This could be achieved by either calculating these rates as functions of difference between vapor pressure and temperature-dependent saturation pressure (cf. Section~\ref{subsec:phasechange}), or by utilizing the local pressure-dependent saturation temperature to calculate superheating or subcooling when calculating temperature-dependent rates. To our knowledge, apart from the present MoF model, no other published framework currently incorporates a coupling of evaporation- and condensation rates with local vapor pressure.

\subsubsection{Pulsed Laser Processing}
In the context of simulating pulsed laser material processing at the meso-scale, the first major category of models available in the literature comprises hybrid models, which combine Molecular Dynamics simulations with the Two-Temperature Model (TTM), which was first introduced by Anisimov et al. in~\cite{Anisimov1974}. MD simulations are employed to calculate interatomic forces induced by rapid heating, phase transitions, and ejection dynamics, resulting in detailed distributions of temperature, pressure, and density within the material. This information is coupled with the TTM, which models the interaction between the electronic subsystem and the vibrational modes of the lattice structure via electron-phonon relaxation mechanisms~\cite{Schafer2002, Foumani2018, Foumani2020}. The scope of these models varies significantly depending on the process being investigated. For instance, some studies incorporate a full description of the laser beam propagation using the Helmholtz wave equation~\cite{Povarnitsyn2015}, while others focus more on the ablation dynamics and treat the laser as a simple heat source~\cite{Wu2018}. However, a key limitation of MD simulations is their high computational cost, which restricts the domain size and typically confines the models to quasi-one-dimensional simulations.

\begin{figure}[h]
\centering
\includegraphics[width=0.99\textwidth]{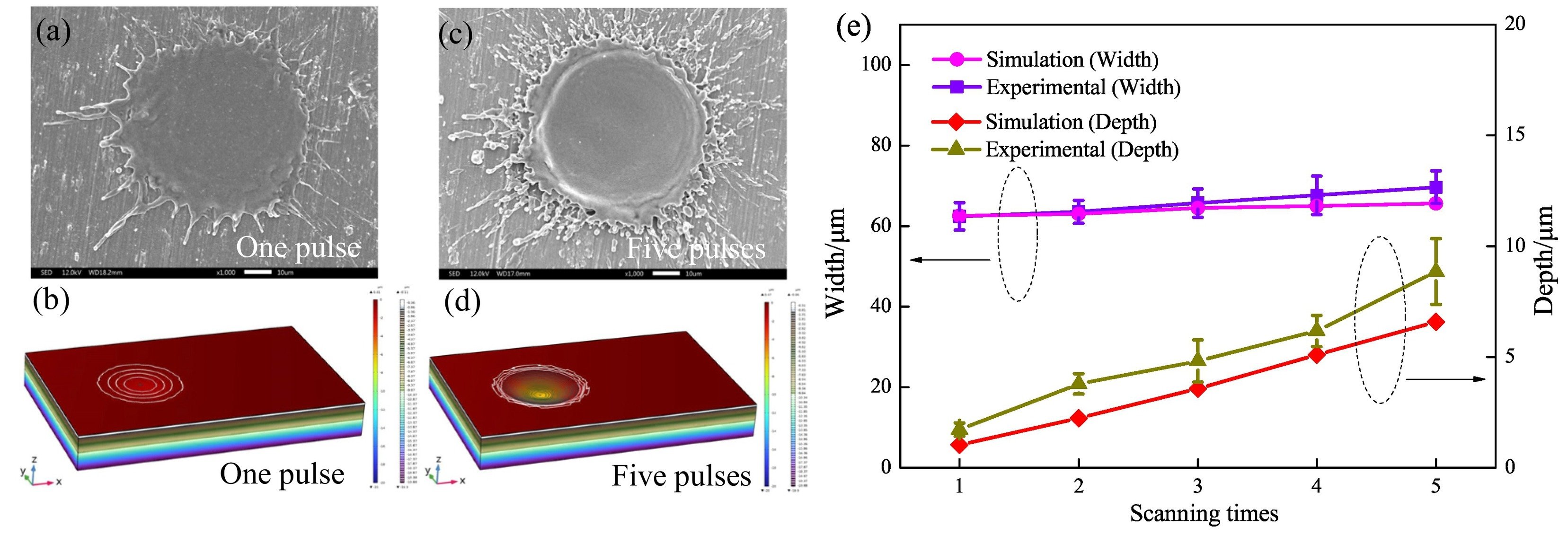}
\caption{Simulation of ablation of Titanium alloy with nanosecond pulses by Zhao et al.~\cite{Zhao2023} showing experimentally obtained (a, c) and simulated (b, d) crater morphologies after one and several pulses, as well as depth and width comparison between experiment and simulation (e) (Reprinted from~\cite{Zhao2023}, J. Manuf. Processes, Vol \textbf{99}, Zhao et al., Numerical simulation and experimental analysis on nanosecond laser ablation of titanium alloy, 138--151, Copyright (2023), with permission from Elsevier)}\label{fig:intro:Zhao2023}
\end{figure}

The second category of simulation models involves hydrodynamic approaches, which efficiently calculate energy, mass, and momentum transport on larger spatial scales. In pulsed laser processing with temporal scales shorter than a picosecond, these models often begin by coupling a laser source and a TTM to the hydrodynamic calculations~\cite{Christensen2008, Feng2024}. Depending on the material, additional mechanisms, such as multiphoton absorption and electron-hole dynamics in semiconductors and dielectrics, may be included~\cite{vanDriel1987, Bulgakova2004, Mao1987}. For longer pulse durations, the TTM can be omitted, assuming direct thermal coupling of the laser energy to the lattice. This simplification allows to model the heat input in a similar way as in continuous wave applications, with added non-linear terms in cases of high-fluence applications~\cite{Zhao2023}. An example of such an approach is shown in Figure~\ref{fig:intro:Zhao2023}.
\begin{figure}[h]
\centering
\includegraphics[width=0.8\textwidth]{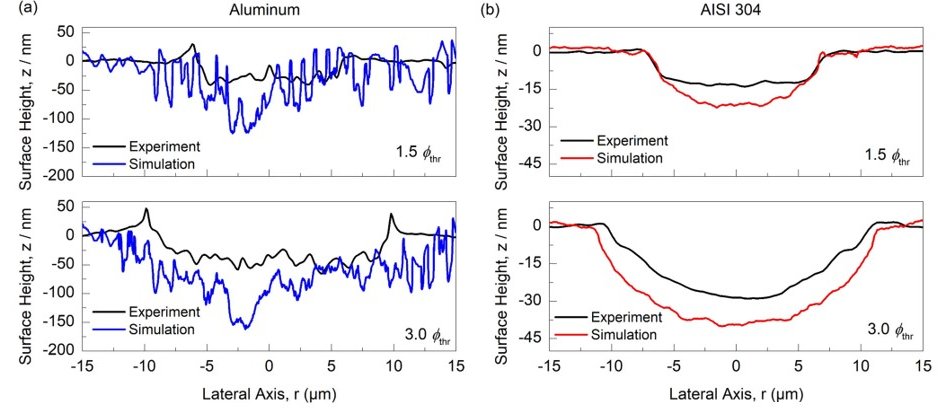}
\caption{Simulation of ablation of aluminum and steel with femtosecond laser pulses at 1.5 and 3 times threshold fluence by Thomae et al.~\cite{Thomae2025}. The simulated crater geometries are compared to the respective experiments of~\cite{Winter2021} (Reprinted and cropped, showing a part of the original Figure, from~\cite{Thomae2025}, Copyright (2025) under Creative Commons BY 4.0 license)}\label{fig:intro:Thomae2025}
\end{figure}

While several models exist in this category, they often omit gaseous phases or treat evaporation as a simple material removal criterion~\cite{Moser2018}, effectively making them CP models. Additionally, due to the characteristics of most pulsed processes, these models frequently adopt a simplified two-dimensional (2D) framework, assuming rotationally symmetric laser beams. A recent example of a multiphase model applied in 2D for the simulation of ultrashort pulse ablation of aluminum and steel is the work of Thomae et al.~\cite{Thomae2025}, for which the simulated crater geometries are compared to experimental results in Figure~\ref{fig:intro:Thomae2025}. The exclusion of the third dimension imposes limitations on accurately representing the processed geometry, such as the borehole topology.

\FloatBarrier
\section{Model Description}\label{sec:model}
Laser material processing poses a coupled multiphysical problem. When formulating a model, it is important to build on general physical  principles and to limit the number of simplifications and assumptions as much as possible, especially within the core model equations. Therefore, we start with the basic principles of conservation of mass, momentum and energy. As laser material processing typically involves a number of segregated and dispersed phases, with the possibility of one phase converting into another via phase changes, we employ a mixture formulation, i.e., phases locally (in space) share a common velocity, pressure and temperature. This allows for the conservation of momentum to be formulated for the mixture alone, greatly reducing the complexity of the model and omitting the need for inter-phase momentum coupling. The error introduced via this assumption reduces with increased spatial resolution (as opposed to other assumptions, such as incompressibility, which will have the same effect regardless of the chosen spatial or temporal discretization). While assuming a mixture temperature, the conservation of mass and energy is formulated separately for each phase, allowing, e.g., for the introduction of non-equilibrium conditions in phase change scenarios. Additionally, we introduce a two temperature formulation whereby we can distinguish between electron and lattice subsystem temperature locally, which can be in thermal non-equilibrium on very short time scales.

In the following Sections we formulate a mathematical model, while aiming to present it in a way that is independent of the chosen numerical solution method or discretization procedure. Then, we elaborate how these governing equations can be solved using a pressure-based segregated Finite Volume Method. The following exposition is mainly based on our previous work related to the Mass-of-Fluid (MoF) method~\cite{Zenz2024a}. Additional model aspects relevant on ultrashort time scales, such as the Two-Temperature-Model (TTM) or non-linear laser-material-interaction phenomena are based on our previous work on modeling of ultra-short pulsed laser processing~\cite{Tatra2016,Florian2024}. Further references are provided where applicable.

\subsection{Conservation of Mass}\label{subsec:massconserv}
Continuity of the mixture of phases is ensured through
\begin{equation} \label{eq:globalContinuity}
    \frac{\partial \rho}{\partial t} + \nabla \cdot \left( \boldsymbol{u} \rho \right)=0,
\end{equation}
where $\rho$ and $\boldsymbol{u}$ denote the mass density and velocity of the mixture of phases. Considering that, in laser material processing, we usually encounter multiple phases undergoing various phase changes, we formulate an additional continuity equation for each phase $i$ following the MoF approach, reading
\begin{equation} \label{eq:alphaEqn}
    \frac{\partial \rho_{i}}{\partial t}+\nabla \cdot \left( \boldsymbol{u} \rho_{i} \right)=S_{M,i},
\end{equation}
with $\rho_{i}$ being the local mass density of phase $i$, defined as $\rho_{i}=m_{i,CV}/V_{i,CV}$, with $m_{i,CV}$ being the mass of phase $i$ encompassed in an arbitrary control volume of size $V_{CV}$. The term $S_{M,i}$ contains any changes in $\rho_{i}$ due to phase changes. For any given means of phase change, we always consider a phase pair $i,j$ exchanging mass via phase change. The gain in phase $i$ is exactly matching the loss in phase $j$, hence $\sum_{i}S_{M,i}=0$. Furthermore, we define $\rho=\sum_{i}\rho_{i}$, and hence global continuity, Eq.~\eqref{eq:globalContinuity}, is always satisfied by Eq.~\eqref{eq:alphaEqn}. A detailed description of $S_{M,i}$ is provided in Section~\ref{subsec:phasechange}.

\subsection{Conservation of Momentum}\label{subsec:momentumconserv}
Conservation of momentum for the mixture of phases is modeled via the Navier-Stokes equations in the form
\begin{equation}
    \frac{\partial \left( \rho \boldsymbol{u} \right)}{\partial t} + \nabla \cdot \left( \rho \boldsymbol{u} \boldsymbol{u} \right) = - \nabla p
    + \nabla \cdot \boldsymbol{\tau}
    - \boldsymbol{S}_{B}
    + \boldsymbol{S}_{S} 
    + \boldsymbol{S}_{D} \label{eq:NSE},
\end{equation}
with $p$ denoting pressure, $\boldsymbol{\tau}$ being the viscous stress tensor and $\boldsymbol{S}_{B}$, $\boldsymbol{S}_{S}$, and $\boldsymbol{S}_{D}$ containing sources of momentum due to gravity, surface tension, and restriction of movement in the mushy zone and solid regions, respectively.

Gravitational body forces are modeled as
\begin{equation}
    \boldsymbol{S}_{B} = \boldsymbol{g}\cdot\boldsymbol{h} \nabla\rho,
\end{equation}
with $\boldsymbol{g}$ denoting gravitational acceleration and $\boldsymbol{h}$ being the location vector with respect to an arbitrary reference point in space.

Surface tension forces are formulated as volumetric forces via Brackbill's Continuum Surface Force Model~\cite{Brackbill1992} leading to
\begin{equation}\label{eq:surftensionforce}
    \boldsymbol{S}_{S} = \sum_{i,j \forall i < j}\nabla \cdot \left[ \sigma_{i,j} \left( \alpha_{j}\nabla \alpha_{i}-\alpha_{i} \nabla \alpha_{j} \right) \right],
\end{equation}
where the volume fraction $\alpha_{i}$ has been introduced, denoting the fraction of volume locally occupied by phase $i$ and notably fulfilling the requirement $\sum_{i}\alpha_{i}=1$. Here, $\sigma_{i,j}$ is the surface tension at the interface of phases $i$ and $j$. For a given phase pair, this quantity can, if for example assuming non-polar solid surfaces, be calculated as~\cite{Girifalco1957} $\sigma_{i,j}=\sigma_{i}+\sigma_{j}-2\sqrt{\sigma_{i}\sigma_{j}}$, with $\sigma_{i}$ being the surface energy (i.e., the surface tension at a phase-vacuum interface) of phase $i$. It is worth noting that Marangoni forces are already included in Eq.~\eqref{eq:surftensionforce} as $\sigma_{i,j}=f(T)$. It is possible to rearrange Eq.~\eqref{eq:surftensionforce} to recover the individual contributions due to curvature, Marangoni forces and triple line forces, as for example shown in~\cite{Ruiz-Gutierrez2024}.

Furthermore, $\boldsymbol{S}_{D}$ is formulated using the widely-used Darcy porosity model as~\cite{Voller1987}
\begin{equation}
    \boldsymbol{S}_{D}=-\frac{\mu}{A_{perm}}\frac{\alpha_{s}^{2}}{\left( 1-\alpha_{s} \right)^{3}+\delta}\boldsymbol{u},
\end{equation}
with $\alpha_{s}$, $\mu$ and $A_{perm}$ being the volume fraction of solid material, the viscosity of the mixture, and the mushy zone permeability, respectively. The latter can be calculated from the material's secondary dendrite arm spacing, $\lambda_{2}$, via the Carman-Kozeny approximation~\cite{Dantzig2016} as $A_{perm} = \lambda_{2}^{2}/180$, leading to values around $A_{perm} \approx 10^{-13}$ as a reasonable estimate. The small numerical constant $\delta$ prevents division by zero when $\alpha_{s}=1$, where any flow is blocked and $|\boldsymbol{u}|\rightarrow 0$. In practice, it should be checked that the choice of $\delta$ value does not affect the solution (different values may be needed for a given value of $A_{perm}$, the employed spatial discretization, etc.).

Any additional, process-specific momentum sources, such as a magnetically introduced Lorentz force, if for example magnetic weld support is of interest, could be added to Eq.~\eqref{eq:NSE}. Considering the aim to formulate a universally applicable model, the above formulation contains the most relevant momentum sources at play in a very large number of laser-material processing scenarios.

Notably, evaporation-induced recoil pressure is not explicitly included in Eq.~\eqref{eq:NSE} as a source term. The mixture governed by Eq.~\eqref{eq:NSE} fills the entire domain of interest and consists of all phases included in the model, i.e., the material(s) to be processed in their solid, liquid, gaseous, and potentially supercritical or decomposed states, as well as any other phases such as ambient gas, shielding gas, or process fluids such as lubricant or water. Therefore, no inter-phase momentum coupling sources such as a phenomenological recoil pressure have to be accounted for: Any local change in density results in a change in pressure if expansion or compression cannot occur freely (e.g., at an interface). A typical example is the drastic change in density when liquid undergoes phase change via evaporation, but even rapid melting can issue a recoil pressure between the melt and the solid material in ultrashort pulsed processing (as shown in~\cite{Florian2024}). 

Thermophysical properties of the mixture, e.g., its thermal conductivity or viscosity, are calculated as a weighted average of the individual phases' properties. Hence, to calculate a property $\phi$, the most straightforward method is via a volume-weighted arithmetic mean, $\phi=\sum_{i}\alpha_{i}\phi_{i}$, where $\phi_{i}$ is the corresponding property of phase $i$. However, depending on the mixture components, their physical distribution, i.e., segregated or dispersed (either of these behaviors can be achieved via employment or lack of surface tension forces and interface compression, cf. Section \ref{sec:num:adv}), the underlying physical phenomena and feature resolution, other averaging methods are necessary, e.g., a harmonic average for the thermal conductivity at the workpiece-air interface to limit heat flux.

There is still the problem of obtaining volume fractions of unitary sum for a given distribution of $\rho_{i}$. This is handled via a procedure called phase-norming, in which the pressure difference $\Delta p$ locally necessary to fit a given distribution of $\rho_{i}$ inside a control volume is found. For the general case of phases with different bulk moduli $K_{i}$ or even different equations of state (EoS), this problem requires an iterative solution procedure.

\begin{figure}[h]
\centering
\includegraphics[width=0.99\textwidth]{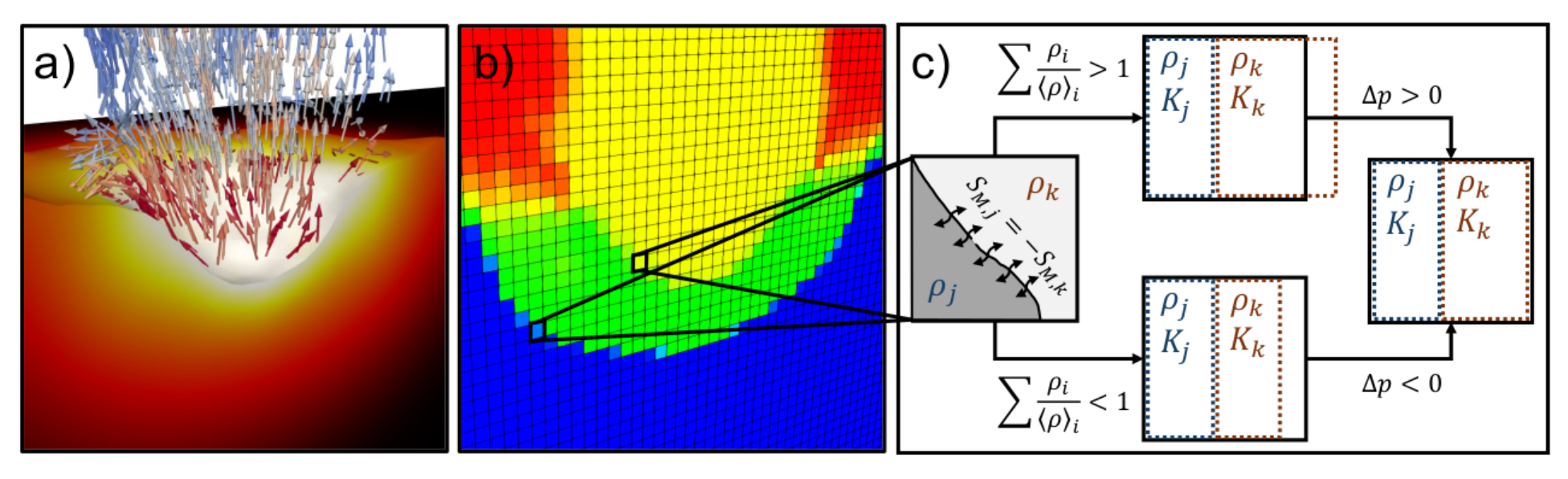}
\caption{Illustration of MoF approach at the example of phase change: a) Exemplary laser-induced phase change process. b) Phase distribution in control volumes. c) Schematic illustration of phase transitions via exchanges of $\rho_{i}$ leading to change in volumetric phase distributions; either compression ($\Delta p>0$) or expansion ($\Delta p<0$) is necessary for given resulting $\rho_{i}$, obtained via phase-norming and each phases' equation of state}\label{fig:MoF}
\end{figure}

A schematic overview of the key components of the MoF approach is provided in Figure~\ref{fig:MoF} for the example of phase transitions, where a mass-conservative change in distribution of $\rho_{i}$ during arbitrary phase change between phases $j$ and $k$ generally (if $\langle\rho\rangle_{j} \neq \langle\rho\rangle_{k}$) leads to a change in the phases' local volume fraction. Here,  $\langle\rho\rangle_{i}$ denotes the material density -- the (generally temperature-dependent but otherwise constant) material property -- of phase $i$. Figure~\ref{fig:MoF}c schematically illustrates the procedure of phase-norming. Note that convective mass transport (via the term $\nabla \cdot \left(\boldsymbol{u} \rho_{i} \right)$ in Eq.~\eqref{eq:alphaEqn}) or thermal expansion or contraction (yielding a change in $\langle \rho \rangle_{i}$) also lead to changes in volumetric phase distributions, and hence expansion or contraction. These phenomena are not depicted in Figure~\ref{fig:MoF} for simplicity.

\subsection{Conservation of Energy}\label{subsec:energyconserv}

Conservation of the energy $H=\rho h$ of the mixture of phases can be formulated as 
\begin{equation}\label{eq:energyEqn}
    \frac{\partial (\rho h)}{\partial t} + \nabla \cdot \left( \boldsymbol{u} \rho h \right) +  \frac{\partial \left( \rho k \right) }{\partial t} + \nabla \cdot \left( \rho \boldsymbol{u} k \right) - \frac{\partial p}{\partial t} = \nabla \cdot \left( \lambda \nabla T \right) + G (T_{e} - T).
\end{equation}
Here, $h$, $T$, $k$ and $\lambda$ denote specific energy, (lattice) temperature, (lattice) kinetic energy and thermal conductivity of the mixture, respectively. The last term on the right hand side denotes the energy coupling of the lattice and electronic subsystem, with $T_{e}$ denoting electron average kinetic energy and the coupling factor $G=C_{e}/\tau_{e}$ being the ratio of electronic heat capacity, $C_{e}$, and characteristic electron-phonon relaxation time, $\tau_{e}$. An additional conservation equation is formulated for the electronic subsystem\footnote{If the characteristic time scale of the process, $\tau_{ch}$ (e.g., the pulse duration) is much greater than the electron-phonon coupling time, $\tau_{ch}>>\tau_{e}$, the lattice and electron system can be regarded as in equilibrium. Consequently, $Q_{abs}$ can be directly put into Eq.~\eqref{eq:energyEqn} and~\eqref{eq:alphaEnergyEqn} instead of the coupling term $G(T_{e}-T)$, and Eq.~\eqref{eq:electronEnergyEqn} does not need to be solved.}, reading
\begin{equation}\label{eq:electronEnergyEqn}
    \frac{\partial (\rho e)}{\partial t} + \nabla \cdot \left( \boldsymbol{u} \rho e \right) = \nabla \cdot \left( \lambda_e \nabla T_e \right) + Q_{abs} - G (T_e - T),
\end{equation}
where $e$, $\lambda_e$ and $Q_{abs}$ denote the specific electron energy, the electron thermal conductivity and the laser energy transferred to the electrons, i.e., absorbed by the material. Following the MoF approach, we formulate a separate conservation equation for each phase $i$, analogously to Eq.~\eqref{eq:alphaEqn}, and hence introduce $H_{i}=\rho_{i}h_{i}$ as the energy of phase $i$, where $\sum_{i}H_{i}=H$. As all phases locally share a common temperature in the here-employed mixture formulation, a viable option is to formulate phase-specific conservation equations for convective transport of energy, and a global equation for heat conduction, and we therefore split Eq.~\eqref{eq:energyEqn} into
\begin{equation}\label{eq:alphaEnergyEqn}
    \frac{\partial H_{i}}{\partial t} + \nabla \cdot \left( \boldsymbol{u} H_{i} \right) +  \frac{\rho_{i}}{\rho} \left( \frac{\partial \left( \rho k \right) }{\partial t} + \nabla \cdot \left( \rho \boldsymbol{u} k \right) - \frac{\partial p}{\partial t} \right) =  \frac{\rho_{i}}{\rho} G (T_e - T) + S_{H,i},
\end{equation}
which ensures conservation of energy for each phase, including a source term $S_{H,i}$ which contains any changes in energy of phase $i$ due to phase change, and fulfils (analogously to $S_{M,i}$, cf. Section~\ref{subsec:massconserv}) the requirement $\sum_{i}S_{H,i}=0$. Additionally, we consider heat conduction within the mixture of phases via
\begin{equation}\label{eq:heatConduction}
    \frac{\partial \left( \rho c_{p} T \right)}{\partial t} = \nabla \cdot \left( \lambda \nabla T \right),
\end{equation}
where $c_{p}$ is the specific heat capacity of the mixture. The above procedure of resolving individual phase energies via a set of $i$ equations and decoupling thermal conduction by formulating it for the mixture is analogously performed for Eq.~\eqref{eq:electronEnergyEqn}.

\subsection{Laser Beam Propagation}\label{subsec:beamprop}

To accurately simulate laser-material processing, it is essential to carefully model the laser beam propagation and its interaction with matter, while capturing all relevant physical phenomena with precision. This starts with the beam propagation through empty space. For a Gaussian beam, the intensity distribution using the paraxial approximation along cylindrical coordinates $r$ and $z$ ($z$ being aligned with and $r$ normal to the optical axis, respectively) is defined as~\cite{HuegelGraf2009}
\begin{equation}\label{eq:gaussIntensity}
    I(r,z) = I_{0}\left( \frac{w_{0}}{w(z)} \right)^{2} \exp{\left(\frac{-2r^{2}}{w(z)^{2}}\right)},
\end{equation}
where $I_0$ is the maximum intensity at the beam's waist $w_0 = w(z = 0)$ calculated as 
\begin{equation}\label{eq:gaussMaxIntensity}
    I_0 =  \frac{2 P }{\pi w_0^2},
\end{equation}
where P is the instantaneous laser power. The beam waist radius, $w(z)$, can be calculated as follows
\begin{equation}\label{eq:gaussWaist}
    w(z)=w_{0}\sqrt{1+\left(\frac{z}{z_{R}}\right)^{2}},
\end{equation}
with the Rayleigh length, $z_{R}$, being
\begin{equation}\label{eq:raileighLength}
    z_{R}=\frac{n\pi w_{0}^{2}}{M^{2}\lambda_{0}},
\end{equation}
where we introduced the vacuum wavelength, $\lambda_{0}$, the refractive index, $n$, and a dimensionless quality factor $M^{2} \geq 1$. From this we can calculate the local direction of propagation, i.e., a unit vector normal to the propagating wavefront as
\begin{equation}\label{eq:rayf}
    \boldsymbol{r}(r,z) = \frac{\cos(\beta) \boldsymbol{e}_{z} + \sin(\beta) \boldsymbol{e}_{r}}{\left| \cos(\beta) \boldsymbol{e}_{z} + \sin(\beta) \boldsymbol{e}_{r} \right|},
\end{equation}
with 
\begin{equation}
    \beta(r,z) = \arcsin\left(\frac{r}{z\left(1+\left( \frac{z_{R}}{z} \right)^{2} \right)}\right).
\end{equation}

The above exposition shows the simple example of a Gaussian intensity distribution, but in principle any type of intensity distribution can be used. If a closed analytical description for the beam is available, the procedure is analogous to the one for a Gauss beam. If the intensity distribution is recovered from a set of measurements of a real laser, the analytical expression in Eq.~\eqref{eq:gaussIntensity} is replaced by measured data, and the remaining quantities, such as $w$, $z_{R}$ or $\boldsymbol{r}$ need to be determined numerically from the available dataset.

\subsection{Laser-Material Interaction}\label{subsec:lasermatinter}

Up to this point, the above elucidation only considers laser beam propagation through empty space, interaction with material in the laser beam's path is of particular interest in laser material processing. To this end, the amount of absorbed and reflected energy at each point in space need to be determined. The volumetric flux of laser energy absorbed by matter, $Q_{abs}$, going into Eq.~\eqref{eq:electronEnergyEqn}, is obtained using the Beer-Lambert equation as
\begin{equation}\label{eq:beerLambert1}
    Q_{abs} = \zeta \cdot I(\boldsymbol{r}),
\end{equation}
where $I(\boldsymbol{r})$ and $\zeta$ are the intensity along the laser propagation direction and the absorption coefficient, respectively. The absorption coefficient, $\zeta$, is calculated from Eq.~\eqref{eq:absCoeff}, consisting of linear, multi-photon, plasma and additional absorption coefficients,
\begin{equation}\label{eq:absCoeff}
    \zeta = \zeta_{linear} + \zeta_{MP} + \zeta_{plasma} + \zeta_{additional}, 
\end{equation}
with  
\begin{equation}\label{eq:absCoefflinear}
    \zeta_{linear} = \frac{4\pi\kappa}{\lambda_0},
\end{equation} 
\begin{equation}\label{eq:multi-photon}
  \zeta_{MP} = \sum_{m=2}^{\infty} \zeta_m I^{(m-1)}.
\end{equation}

 Here, $\kappa$, $\lambda_0$ and $\zeta_m$ are the extinction coefficient, the wavelength of the incident light in vacuum, and the $m$-photon absorption coefficient, respectively. The values of higher order non-linearity terms needed for multi-photon absorption, $\zeta_{MP}$ for $m>2$, which become important at high laser intensities and in the absorption by semi-conductive or dielectric materials, are usually taken from empirical values found in literature or calculated through specific models, e.g., following the Keldysh theory~\cite{Keldysh1965}. Also, at elevated laser intensities, it is possible to induce the formation of a plasma in gaseous states of matter. The effect of the subsequent plasma-shielding can be accounted for by calculating the ionization grade, the free electron density and the resulting absorption coefficient, $\zeta_{plasma}$, via
  \begin{equation}\label{eq:plasmaAbs}
    \zeta_{plasma} = \frac{n_e^2 q^6}{6\sqrt{3}n_{p}\epsilon_0^3 c \hbar \omega^2 m_e^2}
    \cdot
    [1-\exp{(-\frac{\hbar \omega}{k_B T})}]
    \cdot
    \sqrt{\frac{m_e}{2 \pi k_B T}}
    \cdot \bar{g},
\end{equation}
with a comprehensive list of symbols introduced in Eq.~\eqref{eq:plasmaAbs} being provided in Table~\ref{tab:variables}. The calculation of the plasma fraction using the free electron density is detailed in~\cite{Tatra2016}.

\begin{table}[h!]
    \centering
    \begin{tabular}{ll}
        \hline
        \textbf{Symbol} & \textbf{Description} \\
        \hline
        $n_e$ & Free electron density \\
        $q$ & Electron charge \\
        $n_{p}$ & Real part of the plasma refractive index \\
        $\epsilon_0$ & Permittivity of vacuum \\
        $c$ & Speed of light in vacuum \\
        $\hbar$ & Reduced Planck constant \\
        $\omega$ & Angular laser frequency \\
        $m_e$ & Electron mass \\
        $T$ & Thermal equilibrium temperature \\
        $\bar{g}$ & Gaunt factor \\
        \hline
    \end{tabular}
    \caption{List of symbols in Eq.~\eqref{eq:plasmaAbs}}
    \label{tab:variables}
\end{table}
\FloatBarrier
 Additional non-linear effects $\zeta_{additional}$ include, e.g., the ballistic electron phenomenon calculated in Eq.~\eqref{eq:ballElectrons} \cite{Cheng2016} arising from the interaction of ultrashort, highly intense laser pulses and metal

 \begin{equation}\label{eq:ballElectrons}
    \zeta_{additional} = \sqrt{\frac{2k_BT_e}{m_e}}\cdot t_B,
\end{equation}
where $k_B$, $T_e$, $m_e$ and $t_B$ are the Boltzmann constant, the electron average kinetic energy, the electron mass and the Drude relaxation time, respectively.

The fraction of laser light reflected at phase interfaces, $R$, is calculated via the Fresnel equations,
\begin{align}
 R &= w_{S}R_{S}+w_{P}R_{P}, \label{eq:fresnel3}\\
 R_S &= \left| \frac{cos(\theta)-\underline{n}\sqrt{1-sin^{2}(\theta)/\underline{n}^2}}{cos(\theta)+\underline{n}\sqrt{1-sin^{2}(\theta)/\underline{n}^2}} \right|\label{eq:fresnel1},\\
    R_P &= \left| \frac{\underline{n} cos(\theta)- \sqrt{1-sin^{2}(\theta)/\underline{n}^2}}{\underline{n} cos(\theta)+ \sqrt{1-sin^{2}(\theta)/\underline{n}^2}} \right|,\label{eq:fresnel2}\\
    \theta &= \arccos{(\boldsymbol{n}_{surf}\cdot\boldsymbol{r})},\label{eq:angleIncidence} \\
    \boldsymbol{n}_{surf} &= \frac{\nabla \alpha_{cond}}{|\nabla \alpha_{cond}|}
\end{align}
where $\underline{n}= n + \text{i}\kappa$ is the complex refractive index, comprised of the refractive index $n$ and the extinction coefficient $\kappa$, $\text{i}$ being the imaginary unit and $\theta$ is the local angle of incidence. Furthermore, $\alpha_{cond}$ is the volume fraction of condensed matter and $\boldsymbol{n}_{surf}$ hence represents a unit vector normal to the surface of condensed matter. The weights $w_{S}$ and $w_{P}$ are, in the case of polarized light with polarization direction $\boldsymbol{n}_{pol}$, calculated as
\begin{align}
    w_{S} &= \frac{p_{S}}{p_{S}+p_{P}}, \\
    w_{P} &= \frac{p_{P}}{p_{S}+p_{P}}, \\
    p_{S} &= \left| \boldsymbol{n}_{pol} \cdot \boldsymbol{n}_{normal} \right|, \\
    p_{P} &= \sqrt{1-p_{S}^{2}}, \\
    \boldsymbol{n}_{normal} &= \frac{\boldsymbol{r} \times \boldsymbol{n}_{surf}}{\left| \boldsymbol{r} \times \boldsymbol{n}_{surf} \right|},
\end{align}
with $\boldsymbol{n}_{surf}$ being the unit vector normal to the interface surface. In the case of unpolarized light, $w_{S}=w_{P}=0.5$.\\
To solve Eqs.~\eqref{eq:beerLambert1} to~\eqref{eq:fresnel3}, $\underline{n}$ needs to be determined, whose values usually depend on temperature and can be found in literature or be obtained through measurements. In specific cases, as for the absorption of ultra-short pulses, specific models have been developed to calculate these properties from the material's dielectric function (i.e., the Drude model employed in~\cite{Florian2024}). For more specific applications, such as the simulation of semi-transparent materials, i.e., many glasses and other materials relevant to the laser manufacturing application, such as Si and SiC, refraction at the interfaces needs to be accounted for. This can be done by following Snell's law of refraction, reading
\begin{equation}\label{eq:snell}
    n_1 \sin{\theta_1} = n_2 \sin{\theta_2},
\end{equation}
where $n_i$ and $\theta_i$ are the refractive indices of the materials involved in the investigated interface and the angle of incidence and emergence respectively.

\subsection{Phase Changes}\label{subsec:phasechange}
Within the MoF framework, phase changes are realized via the sources $S_{M,i}$ and $S_{H,i}$ in Eqs.~\eqref{eq:alphaEqn} and~\eqref{eq:alphaEnergyEqn}, respectively. Any mass undergoing a phase transition is removed from the source phase and added to the target phase, together with its corresponding energy. To this point, the framework is extremely flexible and allows for the inclusion of arbitrary physical models to calculate the values of $S_{M,i}$ and $S_{H,i}$. For most laser-material-processing scenarios, accurate modeling of melting, solidification, evaporation and condensation are necessary. We therefore formulate mass transfer rates for melting, solidification, evaporation and condensation, $\dot{\rho}_{i,m}$, $\dot{\rho}_{i,s}$, $\dot{\rho}_{i,e}$ and $\dot{\rho}_{i,c}$ from which we can calculate $S_{M,i}=\dot{\rho}_{i,m}+\dot{\rho}_{i,s}+\dot{\rho}_{i,e}+\dot{\rho}_{i,c}$. Depending on the material and the laser specifications, other mechanisms such as sublimation or decomposition could also be relevant, but are not elaborated in detail here. 

\subsubsection{Melting and Solidification}
Considering phase transition via melting and solidification between a solid and liquid phase, denoted $s$, $l$, we define the fraction of superheated solid, $\Delta_{s} \alpha_{s}$, and subcooled liquid, $\Delta_{s} \alpha_{l}$, as
\begin{align}
    \Delta_{s} \alpha_{s} &= \frac{H_{s}-\rho_{s}h_{s}(T_{solidus})}{L_{fus,m}},\label{eq:superheatedSolid}\\
    \Delta_{s} \alpha_{l} &= \frac{\rho_{l}h_{l}(T_{liquidus})-H_{l}}{L_{fus,s}},\label{eq:subcooledLiquid}
\end{align}
where we account for the general case of a solidification interval between liquidus and solidus temperatures, $T_{liquidus}$ and $T_{solidus}$. In the simple case of a pure element, those coincide to $T_{melting}$. Furthermore, it is possible to account for non-equilibrium conditions by specifying $T_{solidus}$ as a function of liquid fraction (or cooling rate), or to specify different values for $T_{liquidus}$ and $T_{solidus}$ for melting and solidification. For example, the implementation of Gulliver-Scheil solidification within this framework was recently demonstrated and validated in~\cite{Zenz2024b}. Furthermore,
\begin{equation}
    L_{fus,m}=\rho_{s}\left(h_{l}(T_{liquidus})-h_{s}(T_{solidus})\right)
\end{equation}
and
\begin{equation}
    L_{fus,s}=\rho_{l}\left(h_{l}(T_{liquidus})-h_{s}(T_{solidus})\right)
\end{equation}
are the latent heat of fusion required for melting and solidification, respectively. Depending on the amount of superheated solid or subcooled liquid, we can formulate a rate of melting and solidification, respectively as
\begin{align}
    \dot{\rho}_{l,m} = -\dot{\rho}_{s,m} &= \frac{\rho_{s}\Delta_{s} \alpha_{s}}{\tau_{prop}},\label{eq:meltMassRate}\\
    \dot{\rho}_{s,s} = -\dot{\rho}_{l,s} &= \frac{\rho_{l}\Delta_{s} \alpha_{l}}{\tau_{prop}}.\label{eq:solidMassRate}
\end{align}

Here, a propagation time $\tau_{prop}$, is introduced that determines the speed at which the actual phase transition during melting or solidification can occur (which is different from the macroscopic movement of the liquid-solid interface which is determined by the local thermal conditions and the material properties). This propagation time could be chosen related to the speed of sound, to provide an upper limit, which becomes relevant only in extreme scenarios. For most scenarios, this is irrelevant, and the entire amount of subcooled liquid or superheated solid determined by Eqs.~\eqref{eq:subcooledLiquid} and~\eqref{eq:superheatedSolid} will immediately undergo phase change. Any non-equilibrium conditions during solidification are rather a result of a non-constant value of $T_{solidus}$, and melting will rather be limited by thermal conductivity, heat capacity and latent heat of the material than by Eq.~\eqref{eq:meltMassRate}. Therefore, any fraction of superheated or subcooled material will immediately be transferred to the respective target phase in an explicit update during the numerical solution algorithm (cf. Section~\ref{subsec:numerics}).

\subsubsection{Evaporation and Condensation}
Evaporation and condensation is modeled as pressure-driven process via the Hertz-Knudsen equation, where we define a surface mass flux across the liquid-gas interface, $\dot{m}''_{ec}$, as
\begin{equation}\label{eq:HertzKnudsen}
    \dot{m}''_{ec}=\sqrt{\frac{M}{2 \pi R T}}\left(p_{sat}-p\right).
\end{equation}
Here, $M$, $R$, $T$ and $p$ denote molar mass, universal gas constant, temperature and pressure, respectively. Furthermore, $p_{sat}$ is the saturation pressure which we calculate using the Clausius Clapeyron relation, governed by
\begin{equation}\label{eq:clausiusClapeyron}
    \frac{dp}{dT}=\frac{pL_{vap,eff}}{T^{2}R},
\end{equation}
with $L_{vap,eff}$ being the effective latent heat of vaporization, which we calculate from its nominal value, $L_{vap}$, using the Watson correction~\cite{Watson1943}, as
\begin{equation}
    L_{vap,eff}=L_{vap} \left( \frac{T_{c}-T}{T_{c}-T_{b}} \right)^{0.38},
\end{equation}
with $T_c$ and $T_{b}$ being the critical and boiling temperature, respectively. Note that Eq.~\eqref{eq:clausiusClapeyron} only possesses a closed-form analytical integral solution for a constant latent heat, i.e., for moderate degrees of superheating. In the general case (and especially when approaching the critical temperature), Eq.~\eqref{eq:clausiusClapeyron} needs to be integrated numerically. Depending on the sign of $\dot{m}''_{ec}$, Eq.~\eqref{eq:HertzKnudsen} yields either evaporation or condensation. A volumetric mass flux is then calculated in the manner of continuum surface flux methods by multiplying with a suitable measure for the local interface density, which is taken as the inverse of the liquid-vapor interface thickness\footnote{In practice, the numerical value of $\xi$ is taken as the Finite Volume cell size $\Delta x$ (or its projection normal to the local liquid-vapor interface in the case of non-square cells), as in~\cite{Palmore2019}. Tests on the benchmark problem of Section~\ref{subsec:cunningham} have shown negligible difference in results when instead using the formulation $1/\xi=\alpha_{l}|\nabla\alpha_{l}|+(1-\alpha_{l})/\Delta x$, with $|\nabla\alpha_{l}|$ representing another commonly used approximation of the local interface density~\cite{Shang2022}.}, $1/\xi$, leading to
\begin{align}
    \dot{\rho}_{v,e}=-\dot{\rho}_{l,e}&=\rho_{l}\sqrt{\frac{M}{2 \pi R T}}\frac{(p_{sat}-p)^{+}}{\xi \langle\rho\rangle_{l}},\label{eq:evapMassRate}\\
    \dot{\rho}_{l,c}=-\dot{\rho}_{v,c}&=\alpha_{cond}\rho_{v}\sqrt{\frac{M}{2 \pi R T}}\frac{(p_{sat}-p)^{-}}{\xi \langle\rho\rangle_{v}}.\label{eq:condMassRate}
\end{align}
The quantity $\langle\rho\rangle_{i}$ is the temperature-dependent density of phase $i$ (cf. \ref{subsec:momentumconserv}), which can be seen as its material property, not to be confused with the partial density $\rho_{i}$, and $\alpha_{cond}$ is the volume fraction of condensed matter, introduced here to promote condensation at condensed surfaces (as homogeneous nucleation would require substantial subcooling in reality). Furthermore, the superscripts $+$ and $-$ imply
\begin{align}
    x^{+} &= 
    \begin{cases}
        |x| &\forall x \geq 0 \\
        0 &\forall x < 0
    \end{cases},\\
    x^{-} &= 
    \begin{cases}
        0 &\forall x \geq 0 \\
        |x| &\forall x < 0.
    \end{cases}
\end{align}

\FloatBarrier
\subsection{Numerical Methods}\label{subsec:numerics}
In this Section, we briefly present the numerical methods required to solve the model equations laid out in the preceding Sections utilizing a segregated Finite Volume Method. In fact, the underlying model equations could in principle be discretized and solved with other methods such as, e.g., the Finite Element Method. The numerical solution algorithm presented here is implemented using the C++ Finite Volume library \texttt{OpenFOAM}~\cite{Weller1998}. A brief overview of the individual calculation steps performed in each time step of the transient solution procedure is laid out in Table~\ref{tab:alg}. Details regarding individual steps of the solution procedure are explained in the subsequent Sections.

\begin{table*}[htb] 
\centering
\begin{tabular*}{120mm}{@{\extracolsep{\fill}}l l l}
\hline
& \multicolumn{2}{l}{Update time increment $\Delta t \leq \text{min}\left(  \rho c_{p} \Delta x^{2}/\lambda, C_{e} \Delta x^{2}/\lambda_e, \Delta x / |\boldsymbol{u}| \right)$} \\
\hline
\textbf{1} & \multicolumn{2}{l}{\textbf{Advection of mass and energy}} \\
 & 1a & Solve Eq.~\eqref{eq:alphaEqn} with $S_{M,i}=0$ \\
 & 1b & Solve Eq.~\eqref{eq:alphaEnergyEqn} and its electron energy equivalent with $S_{H,i}=0$ \\
 & 1c & Calculate $T$ from resulting value of $H$ \\
 & 1d & Calculate $\alpha_{i}$ from resulting value of $\rho_{i}$ and $T$ and the \\
  & & phases' respective equations of state via phase-norming \\
\hline
\textbf{2} & \multicolumn{2}{l}{\textbf{Beam propagation}} \\
& \multicolumn{2}{l}{\textit{Option A: FVM approach}} \\
 & 2a & Solve coupled steady-state RTE, Eqs.~\eqref{eq:RTE1} and~\eqref{eq:RTE2} iteratively \\
 & 2b & Calculate $Q_{abs,I}$ from $I_{cond}$ \\
 & 2c & Calculate $Q_{abs,MR}$ via ray tracing algorithm using \\
 & 2d & Calculate absorbed laser energy $Q_{abs}=Q_{abs,I}+Q_{abs,MR}$ \\
 & \multicolumn{2}{l}{\textit{Option B: Photonic Parcel approach}} \\
 & 2a & Discretize laser beam into "photonic parcels" \\
 & 2b & Track particle movement on highly parallelized \\
  & & Graphical Processing Unit (Eq.~\eqref{eq:particleVelocity} $\And$  Eq.~\eqref{eq:rayf})\\
 & 2c & Calculate refraction (Eq.~\eqref{eq:snell}) $\And$ reflection (Eq.~\eqref{eq:fresnel3})\\
 & 2d & Keep track of phase, polarization $\And$ residual energy \\
  & & after absorption (Eq.~\eqref{eq:absorptivity}) \\
\hline
\textbf{3} & \multicolumn{2}{l}{\textbf{Heat conduction}} \\
 & 3a & Solve Eq.~\eqref{eq:heatConduction} and its electron equivalent to update $T$ and $T_{e}$ \\
 & 3b & Update $H_{i}$ from new value of $T$ \\
\hline
\textbf{4} & \multicolumn{2}{l}{\textbf{Phase change}} \\
 & 4a & Evaluate Eqs.~\eqref{eq:meltMassRate},~\eqref{eq:solidMassRate},
 ~\eqref{eq:evapMassRate} and~\eqref{eq:condMassRate} \\ 
 & 4b & Calculate $S_{M,i}$ and $S_{H,i}$ and explicitly update $\rho_{i}$ and $H_{i}$ \\
 & 4c & Calculate $T$ from new value for $H$ \\
 & 4d & Calculate $\alpha_{i}$ from resulting value of $\rho_{i}$ and $T$ and the \\
  & & phases' respective equations of state via phase-norming \\
\hline
\textbf{5} & \multicolumn{2}{l}{\textbf{PISO-Loop}} \\
 & 5a & Solve pressure-momentum coupling via compressible \\
  & & PISO-loop (cf.~\cite{Miller2013}) to obtain $\boldsymbol{u}$ and $p$ \\
\hline
\end{tabular*}
\caption{Overview of solution algorithm within one time step of the transient calculation procedure}\label{tab:alg}
\end{table*}

\subsubsection{Phase- and Energy Advection}\label{sec:num:adv}
Eqs.~\eqref{eq:alphaEqn} and~\eqref{eq:alphaEnergyEqn} are solved explicitly on the current velocity field, excluding any source terms due to phase change (these will be accounted for in an explicit update at a later step). To minimize the problem of (artificial) numerical diffusion when solving advection equations on a grid not aligned with the flow direction, an interface compression term is added to the velocity field which is of the form
\begin{equation}\label{eq:interfaceCompression}
    \boldsymbol{u}_{c} = c_{\alpha} \boldsymbol{u} \frac{\nabla \alpha_{cond}}{\left| \nabla \alpha_{cond} \right|}\alpha_{cond}(1-\alpha_{cond}),
\end{equation}
with $\alpha_{cond}$ being the volume fraction of condensed matter, and $c_{\alpha}$ is a constant to set the degree of interface compression. After solving the advection equations, the new distribution of $\alpha_{i}$ is obtained from $\rho_{i}$ via phase-norming (cf. Figure~\ref{fig:MoF}c for a schematic overview of this concept).

\subsubsection{Laser}
For most processes, due to its computational efficiency, the beam propagation and subsequent laser-material interaction is calculated using the Finite Volume Method (FVM). This is appropriate for interactions with opaque materials, where absorption happens in a comparatively small depth near the illuminated material's surface and refraction can be neglected. Polarization and reflections can be accounted for, the latter by using a ray-tracing algorithm after first incidence.

If the specific process does not fulfill the above criteria or some other non-linear mechanism are of interests, a different approach for the beam discretization is employed. Here, refractions at interfaces, phase front propagation and other non-linear optical phenomena (i.e., the Kerr-effect) can be accounted for. This is accomplished by discretizing the beam into "Photonic Parcels" (bundles of multiple photons or discrete energy units $h\nu$)~\cite{Otto2018a} which are then propagated through the domain using a coupled approach based on both FVM and DEM (Discrete Element Method).

\subsubsection*{FVM}

An initial intensity distribution, e.g., Eq.~\eqref{eq:gaussIntensity}, the local curvature in Eq.~\eqref{eq:rayf}, the reflectivity in Eq.~\eqref{eq:fresnel3} and the absorption coefficients in Eq.~\eqref{eq:absCoeff} are calculated and discretized onto a Finite Volume grid. The intensity distribution itself is initially mapped onto a laser inlet patch (boundary patch of the meshed domain), usually located near the surface of the workpiece against light propagation direction, to provide a Dirichlet boundary condition. Its specific location must be chosen depending on the dominance of the reciprocal influence between the incoming laser energy and the processing zone, i.e., vapor  expansion and absorption, shadowing through melt ejection and plasma absorption. 
Subsequently the two coupled steady-state differential Eqs.~\eqref{eq:RTE1} to~\eqref{eq:RTE2} for $I_{g}$ and $I_{cond}$, which represent the laser intensity propagating through gaseous and condensed phases respectively, are solved iteratively. 
\begin{equation}
\begin{split}\label{eq:RTE1}
    \nabla \cdot \left( \boldsymbol{r}I_{g} \right) &= \\ 
    &- \zeta_{g} I_{g} \\
    &- \frac{\left[ \nabla \cdot \left( \boldsymbol{r} \alpha_{cond} \right) \right]^{+}}{\alpha_{g}} I_{g} \\
    &+ \left(1-R\right) \frac{\left[\nabla \cdot \left( \boldsymbol{r}\alpha_{cond} \right) \right]^{-}}{\alpha_{cond}}I_{cond}
    \end{split}    
\end{equation}
\begin{equation}\label{eq:RTE2}
    \begin{split}
    \nabla \cdot \left( \boldsymbol{r}I_{cond} \right) &= \\
    &- \zeta_{cond} I_{cond} \\
    &- \frac{\left[ \nabla \cdot \left( \boldsymbol{r} \alpha_{cond} \right) \right]^{-}}{\alpha_{cond}} I_{cond} \\ 
    &+ \left(1-R\right) \frac{\left[\nabla \cdot \left( \boldsymbol{r}\alpha_{cond} \right) \right]^{+}}{\alpha_{g}}I_{g}
    \end{split}
\end{equation}
The absorptivities (condensed or gas) of the mixture, exemplarily shown for $\zeta_{cond}$, are calculated as
\begin{equation}\label{eq:absorptivity}
    \zeta_{cond} = \frac{\sum_{i}^{N_{cond}} \alpha_{i}\cdot\zeta_{i}}{\sum_{i}^{N_{cond}}\alpha_{i}},
\end{equation}
where $N_{cond}$, $\alpha_{i}$ and $\zeta_{i}$ are the number of condensed phases, the volume fraction and the absorption coefficient of the respective phases, the latter being calculated from Eq.~\eqref{eq:absCoeff}, taking into account the relevant absorption mechanisms for each phase.

This approach essentially solves two coupled steady-state advection equation for the intensity distribution through the domain following the beam's caustic through the air and accounting for absorption, transmission and first reflection. The approximation of steady-state holds, due to the short time-scale involved in beam propagation over small distances compared to fluid-mechanical mass, momentum and energy transport.

Subsequent reflections and further beam attenuation are calculated through a ray-tracing algorithm. This algorithm discretizes the reflected beam into an array of rays, which are propagated in a straight line from an angle of emergence calculated at first reflection. This angle, and all angles of emergence during subsequent reflections, is calculated taking into account the local surface curvature of the reflecting interface, defined at the iso-surface defined via $\alpha_{cond}=0.5$ and the local angle of incidence defined via Eq.~\eqref{eq:angleIncidence}. At each new incidence absorption is calculated and added to $Q_{abs}$ and this approach is continued until a certain number of reflections is reached, all beams have left the domain or the beam's energy drops below a certain threshold, where it can be neglected.

\subsubsection*{Photonic Parcels}

\begin{figure}[h]
\centering
\includegraphics[width=0.99\textwidth]{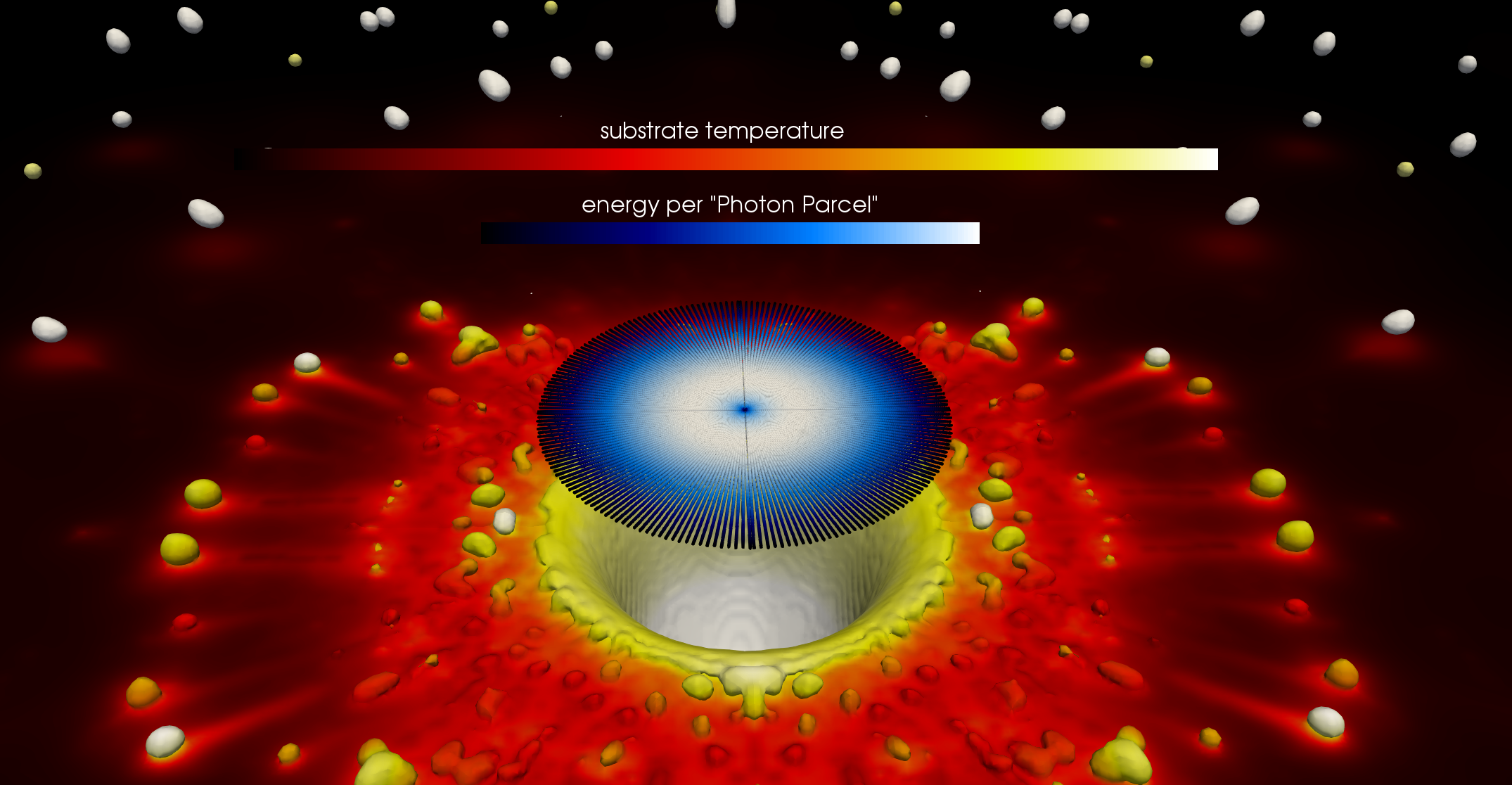}
\caption{Snapshot of propagating Photonic Parcels inbound on substrate during a glass through via drilling process. They represent a Gaussian laser beam, although their energy is lower closer to the center. This is due to the increased particle density in that region}\label{fig:photonCloudPropagation}
\end{figure}

When using the coupled FVM-DEM approach to model beam propagation, the starting point for the calculation is again an intensity distribution at a specified location and the information on  absorptivity and reflectivity of the material. The beam's cross-section is discretized through Lagrangian particles, as seen in Figure~\ref{fig:photonCloudPropagation}, which are modeled as bundles of photons with a specific energy (determined by the local intensity distribution and parcel density, e.g. given in Eq.~\eqref{eq:gaussIntensity}), direction (given by $\boldsymbol{r}$ in Eq.~\eqref{eq:rayf}), speed (given by speed of light in the respective medium) and polarization and phase of propagating electro-magnetic wave. They are then propagated through the meshed domain (substrate and gaseous phases), taking care to allow for proper sub-time-stepping, as to fulfill the condition in Eq.~\eqref{eq:lagrangianCourant}, which arises from the Courant number.
\begin{equation}\label{eq:lagrangianCourant}
    \Delta t_{lagrangian}  < \frac{\Delta l}{|\boldsymbol{v}|}
\end{equation}
Here, $\Delta l$ is the path length of the particle and $|\boldsymbol{v}|$ the speed of light in the considered medium inside a mesh-element. 

 At each spatial point, the volumetric absorbed energy is calculated using the Beer-Lambert law (Eq.~\eqref{eq:beerLambert1}), and deflection due to refractive index gradients is determined using Snell’s law (Eq.~\eqref{eq:snell}). Subsequently, the particle’s energy and velocity are updated, accounting for speed variations caused by differences in refractive indices (Eq.~\eqref{eq:particleVelocity}). 
\begin{equation}\label{eq:particleVelocity}
    \boldsymbol{v} = \frac{c}{n_i} \cdot \boldsymbol{r}
\end{equation}
with c, $n_i$ and $\boldsymbol{r}$ being the speed of light in vacuum, the refractive index of the phase dominant in volume element i and the local particle direction.

When a particle then reaches an interface, determined by the change of the dominant phases' volume fraction in a finite volume cell, refraction using Snell's law and reflection through the Fresnel equation (c.f. Eq. \ref{eq:fresnel3}) are calculated. The original incoming particle is the one being reflected, while the refraction triggers the creation of a new particle, which propagates onward in direction of refraction.

During every time-step of the FVM solution $\Delta t_{FVM}$ all the parcels are propagated until they either leave the domain through the boundaries or have lost enough energy through absorption to be ignored, assuming the $\Delta t_{FVM}$ is large enough compared to the particle velocity and the geometrical constraints arising from the meshed domain size, particle injection location and number of reflections. The absorbed energy at each location is then transferred to the finite volume mesh and the rest of the simulation continues from there according to Table~\ref{tab:alg}.

As the number of individual particles needed for such simulations typically exceeds the order of $10^6$, standard calculation practices on commercial CPU's become unfeasible.    To address this issue, the code for the beam propagation was recently rewritten in CUDA in order to be highly parallelized on modern Graphical Processing Units. For more details on the implementation we refer the reader to~\cite{Heisler2023,Genest2024}.

\subsubsection{Heat Conduction}
As outlined in Section~\ref{subsec:energyconserv}, heat conduction is a temperature-driven process and hence formulated for the mixture of phases sharing a local temperature $T$. While obtaining a temperature $T$ from a given energy $H$ for a single phase and a constant value of $c_{p}$ could be trivially done via the relationship $T =H/(\rho \cdot c_p)$, a direct calculation is not possible for a mixture of $N$ phases $i$ and temperature-dependent values of $c_{p,i}$. In this case, to obtain the temperature $T$ corresponding to a given value of $H$, an iterative procedure is employed, where within each iteration $k$, the energy $H^{(k)}$ corresponding to temperature $T^{(k)}$ is calculated via
\begin{equation}\label{eq:hFromTStart}
    H^{(k)} = \sum_{i}^{N} \rho_{i} \cdot h_{i}(T^{(k)}),
\end{equation}
where $h_{i}(T)$ yields the specific energy of phase $i$ at temperature $T$, taking into account non-constant values of $c_{p,i}$ and latent heats, implemented as a lookup function that is iteratively constructed from the differential relation $dh=c_{p}(T)dT$. Depending on the sign of the error $H - H^{(k)}$, $T^{(k+1)}$ is increased or decreased and Eq.~\eqref{eq:hFromTStart} is solved again.
This process is repeated until the condition $|H - H^{(k)}| < \epsilon$ is satisfied, where $\epsilon \ll 1$ represents a small convergence tolerance, typically close to machine precision. The above procedure of obtaining $T$ is performed at any step in the solution algorithm where $H$ changes, as material properties are usually temperature-dependent and must be kept up to date throughout the simulation. Furthermore, $T$ is obtained prior to solving the heat conduction in Eq.~\eqref{eq:heatConduction}. After obtaining a new value of $T$ by solving Eq.~\eqref{eq:heatConduction}, the values of $H_{i}$ are updated via $H_{i}=\rho_{i} \cdot c_{p,i} \cdot T$.

\subsubsection{Phase Change}

The amount of mass transfer between solid phase $s$ and liquid phase $l$ due to melting or solidification is calculated as
\begin{equation}\label{eq:meltNum}
    \Delta \rho_{l,m} = -\Delta \rho_{s} =  \rho_{s} \Delta_{s} \alpha_{s},
\end{equation}
\begin{equation}\label{eq:solidiNum}
    \Delta \rho_{s,s} = - \Delta \rho_{l,s} =  \rho_{l} \Delta_{s} \alpha_{l},
\end{equation}
with $\Delta \rho_{i,m}$ and $\Delta \rho_{i,s}$ being zero for all phases not involved in the respective phase change (the same holds true analogously for evaporation and condensation).

The amount of mass transfer between liquid phase $l$ and vapor phase $v$ due to evaporation or condensation is calculated as follows. The rate equations~\eqref{eq:evapMassRate} are integrated over a time step via
\begin{equation}\label{eq:evapNum}
    \Delta\rho_{v,e} = -\Delta \rho_{l,e} =
    \rho_{l}\left(1-\exp\left(-\frac{\Delta t c_{1} (p_{sat}-p)}{1+\frac{|c_{1}|\Delta t} {\Psi}}\right)\right),
\end{equation}
\begin{equation}\label{eq:condNum}
    \Delta\rho_{l,c} = -\Delta \rho_{v,c} =
    \rho_{v}\left(1-\exp\left(-\frac{\Delta t c_{1} (p_{sat}-p)}{1+\frac{|c_{1}|\Delta t}{\Psi}}\right)\right),
\end{equation}
with
\begin{equation} \label{eq:c1}
    c_{1} = \begin{cases}
        &  \sqrt{\frac{M}{2 \pi R T}}\frac{1}{\xi \langle\rho\rangle_{l}} \forall (p_{sat}>p) \\
          & -\sqrt{\frac{M}{2 \pi R T}}\frac{\alpha_{cond}}{\xi \langle\rho\rangle_{v}} \forall (p_{sat}<p) \\
    \end{cases}       
\end{equation}
and $\Psi$ being the compressibility of the mixture.
The change of $\rho_{i}$ due to the $S_{M,i}$ in Eq.~\eqref{eq:alphaEqn} is accounted for explicitly to $\rho_{i}^{*}=\rho_{i}+\Delta \rho_{i,m} + \Delta \rho_{i,s} + \Delta \rho_{i,e}+ \Delta \rho_{i,c}$. Then, $\rho_{i}^{*}$ corresponds to the value of $\rho_{i}$ after completion of step 4b in Table~\ref{tab:alg}. When transferring mass between phases, the respective energy contained in said mass also needs to be transferred, hence accounting for the effect of $S_{H,i}$ in Eq.~\eqref{eq:alphaEnergyEqn}. Consequently, the value of $H_{i}$ is updated to $H_{i}^{*}=H_{i}+\Delta H_{i,m} + \Delta H_{i,s} + \Delta H_{i,e}+ \Delta H_{i,c}$ during step 4b in Table~\ref{tab:alg}. Here,
\begin{align}
    \Delta H_{i,m} &= \frac{\Delta\rho_{i,m}}{\rho_{s}}\ h_{s}, \\
    \Delta H_{i,s} &=  \frac{\Delta\rho_{i,s}}{\rho_{l}}\ h_{l}, \\
    \Delta H_{i,e} &=  \frac{\Delta\rho_{i,e}}{\rho_{l}}\ h_{l}, \\
    \Delta H_{i,c} &=  \frac{\Delta\rho_{i,c}}{\rho_{v}}\ h_{v},
\end{align}
where notably all right hand side values refer to the values prior to step 4b.

\subsubsection{PISO-Loop}

Within the here-employed segregated solution algorithm, pressure-velocity coupling needs to be established, i.e., obtaining a set of $\boldsymbol{u}$ and $p$ fulfilling Eqs.~\eqref{eq:NSE} and~\eqref{eq:globalContinuity}. This is achieved via the so-called PISO algorithm (Pressure Implicit with Splitting of Operators). The matrix form of the momentum equation~\eqref{eq:NSE}, $\boldsymbol{M}\boldsymbol{u}=-\nabla p$, is decomposed by introducing a matrix $\boldsymbol{A}$ simply containing the diagonal entries of $\boldsymbol{M}$, yielding
\begin{equation}\label{eq:shortUEqn}
    \boldsymbol{A}\boldsymbol{u} - \boldsymbol{H} =
    - \nabla p.
\end{equation}
Using the approach of~\cite{Miller2013}, a pressure equation is formed analogously to the incompressible pressure corrector equation in the classical PISO loop, but with a non-zero right-hand-side, reading
\begin{equation}\label{eq:pEqn}
    \underbrace{\nabla \cdot \left( \boldsymbol{A}^{-1}\boldsymbol{H} \right)
    - \nabla \cdot \left( \boldsymbol{A}^{-1}\nabla p \right)}_{\nabla \cdot \boldsymbol{u}}
    = \Psi \left( 
    \frac{\partial \left( p \right)}{\partial t} 
    + \nabla \cdot \left( \boldsymbol{u} p \right)
    - p \nabla \cdot \boldsymbol{u}
    \right),
\end{equation}
where $\partial \rho / \partial p = \rho \Psi$. A first guess for $\boldsymbol{u}$ is obtained in a predictor step by solving Eq.~\eqref{eq:shortUEqn}. Then, in a correction step, $p$ is obtained by solving Eq.~\eqref{eq:pEqn}, followed by a correction of $\boldsymbol{u}$ to make it consistent with the result of Eq.~\eqref{eq:pEqn}. Depending on factors such as the time step size, the correction step (solving Eq.~\eqref{eq:pEqn} for $p$ and subsequently updating $\boldsymbol{u}$) might be performed several times to achieve convergence. Outer loops (going back and solving Eq.~\eqref{eq:shortUEqn}) are usually not necessary due to the problem being transient and the time steps being very small.

\FloatBarrier
\section{Benchmark Problems}\label{sec:benchmarks}
It is good practice to carefully assess the accuracy of individual sub-models employed in such a complex multiphysical model, for which a suitable strategy is simulating simple scenarios where an analytical solution to the problem exists. Examples for such test cases are one-dimensional Stefan problems or shock tube problems to assess the accuracy of handling phase transitions and compressibility, respectively. The here-presented model has undergone rigorous tests of this kind~\cite{Zenz2024a}, but the more interesting question is, how to test the accuracy, versatility and predictive capabilities of the model in actual scenarios of laser-material processing. While historically (and in many cases also to date), most simulation models are assessed by comparing the morphology of the resulting weld bead with experimentally obtained ex situ longitudinal and cross sections, and images of the surface morphology (cf. for example~\cite{Buttazzoni2021,Zenz2023}), the advent of in situ observation of laser material processing at high spatial and temporal frequency using synchrotron x-ray radiation opened up unprecedented possibilities of validating even highly dynamic transient process results. The following three benchmark problems aim at validating the model over a broad range of process regimes. In two cases we compare to in situ synchrotron x-ray observations, while in a third case, ex situ measurements of ablation depth are utilized as metric. To assess the predictive capabilities of the model, calibration (in the sense of finding the needed spatial resolution and suitable material properties) is performed to reproduce one parameter set. Then, the exact same model setup is used to perform simulations at different process parameters (changing for example the intensity distribution of the laser beam), with the aim of still matching the corresponding experiment - and hence exclude the possibility of a model overfit. The definition of material properties is usually crucial for the validity of model results and often not straightforward, as information on many important properties is scarce or even not available at all (e.g., the surface tension of an alloy close to its boiling point, or the complex refractive index of a liquid metal containing impurities). For a detailed and holistic discussion on the role of material properties in laser material processing, we refer the reader to Chap. 6 (Mücklich et al.) of this book.

As the results of the following three benchmark problems have already been published in great detail elsewhere, the reader is referred to the respective publications for details on the model setup (e.g., computational mesh, material properties).

\FloatBarrier
\subsection{Keyhole Formation and Collapse under Stationary Illumination}\label{subsec:cunningham}

\begin{figure}[h]
\centering
\includegraphics[width=0.99\textwidth]{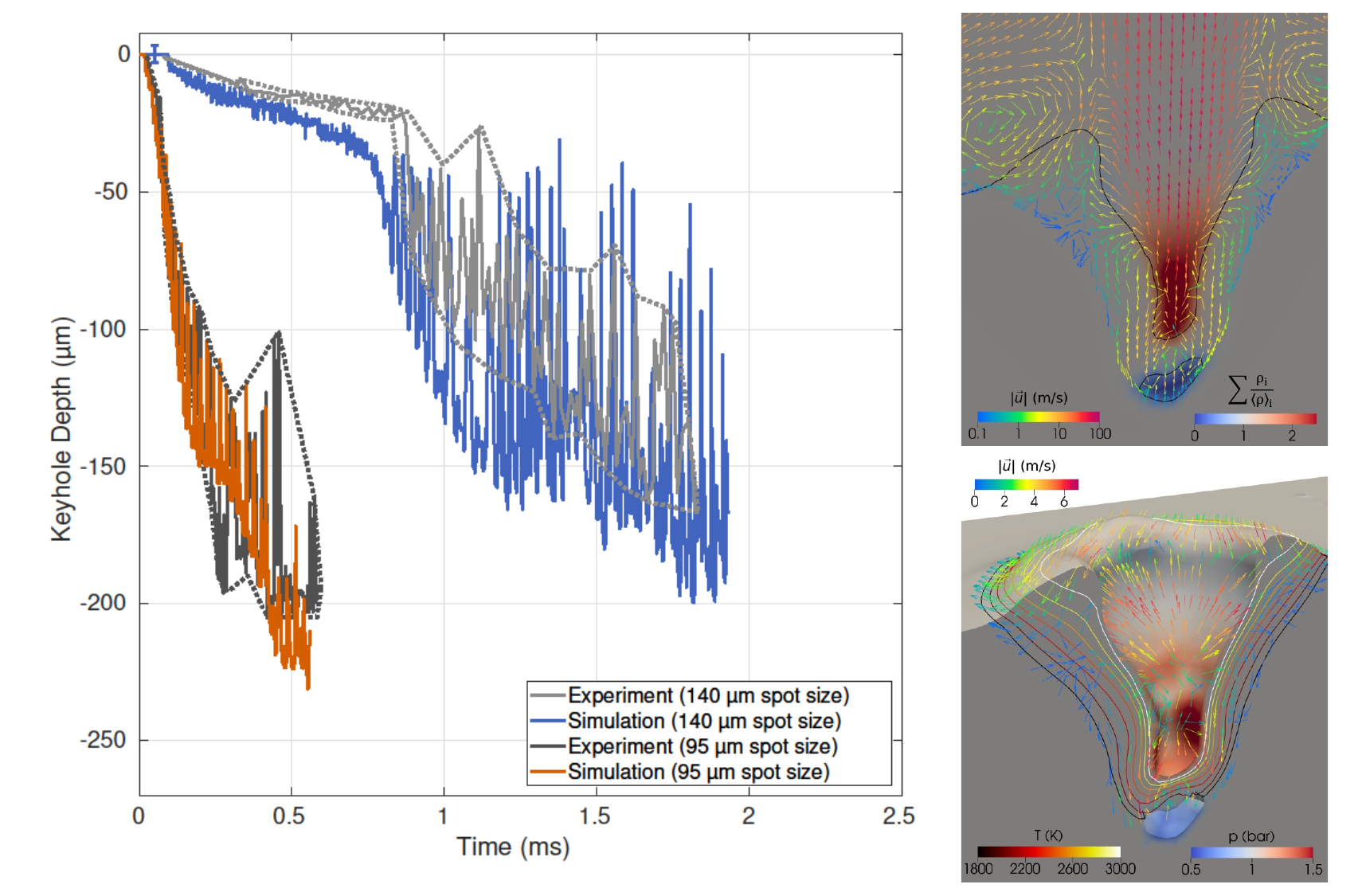}
\caption{Simulation of stationary keyhole fluctuation: Left: Keyhole depth over time for two laser spot sizes, comparing simulation results to experiment of~\cite{Cunningham2019}. Right: Illustrative snapshots of the simulation are shown alongside (Reprinted from~\cite{Zenz2024a} and rearranged, Copyright (2023) under Creative Commons BY 4.0 license)}
\label{fig:bench:cunningham}
\end{figure}

\begin{figure}[h]
\centering
\includegraphics[width=0.99\textwidth]{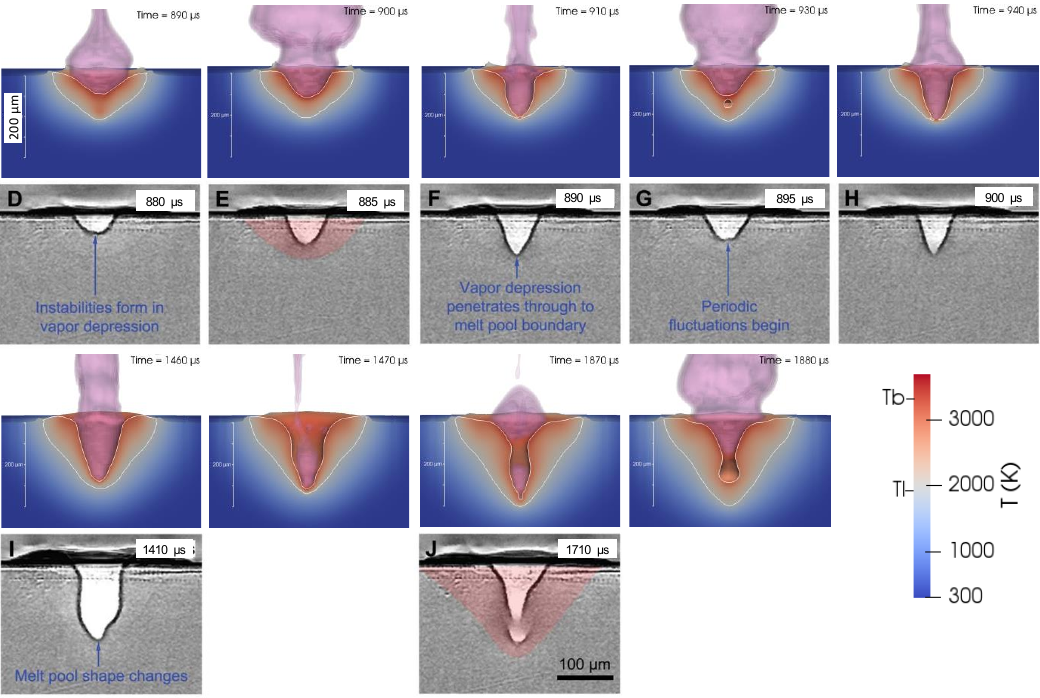}
\caption{Simulation of stationary keyhole fluctuation: Comparison of simulated keyhole morphology (top, colored) to experimental synchrotron x-ray radiographs (bottom, gray-scale, with approx. melt pool colored red)of~\cite{Cunningham2019} (Reprinted from~\cite{Zenz2024a}, Copyright (2023) under Creative Commons BY 4.0 license. Experimental images originally from~\cite{Cunningham2019}. Reprinted with permission from AAAS)}\label{fig:bench:cunninghamQualitative}
\end{figure}

Stationary illumination (drilling) provides a suitable scenario for basic investigations of the multiphysical interactions in laser-material-processing. While the absence of a relative motion between work piece and laser reduces the complexity of the problem, we still encounter the same coupled physical phenomena. One of the key phenomena in laser-material-processing is thermocapillary (in)stability, i.e., the (im)balance of evaporative recoil pressure and surface tension in a vapor capillary (the keyhole). A systematic series of drilling experiments on Ti6Al4V bare plates under conditions relevant for AM was performed by Cunningham et al.~\cite{Cunningham2019}, obtaining in situ synchrotron x-ray radiographs at high spatial and temporal resolution. Of the parameters investigated experimentally in~\cite{Cunningham2019}, the combination of a laser power of $P$=156~W and a beam diameter of $d$=140~µm is particularly interesting, as we can observe a clear separation into two different process regimes: At first, a steady drill rate is observed, without large fluctuations, where surface tension forces, promoting keyhole closure, seem to be in relatively stable balance with the recoil pressure from evaporated vapor acting on the keyhole wall, which opens the keyhole. Then, at a certain point, a shift towards a different regime occurs, where the average drill rate promptly increases and the keyhole depth starts violently fluctuating. The fluctuations occur due to local imbalance of surface tension and recoil pressure, with less laser energy reaching lower parts of the keyhole the further keyhole progresses, hence surface tension becomes dominant, leading to a collapse of the keyhole, immediately followed by a strong increase in keyhole depth. The experimental results are plotted in Figure~\ref{fig:bench:cunningham}, alongside the corresponding simulation using the here-presented model. On the right-hand-side of Figure~\ref{fig:bench:cunningham}, the keyhole is shown during the fluctuating stage of the 140~µm beam diameter process, showing (top) the distribution of normalized phase volume, $\sum \rho_{i}/\langle\rho\rangle_{i}$, with red regions indicating a local excess in volume, and hence an increase in pressure and a compression of matter, and blue regions denoting analogously a volume deficit and hence expansion (cf. Figure~\ref{fig:MoF} for an illustrative explanation). Excess volume in the lower part of the keyhole stems from generated metal vapor that cannot immediately escape the interface and hence results in a recoil pressure acting on the keyhole wall, which can be seen in the lower image, where the pressure distribution on the keyhole wall is plotted. Notably, this local increase in pressure will lead to a reduction in the evaporation rate (i.e., an increase of $p$ in Eq. (\ref{eq:evapMassRate})). Just below the keyhole, a bubble is sealed off due to the keyhole just having collapsed, inside which we can observe a decrease in overall phase volume and hence a pressure. A comparison of simulated and experimentally observed keyhole morphologies during different stages of the process is provided in Figure~\ref{fig:bench:cunninghamQualitative}.

The simulation model is able to correctly predict the transition between the two process regimes, underlining its predictive capabilities. To further test the model's predictive capabilities, another parameter set is chosen (reduced laser diameter of $d$=95~µm at the same power) for comparison. Here, the two regimes are not as distinctively separated. Yet, the simulation model (where the only model parameter changed was of course the laser diameter, as in the experiment) still accurately reproduces the experimental keyhole evolution. A detailed list of material parameters, computational domain and discretization used in the simulations presented here can be found in~\cite{Zenz2024a}.

\FloatBarrier
\subsection{Process Dynamics in Copper Welding}\label{subsec:copperwelding}
In this subsection, we demonstrate the model's capability in capturing the highly dynamic behavior of deep penetration laser welding of copper under varying process regimes by modifying the laser intensity profile. Validation is achieved by comparing the simulated keyhole geometries to those observed through high-speed synchrotron X-ray imaging performed at ESRF in Grenoble. Building upon the validation, this comparative study~\cite{Florian2024b} employed a hybrid approach that combined multiphysical simulations with advanced synchrotron X-ray radiography to uncover four distinct pore formation mechanisms within the process. A detailed investigation of these mechanisms can be found in the original work of Florian et al.~\cite{Florian2024b}.

Here, the aim is solely to highlight the accuracy of the simulation tool. Accordingly, we discuss the experimental and simulation setup, emphasizing the role of the varying intensity distribution. For this purpose, we concentrate on the opposed ramp experiment. In this configuration, a 2~mm-thick copper sheet was welded in a bead-on-plate setup at a feed rate of 10~m/min, utilizing a Coherent ARM laser system. The laser featured a core intensity profile with a diameter of 89~µm and  a ring intensity profile with an outer ring diameter of 208~µm (inner diameter 109~µm). Over the 115~mm weld length, the core intensity was linearly ramped down from 3.5~kW to 0~kW, while the ring intensity simultaneously was ramped up from 0 to 3.5~kW, maintaining a constant total power. To accommodate the X-ray imaging setup, the copper sheet was positioned upright and moved relative to the stationary laser beam.

Due to the high computational demands and the requirement for fine spatial and temporal resolution, only three smaller-scale simulations were conducted rather than modeling the entire experiment. These simulations captured key regimes of the process: one with the core intensity only, one with a 50/50 distribution between core and ring intensities, and one with 100 \% ring intensity. For a comprehensive description of the experimental setup, numerical configuration, material properties, and the modeled laser beam, the reader is directed to the original study~\cite{Florian2024b}.

\begin{figure}[h]
\centering
\includegraphics[width=0.99\textwidth]{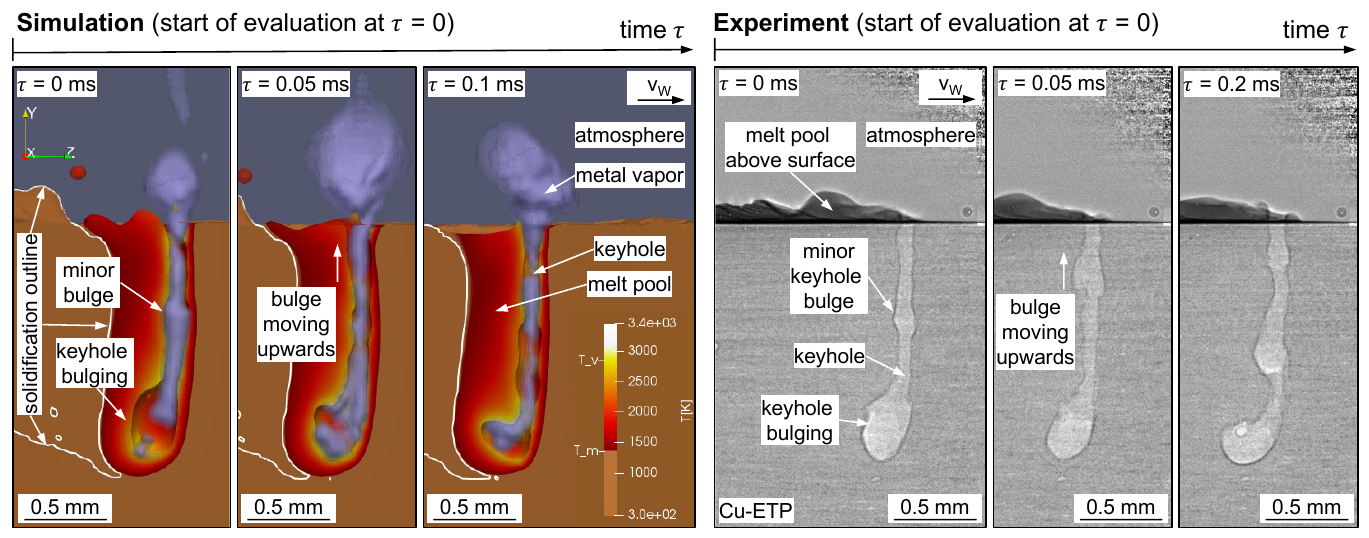}
\caption{Simulation validation based on core-dominated welding process: left: snapshots from the core dominated intensity simulation; right: X-ray images at the beginning of the opposed-ramp experiment ($\approx$ 100\% core intensity)(Reprinted from~\cite{Florian2024b}, Copyright (2024) under Creative Commons BY 4.0 license)}\label{fig:bench:copperWeldingPic1}
\end{figure}

Starting our analysis, we compare the snapshot sequences of the core-dominated intensity simulation with the corresponding X-ray images captured at the beginning of the experiment, depicted in Figure~\ref{fig:bench:copperWeldingPic1}. The simulation provides a longitudinal section of the weld, emphasizing the temperature distribution within the condensed phases. A color-coded map distinguishes between the liquid and solid copper, while a white line marks the solidification front. Additionally, purple contours highlight the presence of metallic vapor both inside and outside the keyhole, with regions lacking vapor indicating the presence of atmospheric air.

The keyhole shapes observed in both simulation and experiment show a strong resemblance in terms of spatial and temporal characteristics. Both reveal a deep and narrow keyhole with a pronounced bulge at the base and smaller bulges that form and migrate upwards. Beyond what is visible in the X-ray images, the simulation provides valuable insights into the melt pool's boundaries, vapor presence, and additional critical details such as velocity and pressure fields, as well as the laser beam's propagation and absorption. This highly dynamic process regime is particularly prone to defects, including melt ejections and pore formation.

\begin{figure}[h]
\centering
\includegraphics[width=0.99\textwidth]{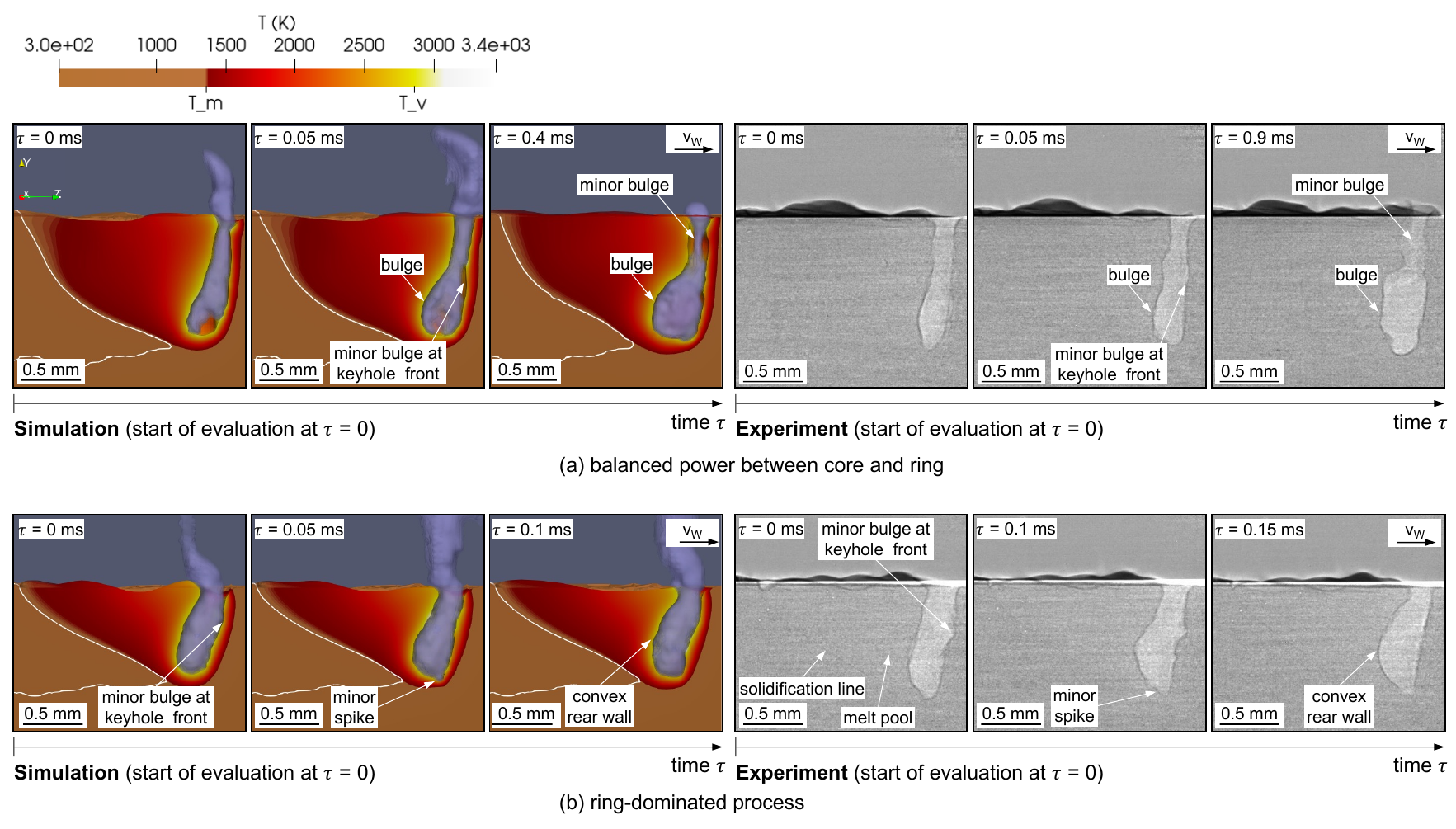}
\caption{Simulation validation based on a) balanced core/ring intensity distribution and b) ring-dominated intensity: left: snapshots from the core simulations; right: X-ray images at a) the middle region and b) end region of the opposed-ramp experiment (Reprinted from~\cite{Florian2024b}, Copyright (2024) under Creative Commons BY 4.0 license)}\label{fig:bench:copperWeldingPic2}
\end{figure}

Figure~\ref{fig:bench:copperWeldingPic2} presents a similar comparison for the balanced core-ring and ring-dominated regimes. The persistent resemblance of the keyhole geometry between simulation and experiment, both spatially and temporally, supports the model's ability to accurately predict process behavior across various regimes. Even finer details, such as minor bulges on the absorbing front or small spiking events, are captured with remarkable precision.

In the balanced intensity distribution regime, the welding process exhibits notable stability, characterized by a reduced penetration depth of approximately 1.5 mm, minimal melt ejections, and negligible pore formation. This stability appears to result from a favorable pressure equilibrium within the keyhole. 

In contrast, the ring-dominated regime achieves a penetration depth of approximately 1 mm. While pore formation remains absent, this regime displays more pronounced melt ejections, as detailed in~\cite{Florian2024b}. These ejections may stem from an increase in keyhole diameter in the upper region, which leads to a reduction in surface tension-related closing pressures.

At this point, we wish to acknowledge the tremendous advancements in synchrotron X-ray imaging of laser material processes, which have unlocked unprecedented opportunities for validating and refining numerical models. Historically, validation efforts primarily relied on comparing ex situ cross-sections and longitudinal sections with numerically obtained solidification outlines. However, the direct comparison of keyhole morphology in both space and time offers a significantly superior validation approach. This is because keyhole geometry lies much closer to the root of the causality chain, whereas many distinct process behaviors can result in the same or at least similar cross-sections.

\FloatBarrier
\subsection{Ultra-short pulse laser ablation of copper}
\label{sec:copperablation}

The subsequent benchmark assesses the model's capability to simulate high-power, single-pulse laser ablation of copper. For an overview of the underlying physics and the resulting process characteristics of ultrashort pulse laser processing, we refer the reader to Chap. 1 (Nolte et al.) of this book. In the context of ultra-short pulse ablation, the Drude model (Section~\ref{subsec:lasermatinter}), the two-temperature model (Section~\ref{subsec:energyconserv}), and the selected equation of state (Section~\ref{subsec:momentumconserv}) are pivotal to the simulation's accuracy.

This case study involves comparing the simulated ablation depths with experimentally measured values across a broad range of pulse energies. The simulation setup closely mirrors the experimental configuration: a single Gaussian laser pulse with a duration of 120~fs (FWHM) and a wavelength of 800~nm, featuring a spatial Gaussian intensity distribution with an effective focused beam diameter of 6.5~µm, is applied to a pure copper foil. With an ablation threshold of 0.5~J/cm², simulations were conducted at fluences of 6.1, 23.2, 45, and 63.4~J/cm². For a detailed description of both the experimental and simulation setups, the reader is referred to the original publication~\cite{Florian2024}.

\begin{figure}[h]
\centering
\includegraphics[width=0.9\textwidth]{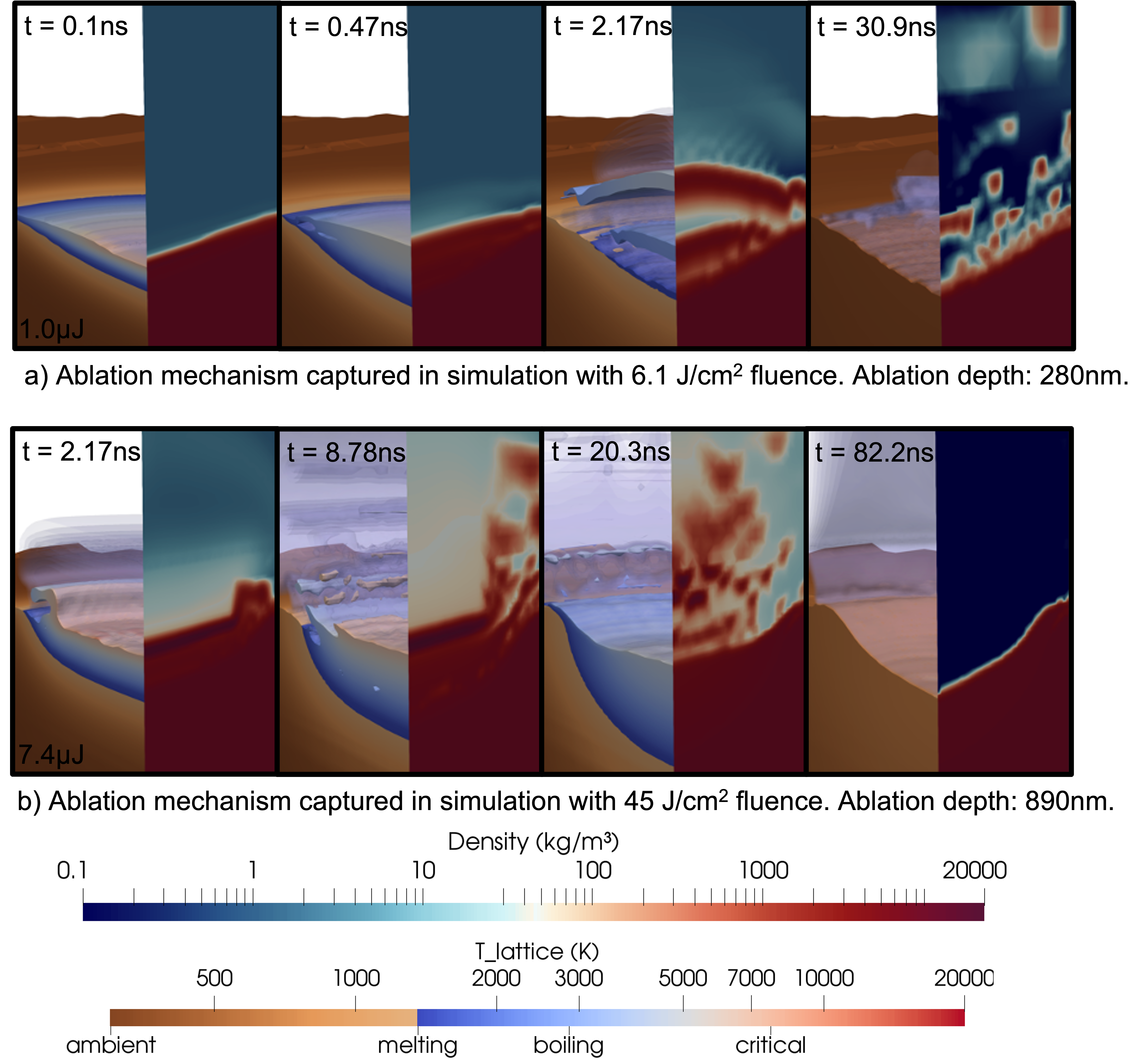}
\caption{Snapshots of the simulation with a) 6.1~J/cm² and b) 45~J/cm². Each snapshot represents the lattice temperature in the condensed phases and the occurrence of vapor on the left side and the density field on the right side (Reprinted in part and rearranged from~\cite{Florian2024}, Procedia CIRP, Vol \textbf{124}, Florian et al., Ultra-short pulse laser ablation of metals: A comprehensive 3D simulation perspective enlightening novel process insights, 602--607, Copyright (2024), with permission from Elsevier)}\label{fig:bench:UKP_Pic2}
\end{figure}

Figure~\ref{fig:bench:UKP_Pic2} illustrates snapshots from simulations conducted at fluences of 6.1~J/cm² and 45~J/cm², highlighting distinct ablation mechanisms. The colormaps display the lattice temperature and vapor presence on the left, while the right side depicts density in logarithmic scale to emphasize shock waves and density variations. These snapshots begin at an advanced stage, after the laser pulse has been fully absorbed. At this point, the energy initially deposited into the electrons has been transferred to the lattice, leading to the formation of a melt pool and the development of a vapor plume. At this stage, two opposing momentum-driven processes dominate the dynamics. The first originates from high evaporation rates at the liquid-vapor interface, generating a recoil pressure that depresses the liquid metal downward. This elevated recoil pressure, in turn, limits the evaporation rate. Simultaneously, the liquid metal absorbs additional energy, reaching temperatures well beyond the boiling point and approaching, or even surpassing, the critical temperature. As a result, the liquid undergoes rapid volumetric expansion, inducing an upward momentum due to the reactive forces from the bulk material. These competing forces result in distinct ablation regimes, as exemplified by the two cases in Figure~\ref{fig:bench:UKP_Pic2}. In Figure~\ref{fig:bench:UKP_Pic2}a, corresponding to 6.1~J/cm², the second snapshot at t=0.47~ns shows the melt pool at its maximum depth. Here, the upward momentum generated by liquid expansion exceeds the downward force from evaporation-induced recoil pressure. This interaction leads to a clean, layer-wise ablation of the liquid metal in a convex manner, leaving a smooth crater without radial ejections. This process is referred to as spallation. In contrast, Figure~\ref{fig:bench:UKP_Pic2}b, representing 45~J/cm², exhibits a different ablation behavior. The first frame at t=2.17~ns reveals a significantly deeper vapor depression at the center, where the downward recoil momentum dominates over the upward momentum from liquid expansion. This allows additional melting of the material, resulting in a deeper melt pool. Initially, layer-wise ablation occurs at the edges, but as the expanding superheated liquid undergoes a swift phase transition, the associated pressure distribution generates a radial momentum. This is evident from the radially ejected melt, forming characteristic radial structures at the crater's edges. In the center, the liquid phase undergoes a phase change, as indicated by the density accumulations visible at t=20.3~ns. This rapid phase transition, triggered by the extreme conditions, is referred to as phase explosion. These findings highlight the interplay of evaporation dynamics, liquid expansion, and phase transitions in determining the ablation mechanism and resulting crater morphology.

\begin{figure}[h]
\centering
\includegraphics[width=0.8\textwidth]{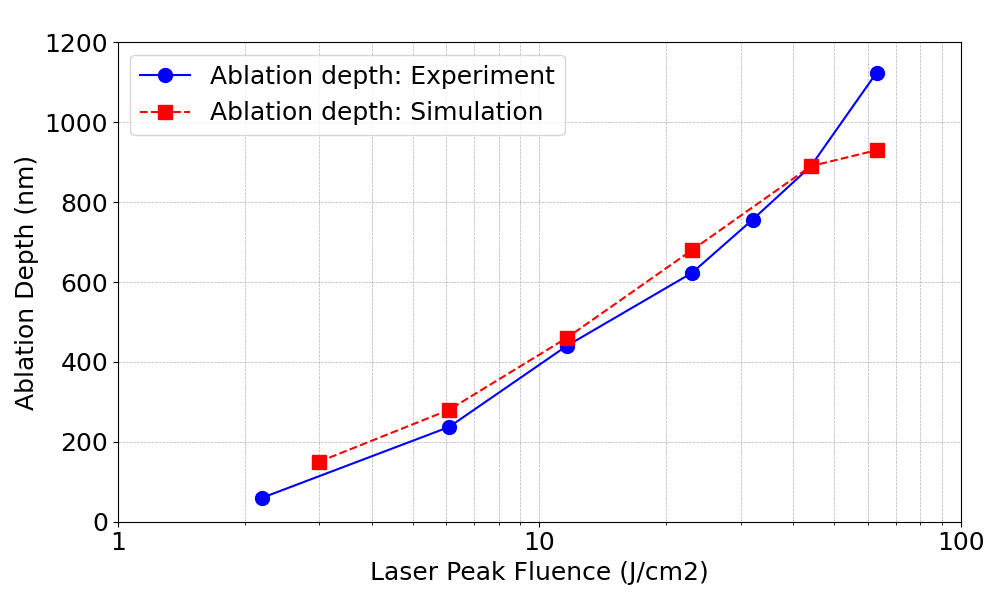}
\caption{Ablation depths of single pulse ablation of copper. Experimental measurements are taken from~\cite{Cheng2016} and~\cite{Byskov_Nielsen2010}. The data set of the simulations compared to~\cite{Florian2024} is extended by two data points, 3 and 11.6~J/cm²}\label{fig:bench:UKP_Pic1}
\end{figure}

Figure~\ref{fig:bench:UKP_Pic1} presents the experimentally measured ablation depths alongside the simulated depths. The results indicate good agreement between the simulations and experimental measurements for fluences up to 45~J/cm². However, a significant discrepancy is observed at the highest fluence of 63.4~J/cm². This divergence is attributed to inaccuracies in predicting the material's behavior near the critical and supercritical regions. As concluded in~\cite{Florian2024}, this mismatch likely arises from either incorrect estimation of material parameters, such as the temperature-dependent isobaric bulk modulus and density of the liquid phase, or the inadequacy of the employed equation of state to accurately describe the fluid dynamics under such extreme conditions. Ongoing efforts aim to refine these aspects of the model.

Despite this limitation, the model demonstrates strong agreement for fluences up to 100 times the ablation threshold, providing valuable insights into the ablation mechanisms and the crater morphology at the meso scale.

\FloatBarrier
\section{Application Examples}\label{section:applications}
The following Sections will illustrate some practical use cases of the simulation model to refine or study processes used in industrial applications.

\subsection{Glass Through Via Drilling} 
\begin{figure}[h]
\centering
\includegraphics[width=0.99\textwidth]{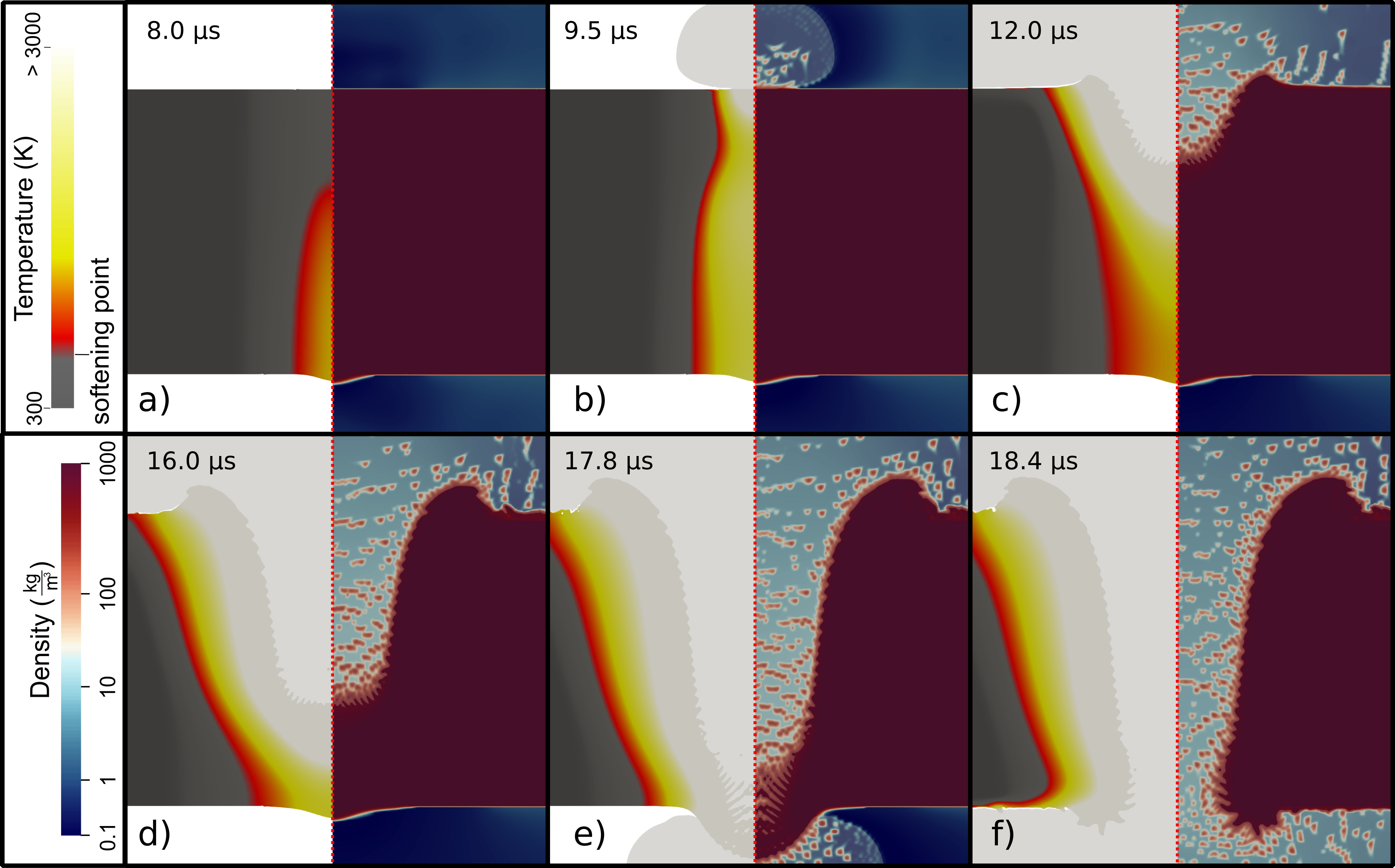}
\caption{Borosilicate glass through via drilling with focus position below the bottom surface shown at different times during the process. The glass layer has a thickness of 50~µm. The left hand side of each tile shows the temperature distribution. The right hand side shows the material's density in logarithmic scale, showcasing the compressive wave following the plume expansion in b) and the ejection of small droplets due to evaporation in b)-f)}\label{fig:BK7}
\end{figure}

In recent years glass through via drilling has become an important processing step for the formation of interconnects in the fabrication of micro-electro-mechanical systems~\cite{Yu2024}. In order to improve throughput and reduce the reliance on multiple processing steps, as is done in laser induced deep etching, the drilling process using quasi-continuous (high repetition rate) pico-second laser pulses without the requirement of chemical pre- or postprocessing is investigated in detail by Matsumoto et al.~\cite{Matsumoto2021} and Schrauben et al.~\cite{Schrauben2024}. 
This investigation has been done with the help of the herein presented simulation software. One of the conclusions of this collaboration is, that it is beneficial (at least for 50~µm thick glass) to focus the beam below the bottom surface of glass for a more efficient process with faster drill times. The reasoning behind this is explained using Figure~\ref{fig:BK7}, where a cross-section of the 50~µm thin borosilicate glass plate is shown at various times during the drilling process. The process uses a green wavelength ps-pulsed laser operating at 30~MHz with an average power of 70~W.

During first stages of investigations, it was found that focusing the beam onto the top surface of the plate induced a temperature accumulation near the focus and led to melting in that same region. This melt is more absorptive to the laser wave length, thereby absorbing even more energy and causing evaporative processes to commence. As the laser fluence combined with the absorptivity and viscosity of the material is not adequate for spallation or even phase explosion, the material, which needs to be removed for the through via, must be entirely vaporized.

In Figure~\ref{fig:BK7} a drilling process is shown, where the laser focus is shifted to a point well below the bottom surface. This makes it possible to have higher laser intensity near the lower region and induce bottom-up melting. This creates a more even temperature distribution throughout the thickness of the glass and by carefully calibrating the focus position it is also possible to design a process where the bottom part does not cool too much and resolidification is suppressed during the later stages of the drilling process. 
After a melt pool has established, most of the absorption happens at the melting front, which is traveling upwards against the direction of light propagation (Figure~\ref{fig:BK7}a). Once the melt pool spans the entire thickness of the part and is further irradiated, evaporation starts at the top surface (Figure~\ref{fig:BK7}b), opening a hole through recoil pressure by pushing the melt to the side (Figure~\ref{fig:BK7}c). This hole then deepens by vaporization and further lateral ejection of molten glass carried by the momentum of the vapor plume escaping to the top (Figure~\ref{fig:BK7}d). This continues until only a small portion of material is left, at which point the pressure of the plume is enough to push the rest of the molten material out to the bottom (Figure~\ref{fig:BK7}e) leaving a clean hole (Figure~\ref{fig:BK7}f).

For this kind of process to work it is crucial to find a balance of out-of-focus-processing, where the intensity is high enough for rapid heating and the beam is sufficiently convergent to induce melting from the bottom. Another crucial factor is that a certain amount of heat energy is input to the lower part, before melting occurs. If this energy is too low, that part will solidify again, once the melt pool shadows that region from the laser energy for long enough time. On the other hand if this energy is too high, evaporation will start at the bottom surface and thus the energy efficiency will drop, as the whole through via is then created by vaporization instead of melt ejection. For a deeper background on similar processes and the underlying physics, an overview of processing of transparent materials with bursts can be found in Chap. 18 (Manek-Hönninger et al.) of this book.

\FloatBarrier
\subsection{Multi-Material Stack Ablation} 

In PCB (printed circuit board) production the ablation of different material layers has become highly relevant in recent times, as their production relies heavily on the fast and precise drilling of blind microvias, while simultaneously reducing their size to increase interconnect density~\cite{Franz2022}. In this Section specifically, we will focus on the punch-drilling of a polymer layer sandwiched between to copper layers. The polymer is a proxy material, as the used material is usually FR4 (a composite material of epoxy resin reinforced with glass fibers; resolving the individual fibers would be computationally too expensive). The task is to analyze and understand the process and optimize the laser parameters in order to get a clean hole through both materials with minimal taper or other defects. This should be done using a single laser source to minimize throughput time. In a subsequent processing step, this via will be electroplated with copper to create an interconnect between the two conducting layers. It is crucial not to damage the bottom copper layer during drilling to allow for the best cohesion in this step.

The simulation for this process must be able to handle multiple materials, each with its own states of matter and possibly different phase change mechanism (in the case of polymer either approximated sublimation or thermal decomposition). Also the heat exchange between the materials has to be modeled correctly to assure consistency and accuracy in the thermal distributions. Furthermore it showcases the true multiphase capabilities of the software, as the simultaneous mass and energy advection and conduction of multiple phases within the same meshed domain must be possible without violating any conservation laws formulated in Section~\ref{sec:model}.

\begin{figure}[h]
\centering
\includegraphics[width=0.99\textwidth]{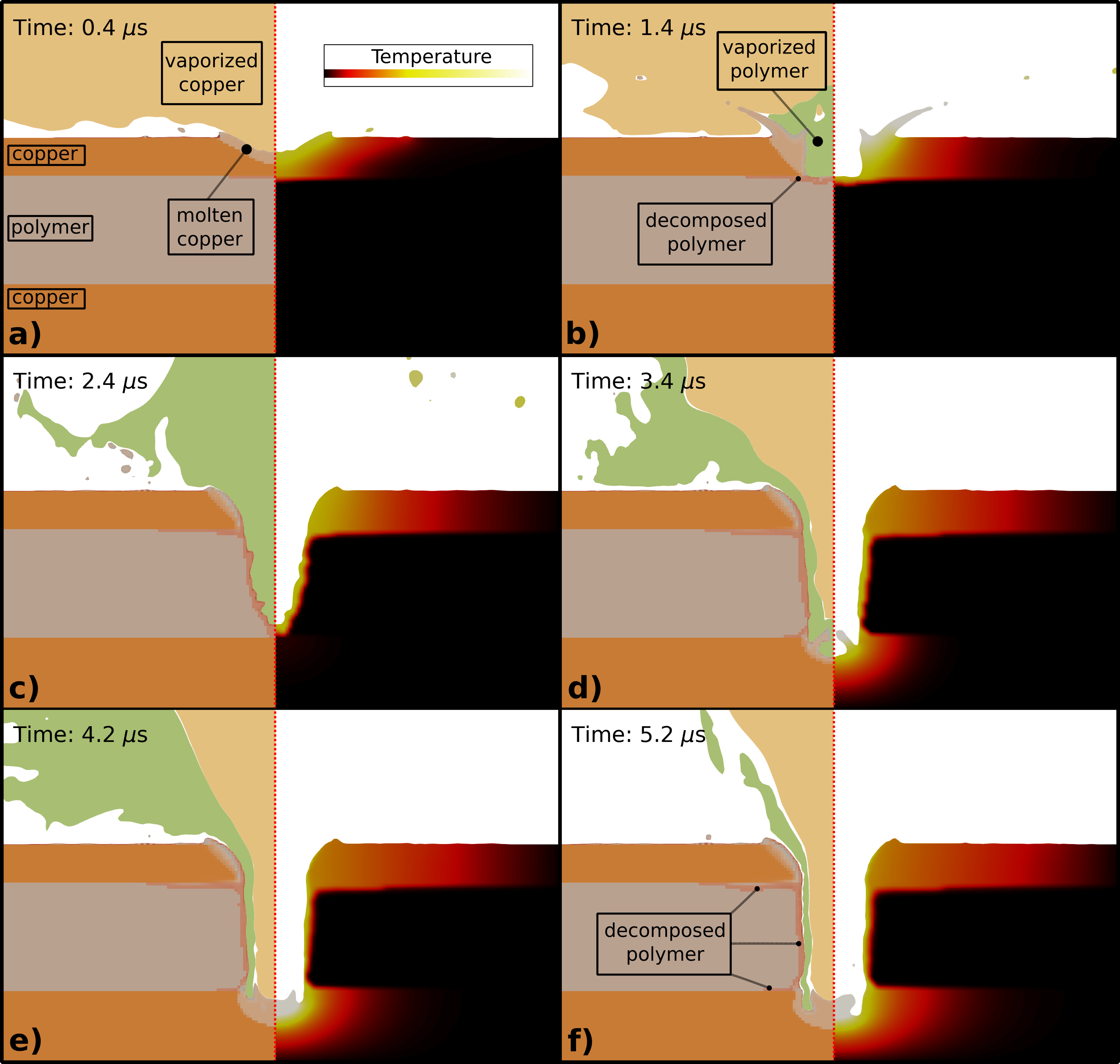}
\caption{Cross-sections through the multi-material stack (copper-polymer-copper of thicknesses 12.5 - 37.5 - 25~µm) at different times between subsequent pulses. The left hand side of each tile shows the distribution of the different materials and states of matter, while the right shows the temperature}\label{fig:CuPoCu}
\end{figure}

Figure~\ref{fig:CuPoCu} shows the drilling process from a cross-sectional perspective, each snapshot taken at different times and separated by the irradiation with a 10~ns UV pulse at a pulse repetition rate of 1 MHz:

After the first pulse is absorbed, the top copper plate begins to melt and vaporize, opening the hole through vaporization pressure (Figure~\ref{fig:CuPoCu}a). The second pulse at 1~µs hits an already heated copper layer, which is also more absorbent due to the curved interface created by the first pulse. This causes an explosive molten copper ejection due to the rapid decomposition and vaporization of the middle layer polymer (Figure~\ref{fig:CuPoCu}b). The third and fourth pulse (Figure~\ref{fig:CuPoCu}c-d) are used to drill through the polymer layer, the latter already impacting on the bottom copper layer. Further pulses (Figure~\ref{fig:CuPoCu}e-f) will only remelt the bottom layer and causing more thermal damage in that region, while widening the hole through heat exchange of the vapor plumes and the polymer.
Also noticeable is the different thermal conductivities of the investigated materials leading to very different heat distribution. The high conductivity of copper causes a decomposition of the polymer near the two shared interfaces (especially the top interface in Figure~\ref{fig:CuPoCu}f).

Using this same model, we can now investigate the influence of changing the laser source to use different wavelengths or change other relevant laser parameters to investigate their effect on important quality requirements. 

\FloatBarrier
\subsection{Ultrashort pulsed laser ablation of ceramic layer on carrier plate}
Due to ever increasing available computational power and the advent of artificial intelligence, a new challenge emerges on how to store the huge amount of data produced and processed. Especially long term storage of cold data (i.e. infrequently accessed data),  is currently relatively resource-intensive and expensive. The task of solving this issue is currently being tackled in the GREEN~\cite{greenproject} project, where one of the goals is to use ultra-short-pulsed lasers to engrave data onto a nanometer-thick ceramic material coated on a glass carrier. The written data can  then be read using high-performance microscopy. Thanks to the resilience of the used materials, and the passive nature of the storage method, it enables long-term data preservation, eliminating the need for frequent copying and drive maintenance (e.g., compared to SSD or HDD storage), thus reducing energy and raw material consumption. Further details on this emerging technology can be found in \cite{Kreuziger2025}. As the processing technology at this scale is still in its infancy, a lot of efficiency and precision is to be gained by properly analyzing and understanding the ablation process. Here, simulations play a major role in aiding the development process to find optimal parameters for clean and efficient ablation.

\begin{figure}[h]
\centering
\includegraphics[width=0.99\textwidth]{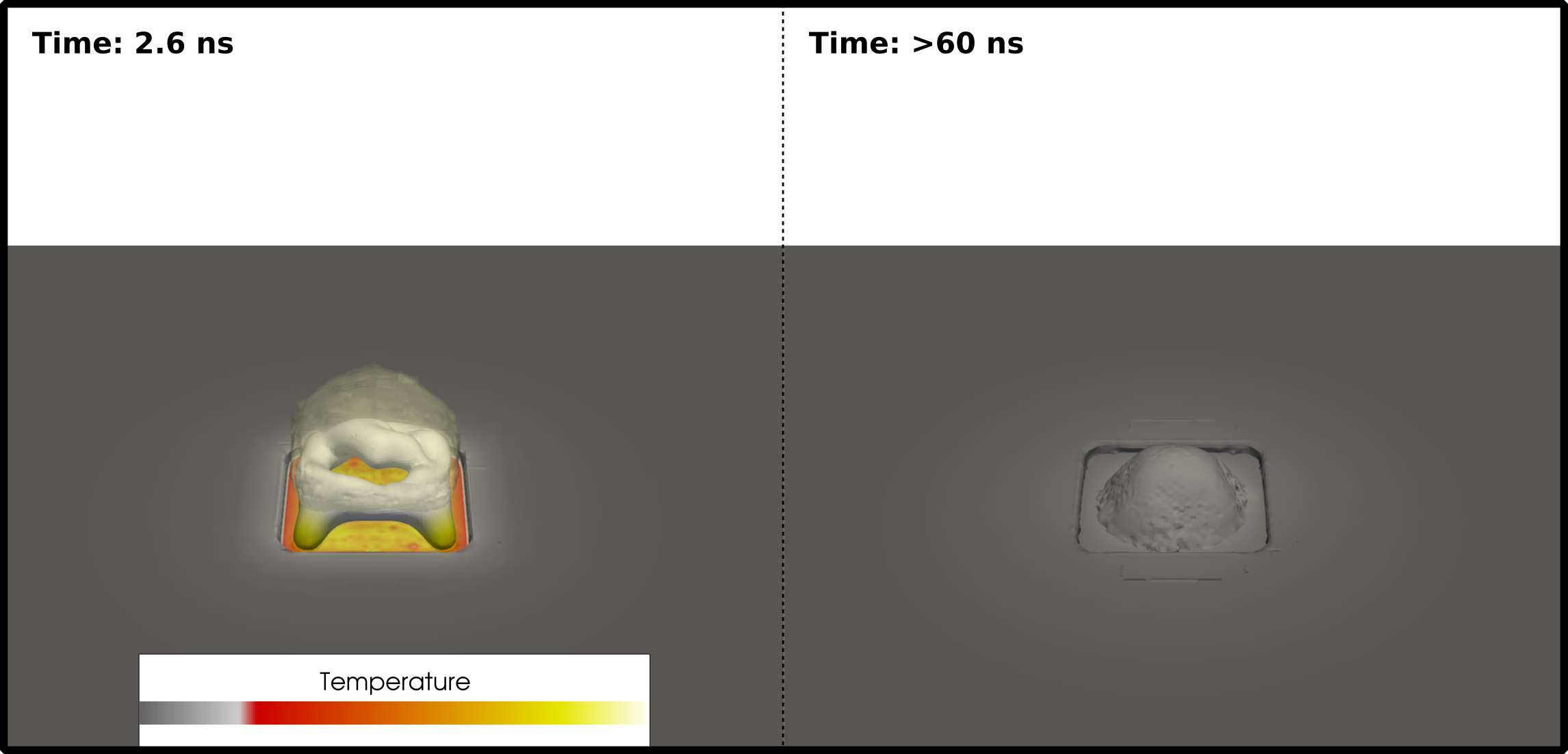}
\caption{Ceramic layer ablation at approx. half the threshold fluence. Laser beam spot size 400nm side length.}\label{fig:ceramicHalf}
\end{figure}

\begin{figure}[h]
\centering
\includegraphics[width=0.99\textwidth]{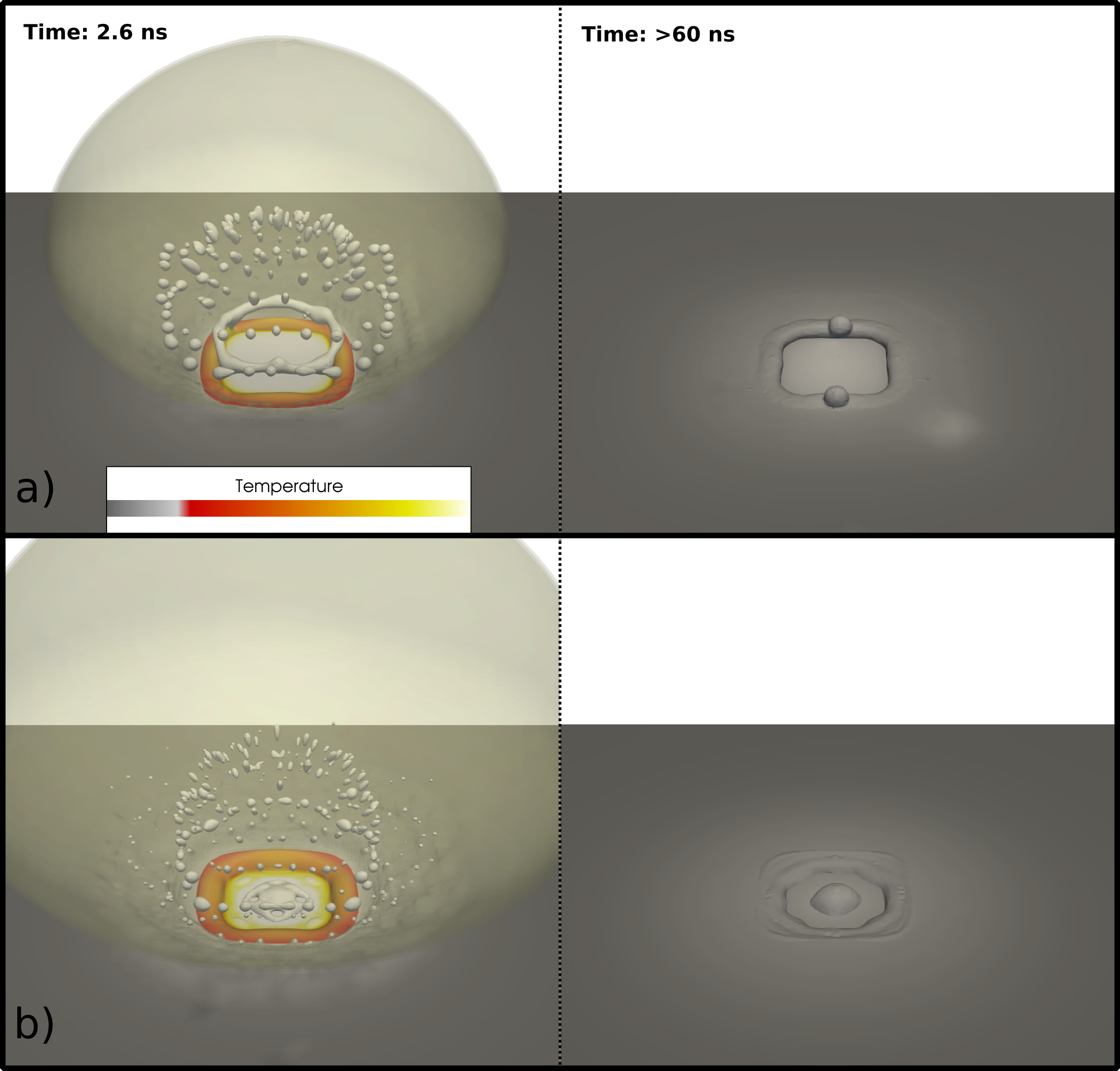}
\caption{Ceramic layer ablation at approx. the threshold fluence, a) at 75\% of the pulse energy of b)}\label{fig:ceramicFull}
\end{figure}

\begin{figure}[h]
\centering
\includegraphics[width=0.99\textwidth]{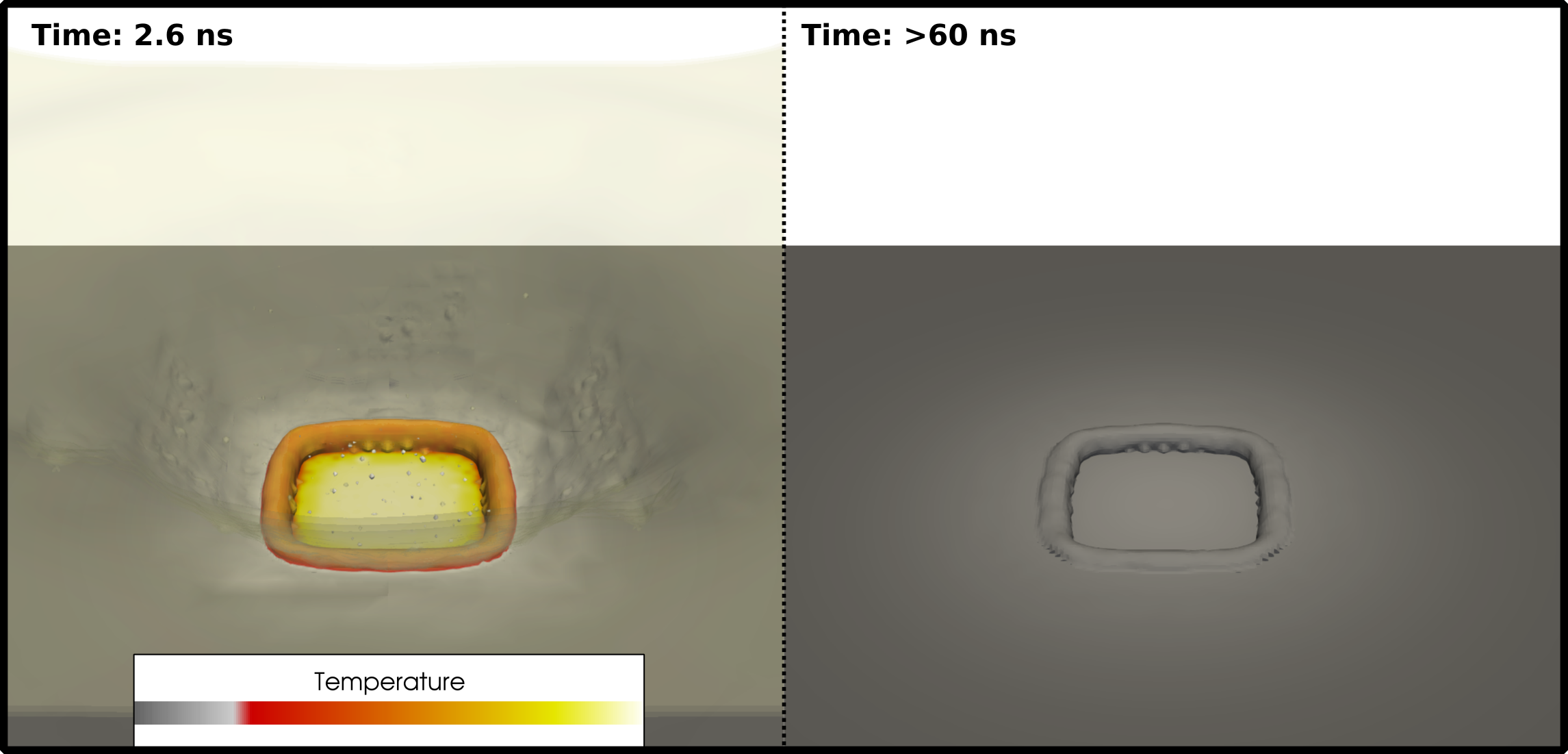}
\caption{Ceramic layer ablation at approx. double the threshold fluence}\label{fig:ceramicDouble}
\end{figure}

During data storage, an ultrashort pulsed laser selectively ablates a ceramic layer in the sub-micrometer range of thickness using a Digital Mirror Device (DMD). This DMD creates an array of rectangular spots, that can be selectively activated or deactivated at a very high switching frequency. The simulations shown in Figure~\ref{fig:ceramicHalf} to \ref{fig:ceramicDouble} show the ablation of one of these spots of side length 400 nm, created by one of the mirrors inside the DMD at different fluences around the ablation threshold for a 250 fs long pulse. The layer thicknesses are taken as 50 nm for both the ceramic and the carrier plate. The goal is to find the optimal fluence for high energy efficiency, while keeping the accuracy and quality of the ablated profile adequate for the large scale processing of data.

Figure~\ref{fig:ceramicHalf} shows the ablation process at approximately half the ablation threshold fluence. The momentum of the rapidly melting and expanding material is not strong enough to eject the melt through spallation. Consequently it is redeposited on the substrate and contracts to the shape seen in the right hand side of Figure~\ref{fig:ceramicHalf} due to surface tension effects during cool-down.

Figure~\ref{fig:ceramicFull} shows the same process at slightly higher pulse energies, around the threshold fluence. In both instances, efficient spallation occurs, ejecting a layer of molten material through rapid melt expansion and sub-surface boiling (more details on the ablation mechanisms can be found in \cite{Florian2024}). In this case, using a higher fluence leads to a larger melt pool, more rapid plume expansion and the ejection of smaller droplets. All of this indicates a less efficient process, as more heat remains on the substrate, and additional thermal energy is required to vaporize the material and overcome its latent heat. The final result in both cases involves the redeposition of some droplets on the surface. To prevent this, proper process control is necessary, such as using lateral gas flow.


Finally, Figure~\ref{fig:ceramicDouble} shows the process at 200\% of the threshold fluence. Here we move out of the spallation regime and into phase explosion, characterized by the rapid vaporization of the ceramic layer with the ejection of a mixture of fine liquid droplets mixed with the expanding vapor plume. The end result is a very clean ablated crater with a relatively high burr caused by the some molten material being pushed to the side by high vapor pressure.

\begin{figure}[h]
\centering
\includegraphics[width=0.99\textwidth]{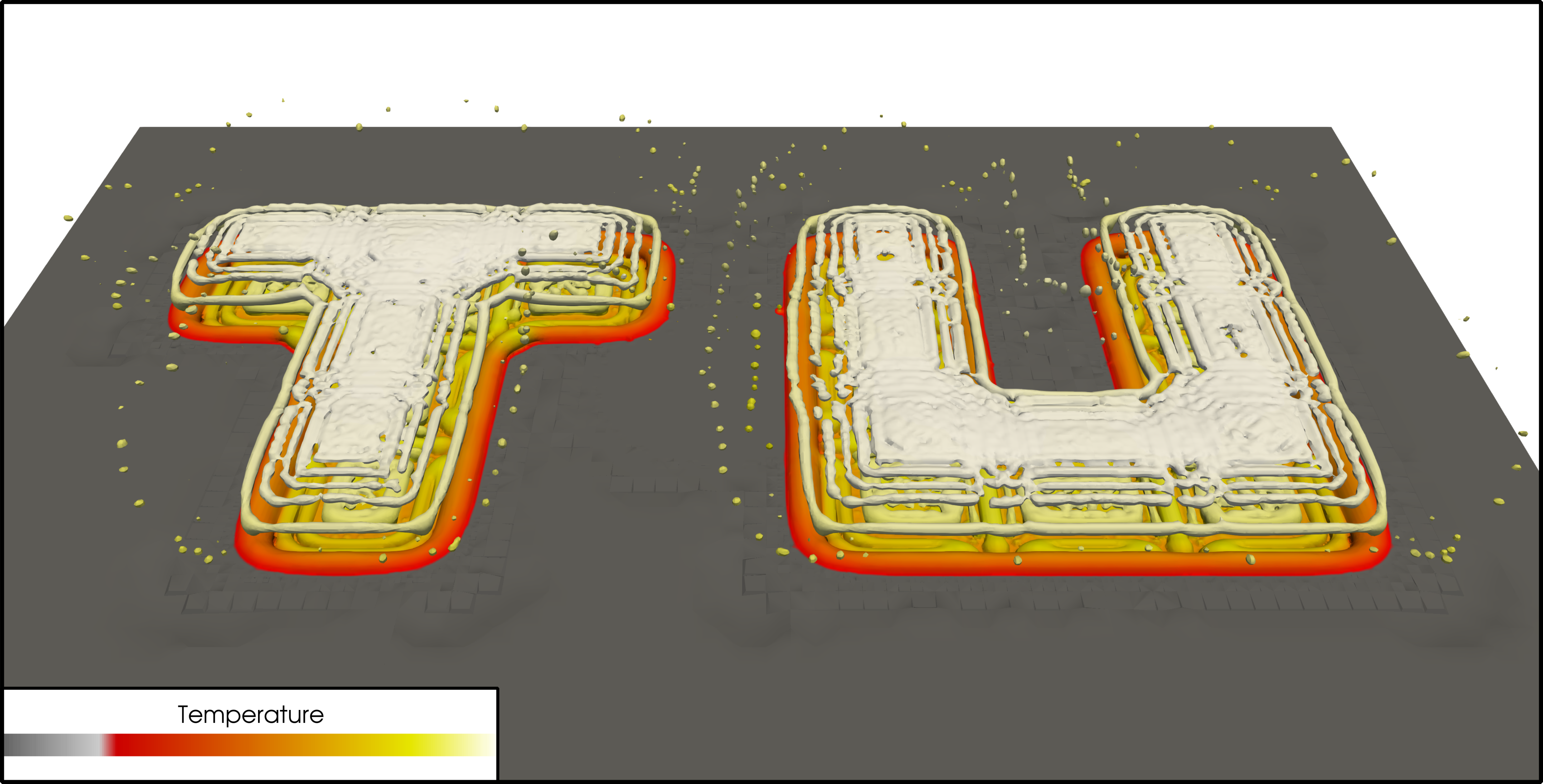}
\caption{Multi-spot ceramic ablation at threshold fluence showing material ejection at $\approx$2~ns after the pulse}\label{fig:TU_logo}
\end{figure}

As this is an ongoing research effort, further investigation is required to identify the optimal laser parameters for efficient ablation. Additionally, since the process involves an array of multiple laser spots, their interactions and mutual influence must be carefully evaluated. An exemplary simulation of multi-spot ablation using 12 spots (cf. Chap. 15 (Haasler et al.) of this book for background on ultrafast laser processing with multi beam setups) is shown in Figure~\ref{fig:TU_logo}. Other laser-based data storage technologies currently explored are discussed in Chap. 42 (Zhang et al.) of this book. 

\FloatBarrier
\subsection{Butt Joint Welding with Filler Wire}
\begin{figure}[h]
\centering
\includegraphics[width=0.6\textwidth]{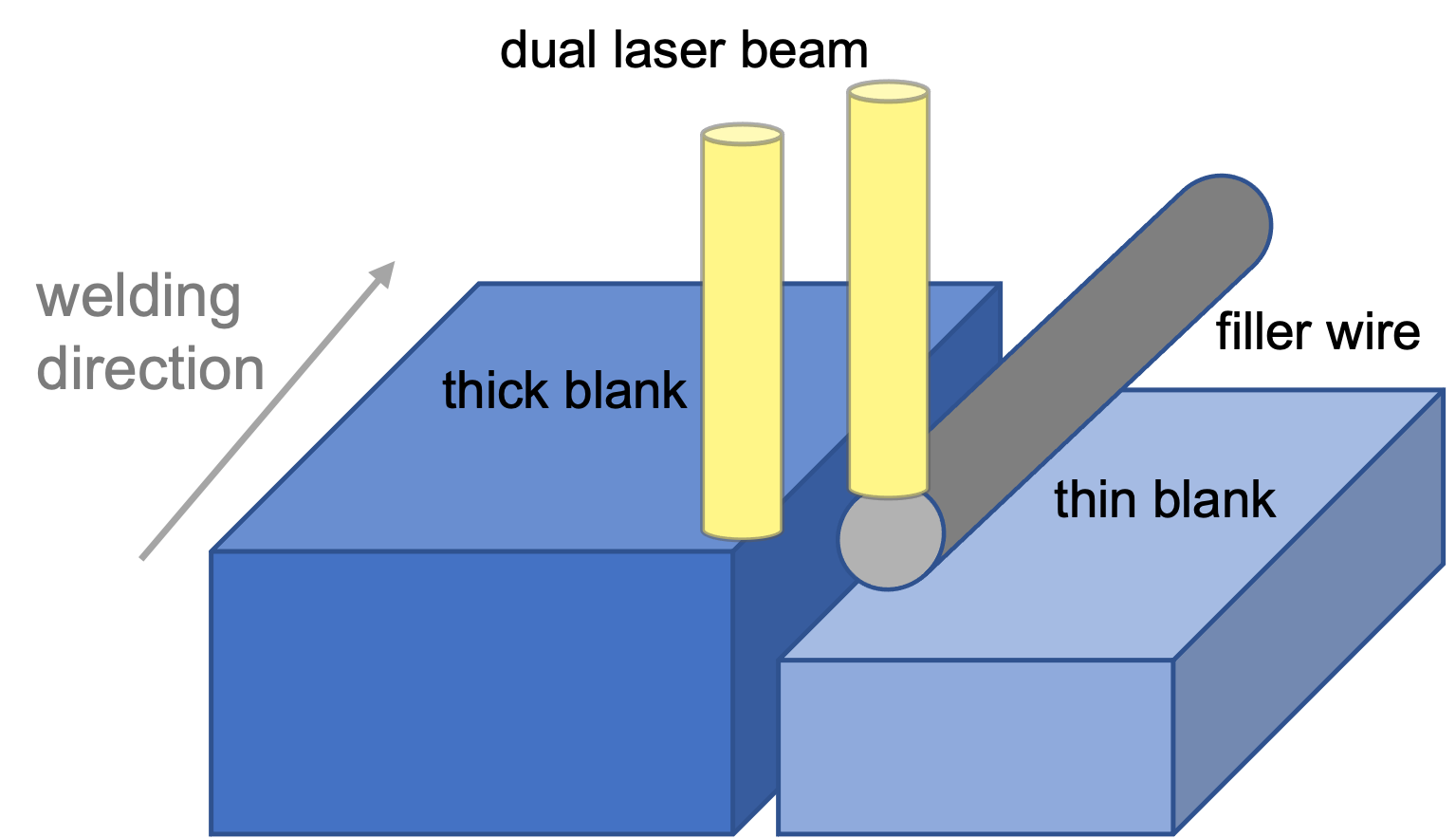}
\caption{Schematic illustration of the experimental set up of butt joint welding with gap and filler wire employing a dual-beam laser configuration}\label{fig:ButtJointWelding_Pic1}
\end{figure}

\begin{figure}[h]
\centering
\includegraphics[width=0.99\textwidth]{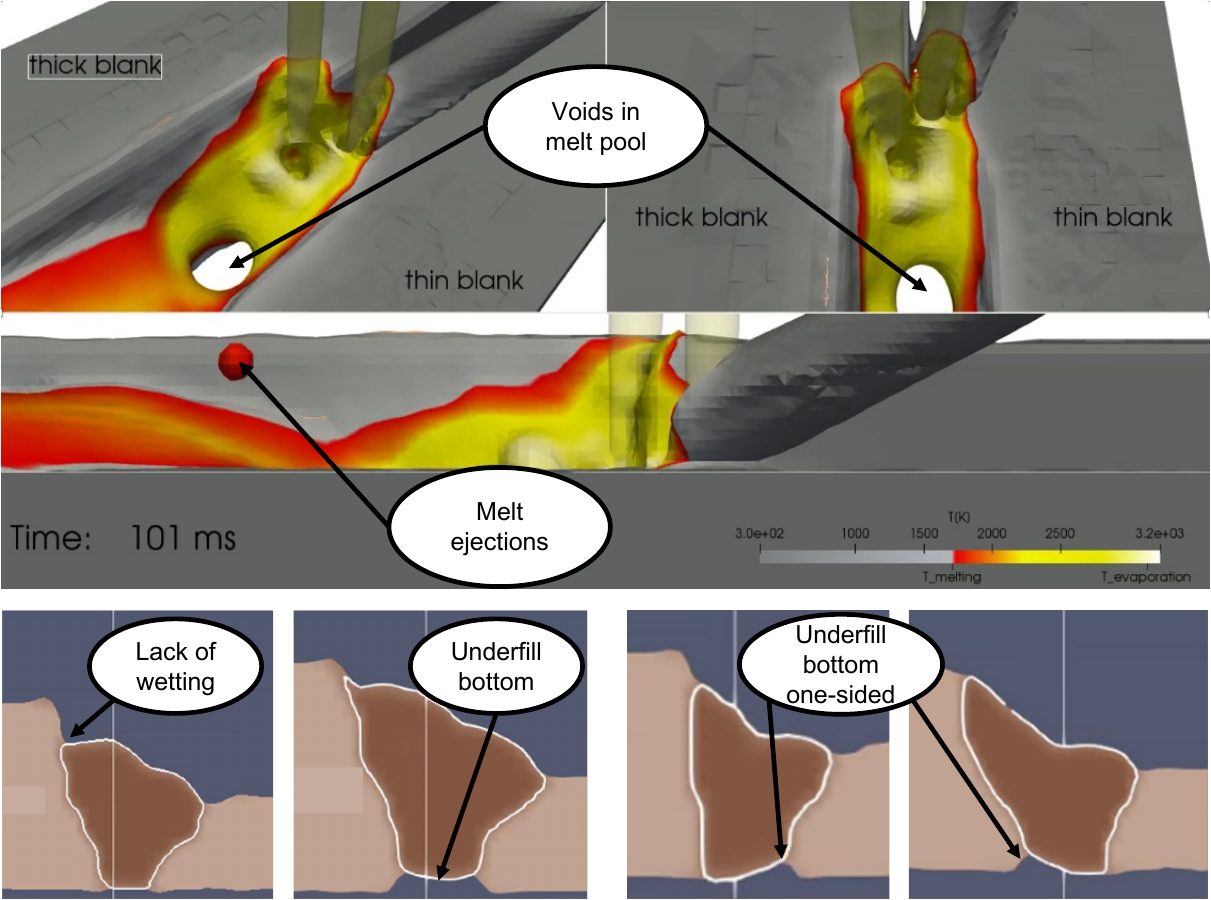}
\caption{Exemplary defects detected in the study of butt joint welding with a gap and filler wire including melt ejections, voids, lack of wetting at the top blank and underfills at the bottom bottom surfaces of the weld seam. These snapshots were recorded at various laser beam configurations}\label{fig:ButtJointWelding_Pic2}
\end{figure}

One significant challenge in industrial manufacturing involves butt joint welding of blanks with varying thicknesses. This task arises in various industries, including automotive manufacturing, aerospace, and shipbuilding, where lightweight design, structural integrity, and material efficiency are critical. The presence of a gap and the need for filler wire to ensure a robust joint add to the complexity of the process. Laser beam welding has emerged as a key enabling technology to tackle this challenge, offering high precision, excellent process stability, and the ability to produce defect-free welds even in demanding geometries. Its capability to achieve consistent results while maintaining high throughput makes it an ideal solution for addressing these complex industrial requirements. This case study investigates the welding of two steel blanks with a considerable difference in thickness, where a gap must be bridged. The approach employed to achieve welds of acceptable quality involves the use of a dual-beam laser configuration in combination with filler wire. The experimental setup is schematically illustrated in Figure~\ref{fig:ButtJointWelding_Pic1}. In this case study, the feed rate was maintained at a relatively high value, while key parameters such as the laser beam distance, position, and diameters were systematically optimized. The primary focus of the optimization was to mitigate common welding defects, including voids within the weld seam, melt ejections, humping, lack of wetting at the top and underfills at bottom surfaces of the weld. Representative examples of these defects, captured in a series of snapshots with different beam configurations, are presented in Figure~\ref{fig:ButtJointWelding_Pic2}.

\begin{figure}[h]
\centering
\includegraphics[width=0.99\textwidth]{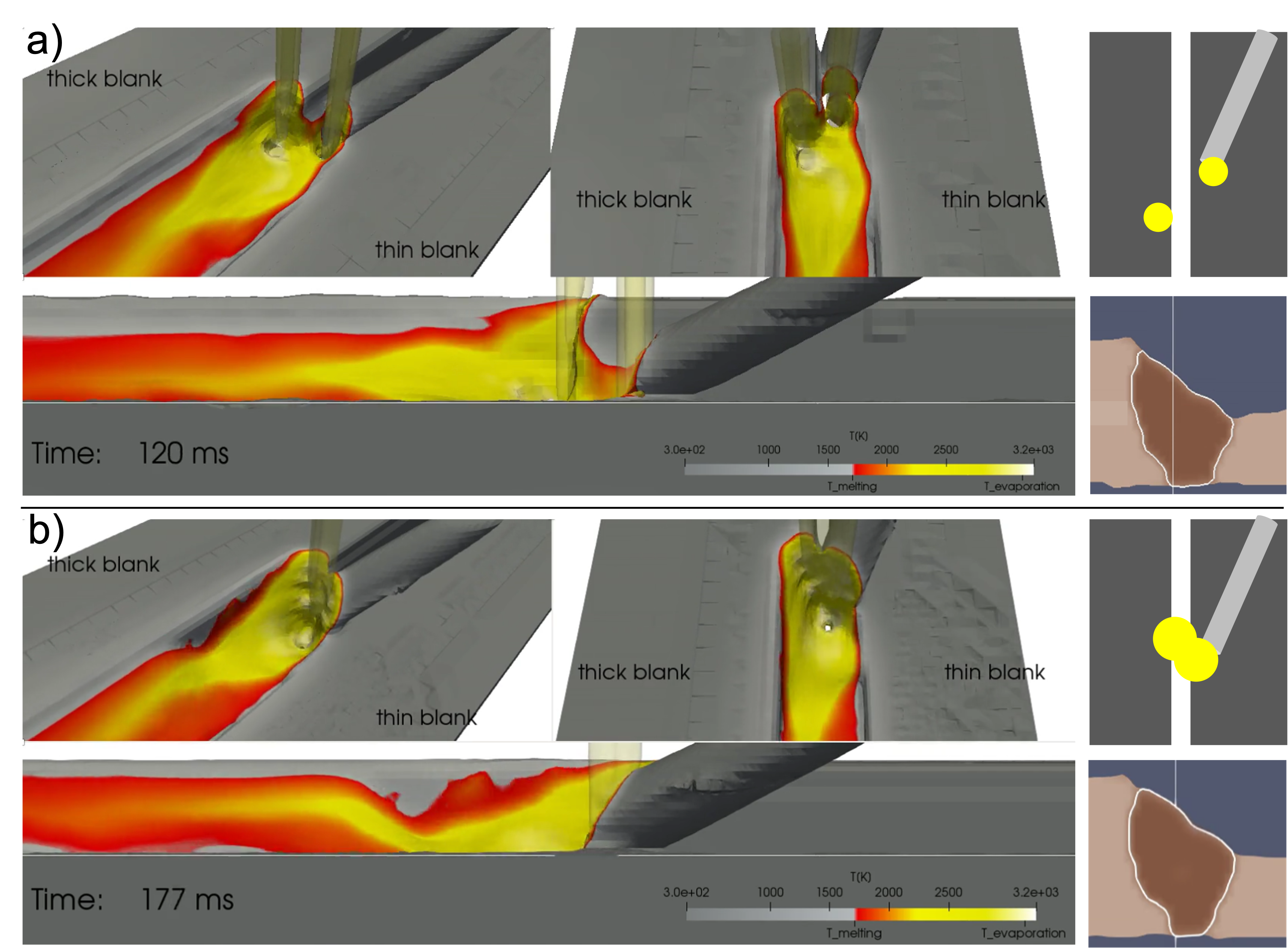}
\caption{Two solution approaches for laser beam configuration depicting snapshots of the welding process, a cross section and a schematic sketch of the configuration. a) Two smaller beams with distant positioning with the leading spot illuminating the filler wire. b) Two overlapping beams with increased diameter with the leading beam illuminating the thick blank}\label{fig:ButtJointWelding_Pic3}
\end{figure}

These observed defects highlight the need for a beam configuration that ensures a stable keyhole formation, resulting in controlled melt flow that avoids humping and prevents voids in the melt pool, as illustrated in Figure~\ref{fig:ButtJointWelding_Pic2}. Such voids can subsequently lead to defects within the weld seam. Additionally, the configuration must guarantee sufficient melting of the filler wire and adequate penetration at both the interface of the thin blank and filler wire, as well as the thick blank. To address these challenges, two distinct approaches were pursued: 

The first approach utilizes two laser beams with relatively small diameters and a significant separation distance. The goal is to create two distinct keyholes positioned far enough apart to prevent collapse into a single keyhole. In this configuration, the leading beam is directed primarily at the filler wire. Figure~\ref{fig:ButtJointWelding_Pic3}a presents a schematic of this setup, along with simulation results and a cross-section of the weld. The cross-section reveals sufficient wetting at the top edge of the thick blank, no underfill at the bottom, and stable separation of the keyholes despite their mutual influence. This method is hypothesized to be particularly advantageous when bridging larger gaps, although caution must be exercised to ensure the keyholes remain distinct, as their collapse could lead to instability and defects.

The second approach, illustrated in Figure~\ref{fig:ButtJointWelding_Pic3}b, employs an overlapping configuration of two beams with relatively larger diameters. In this setup, the leading beam targets the thick blank rather than the filler wire. The overlapping intensity region of the two beams ensures full penetration, while the non-overlapping areas contribute to melting sufficient material to fill the gap. Simulations and cross-sections from this configuration demonstrate good weld quality with an adequate cross-section. This approach appears to yield more stable melt pool dynamics; however, a slight reduction in wetting quality at the top edge of the thick blank is observed, potentially due to accelerated melt dynamics around the leading beam. Additionally, a minor onset of underfill is noted at the lower right edge of the thin blank.

Both approaches successfully address the inherent challenges of the welding process, producing high-quality welds. The optimal choice between these methods will depend on specific configurations and must be re-evaluated and optimized for each application. Nevertheless, these two approaches provide a robust starting point for further optimization efforts.

\FloatBarrier
\subsection{Static Beam Shaping for Enhanced Laser Welding of Copper}

\begin{figure}[h]
\centering
\includegraphics[width=0.99\textwidth]{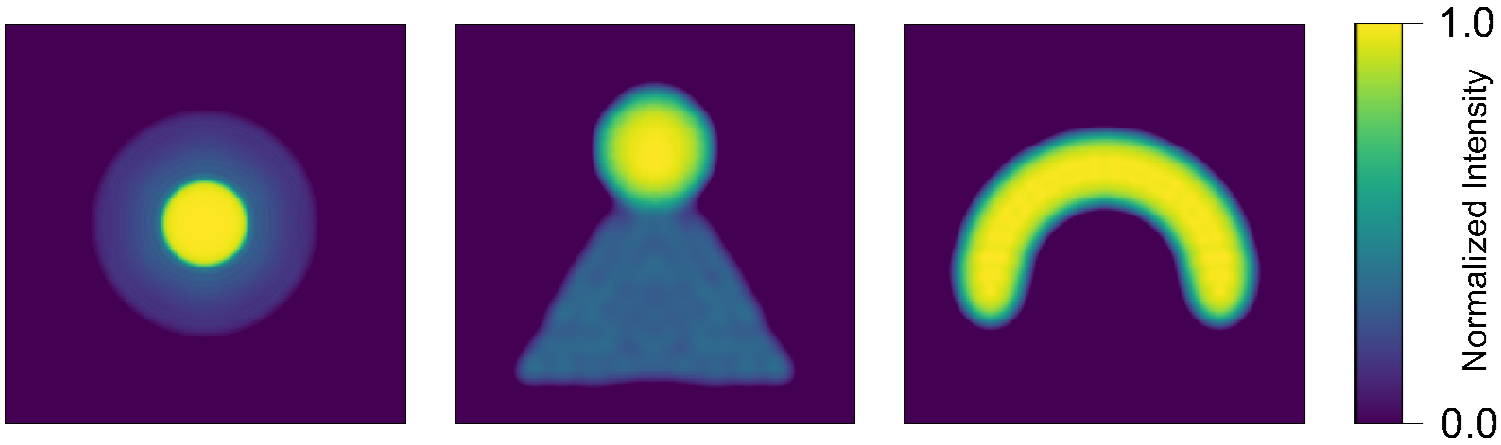}
\caption{Example of static beam shapes achievable with MPLC}\label{fig:exampleStaticShapes}
\end{figure}

Within the spectrum of available optical technologies, static beam shaping provides a method for optimizing laser-assisted manufacturing processes through the application of tailored spatial intensity profiles, as exemplarily illustrated in Figure~\ref{fig:exampleStaticShapes}. For general considerations on how to choose optical setups for shaping the spatial intensity distribution, we refer the reader to Chap. 13 (Schlutow et al.) of this book. Depending on the characteristics of the input laser system, various arbitrary beam shapes can be generated using passive optical elements integrated into customizable devices such as Multi-Plane Light Conversion (MPLC) technology, which can be incorporated into a laser head compatible with various industrial laser sources~\cite{Kumar2024,Pallier2024} . The basic principle of MPLC technology~\cite{MPLCpatent,Morizur2010} comprises a series of phase plates separated by free-space propagation, with each plate modifying the phase profile of the beam or beams passing through them in a stepwise manner until the desired shape is achieved. In the absence of active control or moving parts to dynamically adjust the optical properties of the system, the MPLC provides a beam with an intensity distribution that remains constant during material processing.

When applied to LBW, static beam shaping offers an attractive cost-benefit compromise for meeting demanding industrial tasks, such as welding copper at high feed rates while preserving adequate and consistent joint quality~\cite{Pallier2024}. In general, copper welding encompasses a few challenges originating from the material's high thermal conductivity. This makes it difficult to maintain a non-fluctuating melt pool and a stable vapor capillary, as heat rapidly dissipates away from the laser-material interaction zone. Also, its low absorptivity in the infrared wavelength range, commonly used in high-power industrial applications, makes processing this material difficult. These factors can result in commonly observed defects, including porosity, lack of fusion, excessive spatter, etc. 

To explore the potential of static beam shaping in enhancing productivity for copper welding applications, multiphysical simulations can be systematically employed to assess the performance of various intensity distributions while keeping other process parameters unchanged. The primary objective of the demonstration case presented here is to identify an intensity distribution that produces high-quality weld seams with minimal porosity and spatter at a target feed rate of 20 m/min. The general process conditions for this exercise are outlined in Table~\ref{tab:parametersStaticBeamShapingApp}, where the laser power is adjusted based on the beam shape to ensure consistent penetration depths across simulation tests. For instance, a single spot configuration would require less power than a spot-ring arrangement to achieve a similar penetration depth. 

\begin{table}[h!]
    \centering
    \begin{tabular}{ll}
        \hline
        \textbf{Parameter} & \textbf{Description} \\
        \hline
        Material           & Cu-ETP\tablefootnote{Copper Electrolytic Tough Pitch, $\leq$ 0.04 wt.\% oxygen}\\ 
        Surface treatment  & None\\ 
        Welding speed      & 20 m/min\\
        Power              & 6 - 11 kW\\
        Welding mode       & Continuous Wave\\
        Weld type          & Bead on Plate (BOP)\\ 
        Plate thickness    & 4 mm\\ 
        Plate positioning  & Horizontal\\ 
        Beam orientation   & Perpendicular to plate\\
        Focus position     & 0 mm (focus on surface)\\ 
        Atmosphere         & Air at 298 K and 1 atm\\
        Shielding gas      & Not used\\
        \hline
    \end{tabular}
    \caption{Process parameters employed in copper welding application}
    \label{tab:parametersStaticBeamShapingApp}
\end{table}

\begin{figure}[h]
\centering
\includegraphics[width=0.99\textwidth]{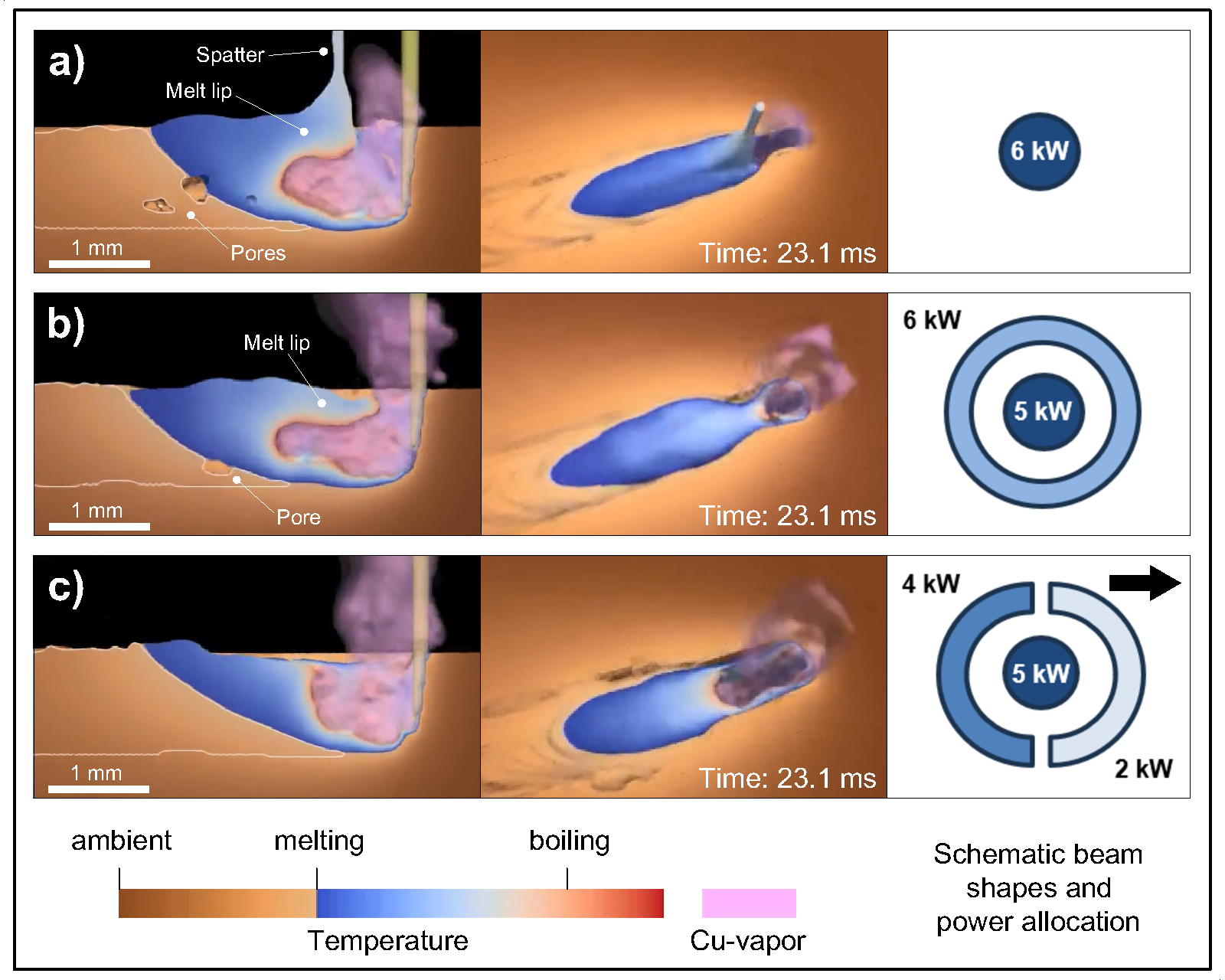}
\caption{Simulations of LBW of copper using different beam shapes: a) spot only, b) spot plus a power-symmetrical ring, and c) spot plus a power-asymmetrical ring. On the left-hand side, both a longitudinal section and a perspective view of the welding process are shown. The yellow contour represents the laser beam, while the white contour outlines the re-solidified material. On the right-hand side, schematic representations of the beam shapes are displayed. For configuration c), an arrow indicates the welding direction (Figure adapted from~\cite{Pallier2024})}\label{fig:processCopperWeldingApp}
\end{figure}

\begin{figure}[h]
\centering
\includegraphics[width=0.99\textwidth]{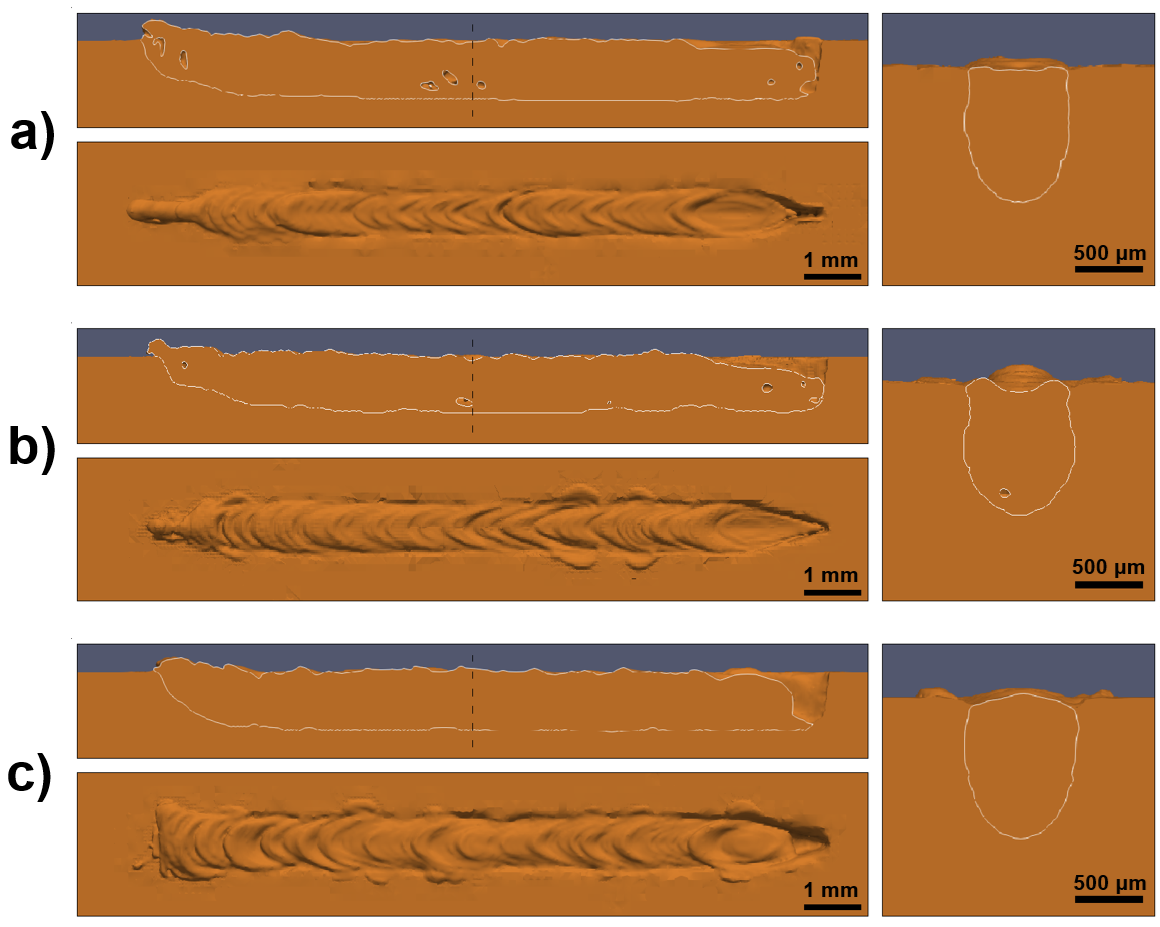}
\caption{Final obtained weld beads from the simulations depicted in Figure~\ref{fig:processCopperWeldingApp}. On the left-hand side, a longitudinal cross-section and a top view of the weld seams are presented. On the right-hand side, a transverse cross-section taken at the center of the weld seams is displayed. The welding direction is from left to right}\label{fig:simulatedCopperWelds}
\end{figure}

After validating the numerical model, e.g., by comparing simulation results with cross-sectional micrographs of experimentally produced weld beads, a variety of beam shapes and their caustic are individually implemented and tested in a BOP welding scenario. Their performance is then evaluated by quantifying attributes of interest, such as porosity and spatter occurrence. Through a systematic iterative approach, a shape with intensity asymmetry relative to the beam's mid-plane and oriented perpendicular to the welding direction is identified as delivering the most favorable results. This is depicted in Figures~\ref{fig:processCopperWeldingApp} and~\ref{fig:simulatedCopperWelds}, which qualitatively compare the simulated welding processes using the asymmetric shape to both a single spot and a configuration combining a spot and a ring with symmetric intensity distribution.

The conditions in Figures~\ref{fig:processCopperWeldingApp}a and~\ref{fig:processCopperWeldingApp}b both result in a pronounced, narrow bead protrusion at the entry region, with pores appearing at different locations along the weld seam. These beam shapes produce an unstable melt lip that obstructs vapor escape, leading to chaotic behavior in the melt pool and vapor capillary. The latter adopts an elbow-like geometry, which eventually collapses at its rear end, leading to pores. Furthermore, the narrow melt pool front induces rearward acceleration of the liquid, contributing to frequent spatter development. Using the asymmetric configuration of Figure~\ref{fig:processCopperWeldingApp}c, the increased power at the rear of the ring contributes to a nearly constant opening of the vapor capillary, enabling the steady and controlled release of vapor, thus reducing process instabilities. The front section of the ring, with lower light intensity, acts as a preheating mechanism, increasing the liquid volume at the melt pool's leading edge and thereby widening the front. This results in reduced rearward liquid acceleration and fewer spatter events.

Simulation of beam shaping combined with multiphysical phenomena provides a powerful alternative to examine the interaction between specific intensity distributions and the welding process, including melt pool and vapor capillary dynamics, as well as the behavior of variables like pressure, temperature, velocity, and phase fields, which cannot be fully captured through experimental observation. As described here, high-fidelity simulations suggest that good quality copper welds could be achieved even at high feed rates when appropriate process parameters are chosen. Static beam shaping technology may serve as an important asset in enabling this objective, providing a reliable and cost-effective means to enhance quality and efficiency in particularly demanding applications. 

\FloatBarrier
\subsection{Simulation of Welding with Dynamic Beam Shaping}

Dynamic beam shaping (i.e., a varying laser intensity distribution over time) provides a virtually infinite parameter space to utilize for process optimization. In this application example, the aim is to obtain defect-free single pass welds of 15 mm thick mild steel sheets. An overview of the challenges involved with welding thick sections and current progress in this field can be found in Chap. 33 (Olschok et al.) of this book.

\begin{figure}[h]
\centering
\includegraphics[width=0.8\textwidth]{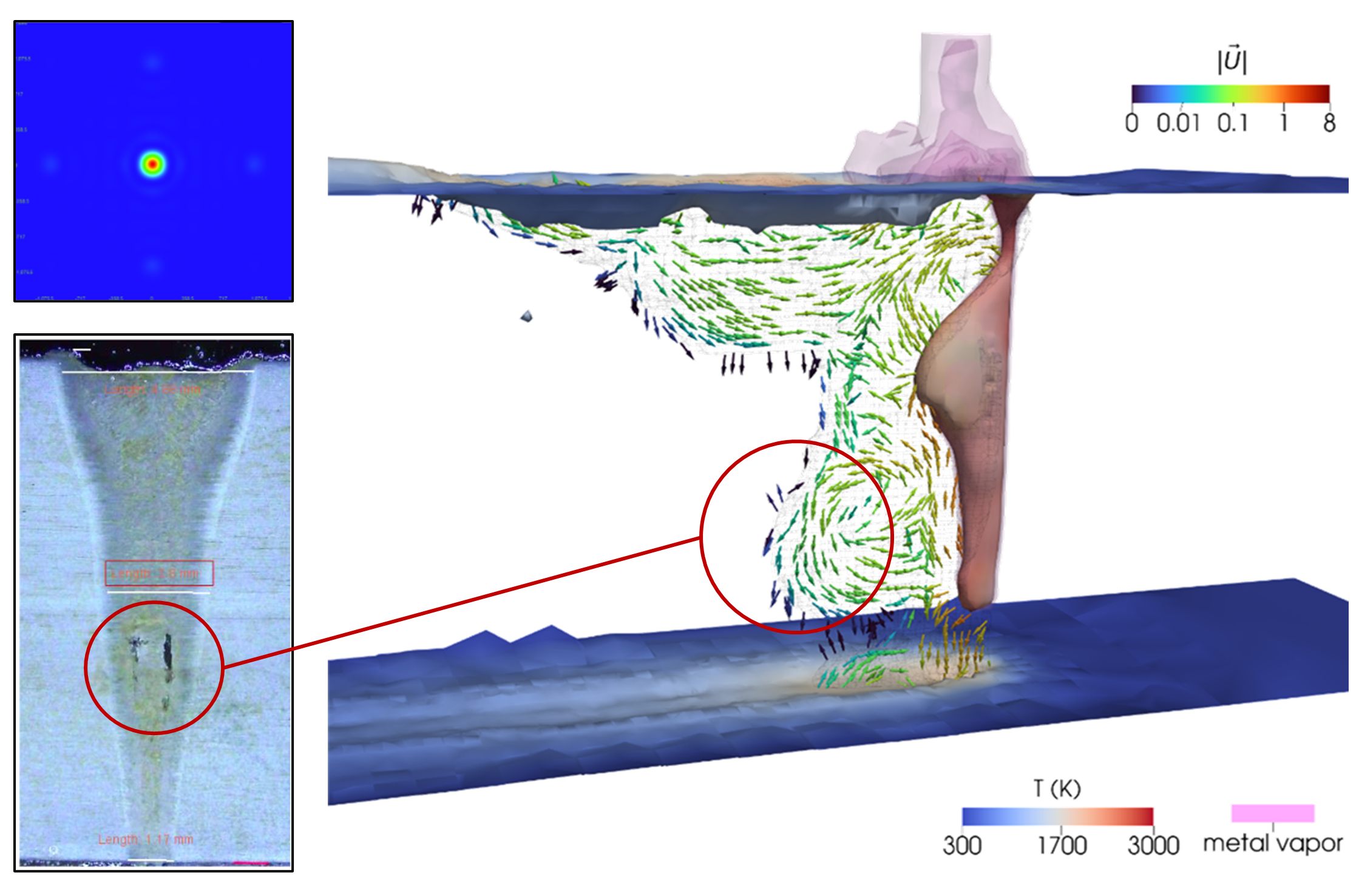}
\caption{Single-pass welding of 15 mm thick mild steel plates: Experimentally obtained cross section and snapshot of corresponding simulation during steady state. The melt pool is denoted by arrows representing the flow field of liquid material. Conventional single spot beam shape leads to cracks and voids due to severe bulging in the lower third of the melt pool tail. Experimental image courtesy of Civan Lasers}\label{fig:appl:dpSteel_singleSpot}
\end{figure}

\begin{figure}[h]
\centering
\includegraphics[width=0.8\textwidth]{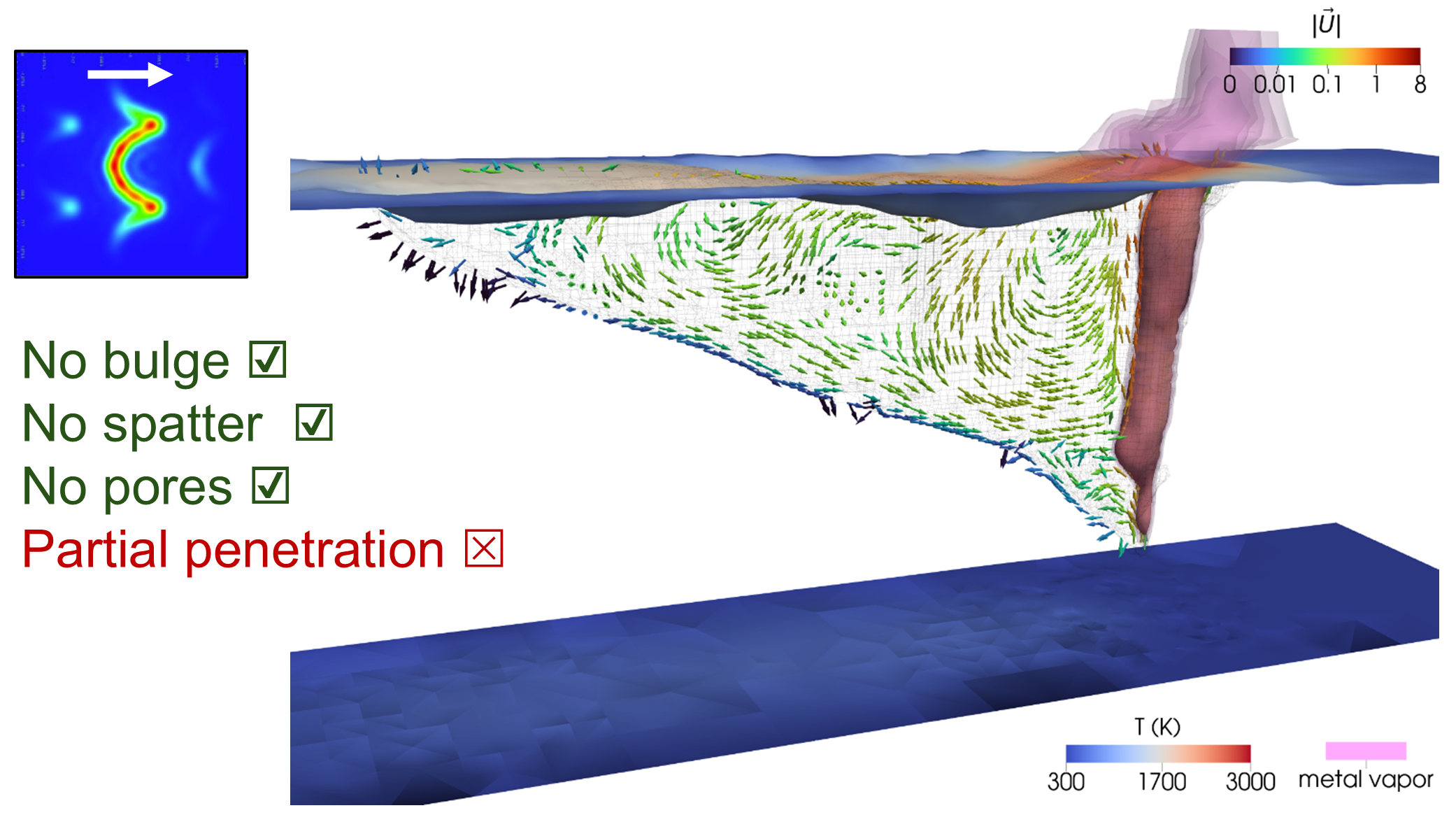}
\caption{Single-pass welding of 15 mm thick mild steel plates: Steady-state snapshot of simulation using dynamic  U-shape (white arrow denotes welding direction). The resulting process is stable and bulge-free, but lacks penetration depth}\label{fig:appl:dpSteel_inwardsU}
\end{figure}

Here, to achieve a high-quality single-pass full-penetration weld of 15~mm, we utilize Coherent Beam Combining via Optical Phase Arrays, a method that allows for freeform beam shaping up to the MHz-regime~\cite{Nissenbaum2022}. While achieving full penetration is not an issue with today's high power lasers when employing a single high intensity laser spot, the process leads to cracks and pores, mainly accumulating at approx. two thirds of the welding depth, as can be seen in the experimentally obtained cross section in Figure~\ref{fig:appl:dpSteel_singleSpot}. A simulation of the process (cf. right part of Figure~\ref{fig:appl:dpSteel_singleSpot}) helps identifying the cause, which is the formation of a bulge in the melt pool tail. The presence of a bulge in deep penetration laser beam welding is known to promote the occurrence of cracks~\cite{Artinov2020}. By employing a different U-shaped beam profile, which is itself modulated in the kHz-regime (the shape itself is made up of a series of discrete point-like shapes, activated in a sequence, cf.~\cite{Weber2025} for details on these types of dynamic beam shapes), the melt pool can be stabilized and bulging can be omitted, cf. Figure~\ref{fig:appl:dpSteel_inwardsU}. However, the overall size of this shape is much larger than the single spot, leading to lower average intensity and hence a reduced penetration depth, not fulfilling the requirement of full penetration.

\begin{figure}[h]
\centering
\includegraphics[width=0.99\textwidth]{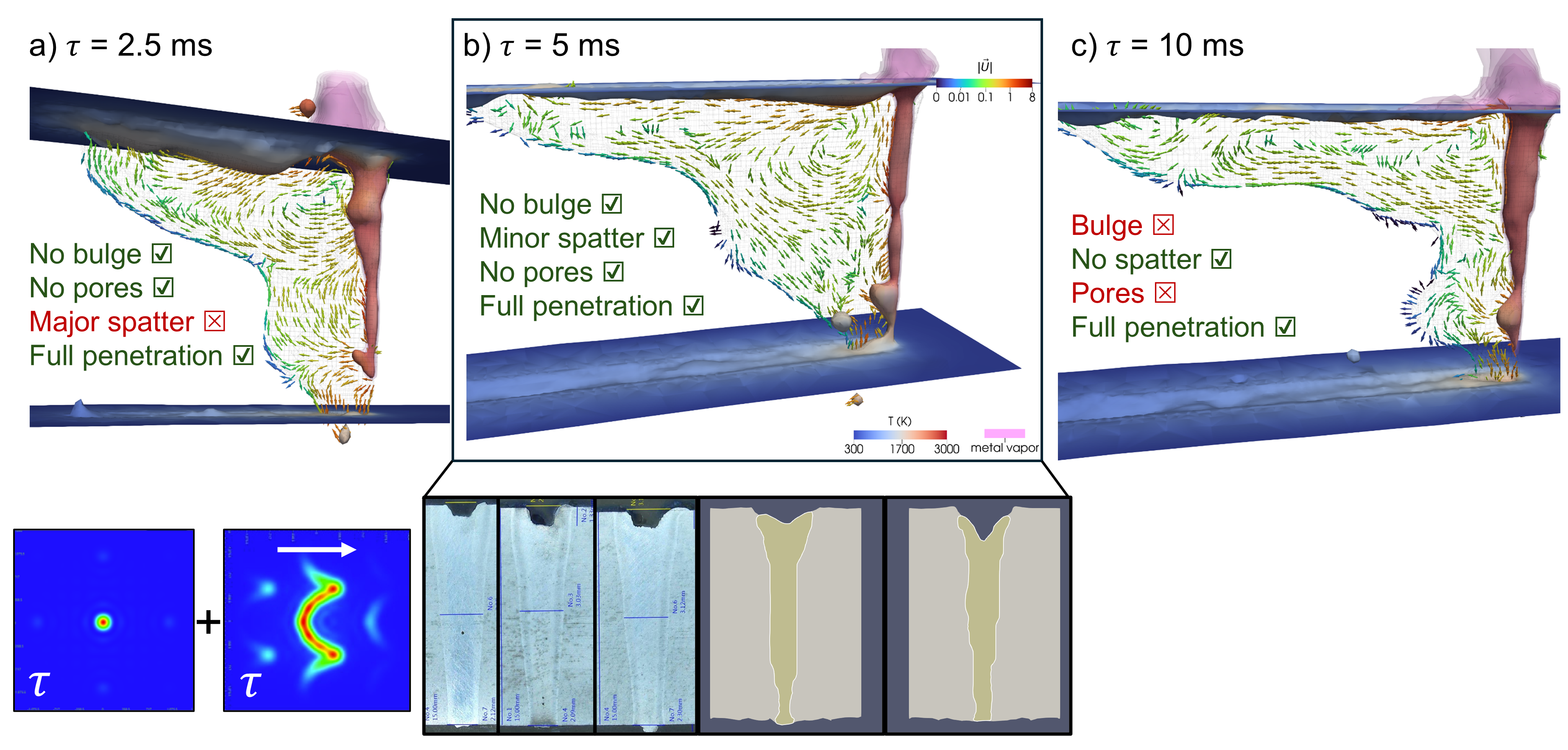}
\caption{Single-pass welding of 15 mm thick mild steel plates: Simulation results of different periods for switching between "inwards-U" and single spot beam shapes. The most promising parameter identified in simulations ($\tau$=5 ms) leads to crack- and pore-free, stable welds in experiment (bottom row). Experimental images courtesy of Civan Lasers}\label{fig:appl:dpSteel_switching}
\end{figure}

By introducing a shape-switching strategy~\cite{Bernhard2024}, where the laser intensity distribution switches between a single spot (to obtain a sufficiently deep keyhole) and the U-shape to stabilize the keyhole and establish favorable melt pool flow patterns, a good compromise can be obtained (cf. Figure~\ref{fig:appl:dpSteel_switching}). Depending on the time $\tau$, for which each shape is active, the process either becomes unstable leading to a lot of spatter (Figure~\ref{fig:appl:dpSteel_switching}a), or bulging cannot be avoided, leading to porosity and (potentially) cracks (Figure~\ref{fig:appl:dpSteel_switching}c). In between (at $\tau$=5~ms), good results are obtained, with almost no bulging and stable full penetration. These parameters, identified through simulation, also lead to a crack-free and stable result in experiments (Figure~\ref{fig:appl:dpSteel_switching}b).

\FloatBarrier
\subsection{Multi-Scale Simulation of PBF-LB/M}\label{sec:appl:pbf}

\begin{figure}[h]
\centering
\includegraphics[width=0.8\textwidth]{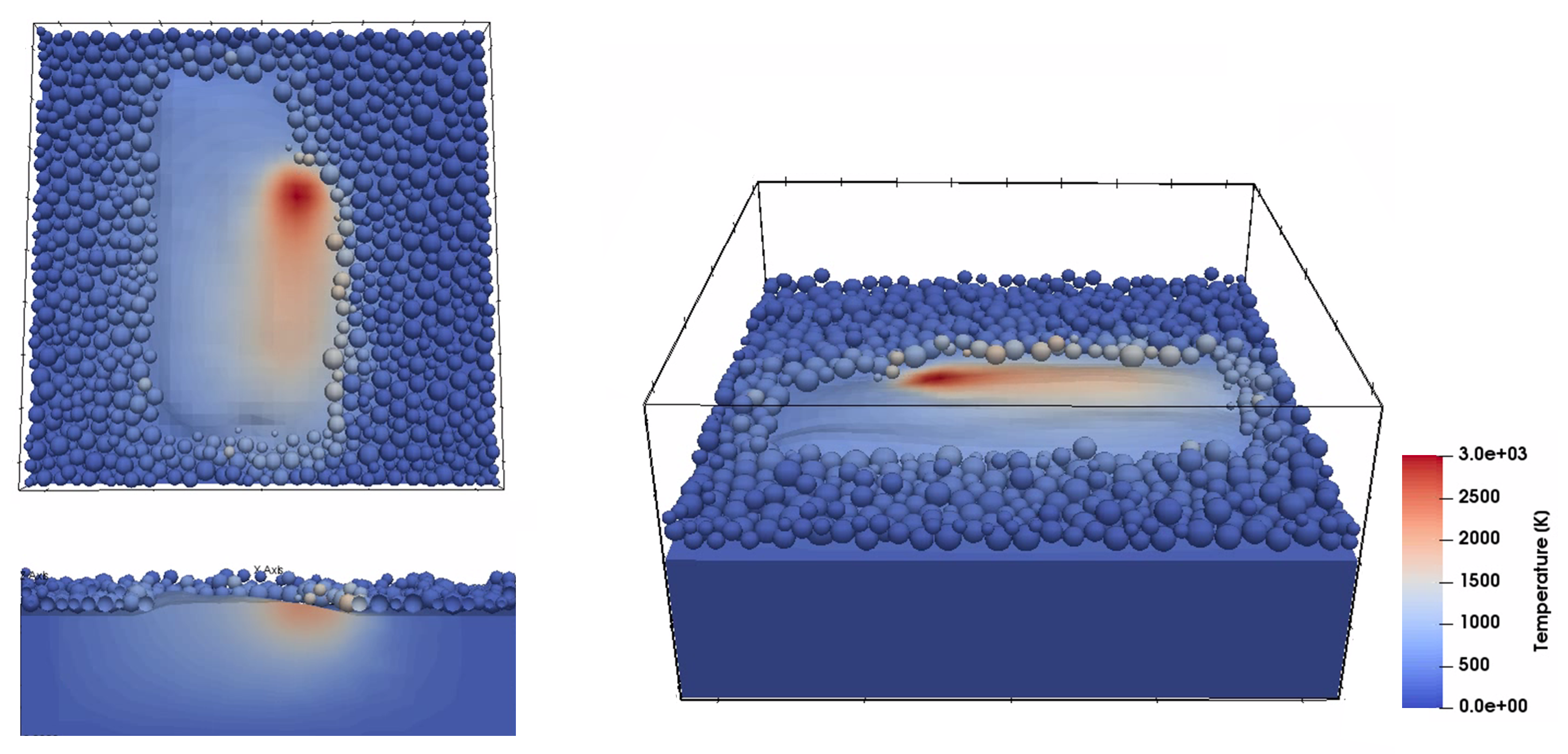}
\caption{Single-layer simulation of PBF-LB/M using coupled FVM-DEM model: Realistic representation of powder particle distribution is achieved via modeling individual particles as discrete elements, which are coupled to the CFD domain and exchange mass and energy when undergoing melting. This comes at a trade-off with higher computational costs}\label{fig:appl:pbfDEM}
\end{figure}

PBF-LB/M poses a very complex simulation and modeling challenge due to its multi-scale aspect. An overview of the state of the art and emerging trends of PBF-LB/M among other laser-based additive manufacturing processes is provided in Chap. 23 (Lasagni et al.) of this book.  In principle, PBF-LB/M can be seen as a series of small scale welds involving powder material. One possible modeling strategy consists of resolving individual powder particles via the discrete element method (DEM) and coupling these elements with the remaining continuum mechanical problem through a mass- and energy transfer once the particles melt, yielding a coupled FVM-DEM model. An example of such a simulation is given in Figure~\ref{fig:appl:pbfDEM}, where multiple adjacent tracks of a PBF-LB/M process are simulated. While such an approach provides advanced predictive capabilities regarding the influence of particle distributions on process outcomes, and enables the simulation of particle-related phenomena such as denudation, the computational costs involved with resolving individual particles become prohibitive when surpassing the scale of several short tracks. While the timescales of interest at the single track level are in the microsecond to millisecond regime, the entire process spans much larger timescales, with processing of entire parts usually taking many hours. Therefore, often, the smallest building block of a PBF-LB/M process is considered when simulating this process: A single scan track. However, it is often needed to go beyond the single track or even single layer scale. To facilitate simulations at a larger scale, a viable alternative to CFD-DEM approaches involves modeling the powder as a continuum with average powder-like properties~\cite{Zenz2023,Zenz2024b}. Consequently, it is possible to simulate several layers of many scan tracks each, where each track bares the entire multiphysical complexity of any welding process.

\begin{figure}[h]
\centering
\includegraphics[width=0.99\textwidth]{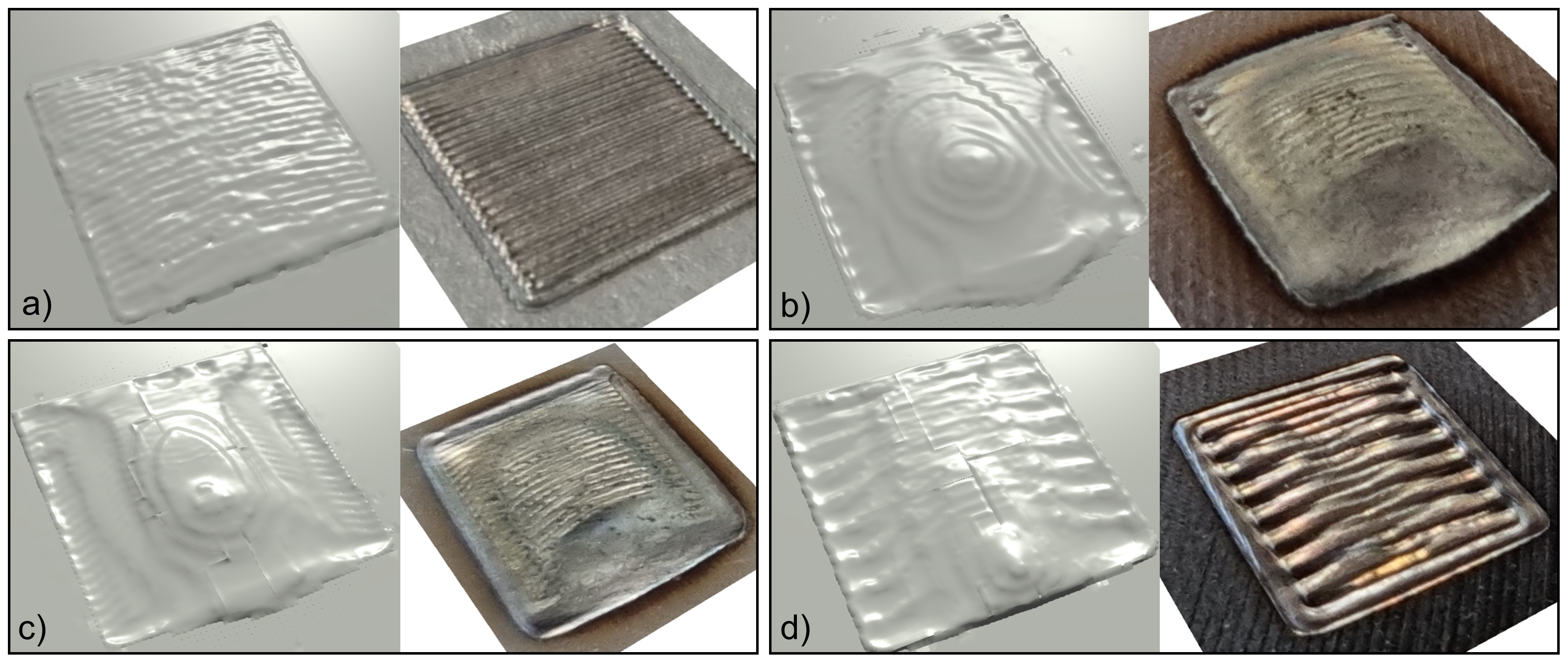}
\caption{Single-layer simulations (left) of PBF-LB/M using continuum powder model: Solidified layer (powder removed) compared to corresponding experiments (right) for different dual beam approaches: a) Single beam reference result, b) -- d): Addition of various secondary large diameter tophat beams and variation of scan speed and hatching distance, all of which lead to unwanted heat accumulation- and surface tension-driven defects. Experimental images courtesy of AIDIMME, Spain\protect\footnotemark}\label{fig:appl:pbfDualBeams}
\end{figure}

Figure~\ref{fig:appl:pbfDualBeams} highlights an example of the multi-scale nature of the problem: Process parameter combinations that work well on the level of a single track can lead to unwanted defects at the level of an entire part. Figure~\ref{fig:appl:pbfDualBeams}\nolinebreak a shows the top surface of a 5~mm x 5~mm layer of a solidification crack-prone nickel-based-superalloy, fabricated using a conventional PBF-LB/M process. Nickel-based superalloys are particularly interesting for many applications due to their high strength at high temperatures, making them a widely used class of materials for turbine blades, as an example. However, many of these alloys are classified as ``non-weldable" as they exhibit high crack susceptibility when processed with a laser~\cite{Sanchez2021}. To avoid the formation of cracks, a new process was designed where a large, low-intensity secondary laser beam was added to provide sufficient pre- and post heating to the process zone in order to lower cooling rates~\cite{Zenz2023}. Figure~\ref{fig:appl:pbfDualBeams}\nolinebreak b-d show exemplary results using different secondary beams (that lead to good results in both experiments and simulations of a single track), where the scan speed and/or hatching distance was already increased to compensate for the additional heat input of the secondary beam. However, both experiments and simulations in b-d exhibit unwanted heat accumulation- and surface tension-driven defects that only develop after multiple tracks. Towards the end of the processed layer in b and c we see a region where melt accumulated, that did not fully solidify between processing of individual scan tracks, and accumulated driven by Marangoni currents, even reaching over the defined area to be processed. In d the hatch spacing was increased. While heat accumulation poses a less significant problem here, the structure of the scan tracks can easily be seen, where the melt tracks clearly agglomerated to tube-like structures, yielding unwanted surface finish and inhomogeneous part density. Furthermore, we can see the onset of heat accumulation-driven defects in the form of two slightly elevated regions, on both sides of the center. In the case study presented here, the secondary beams were introduced via MPLC technology~\cite{MPLCpatent,Morizur2010,Bayol2022}, where the approximate shape needs to be defined prior to manufacturing the beam shaping device. Hence, experiments cannot be conducted at the stage of optics design and simulations are essential. However, a PBF-LB/M process leading to good results at the level of a single track might not work at larger scales.

\begin{figure}[h]
\centering
\includegraphics[width=0.99\textwidth]{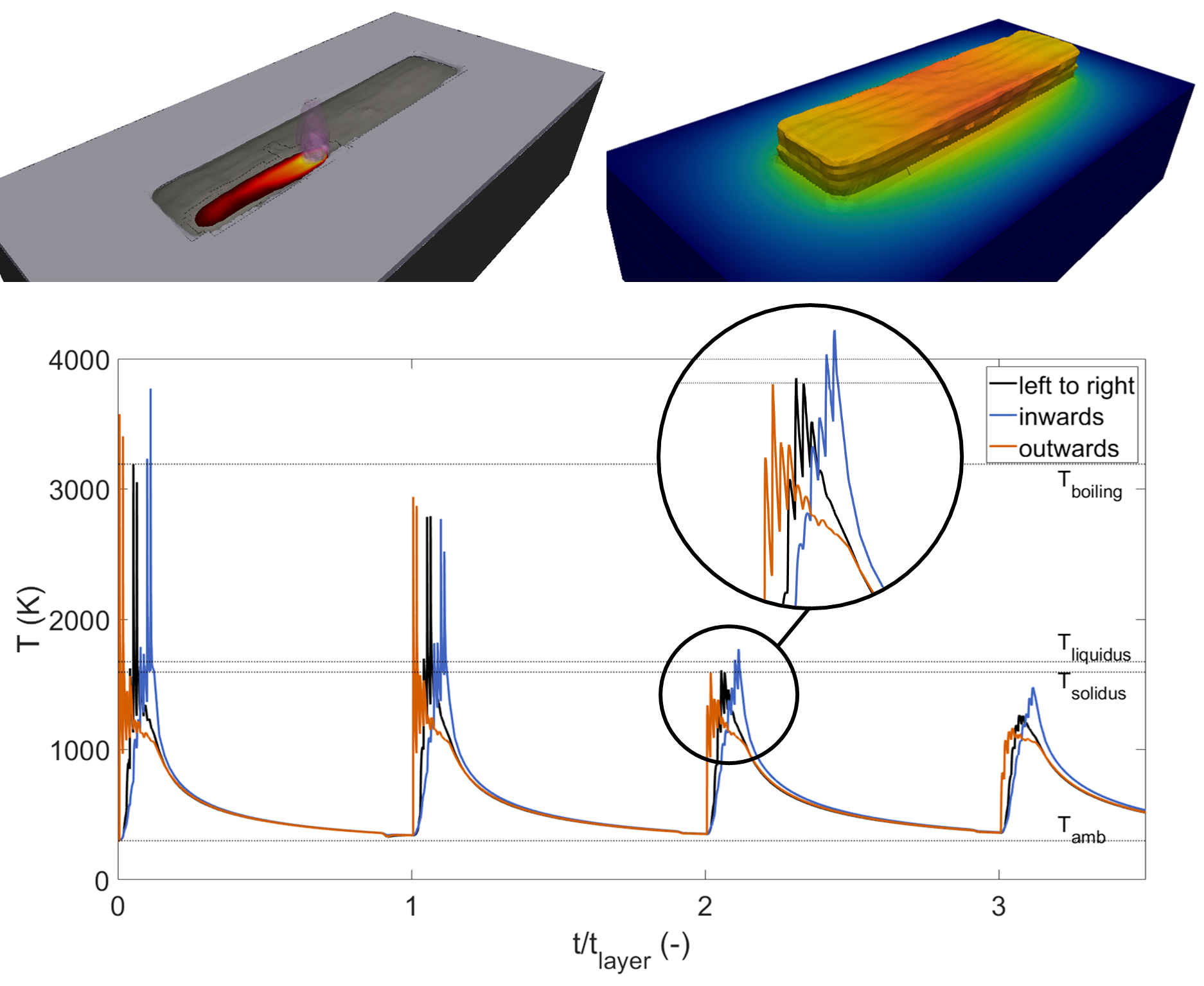}
\caption{Multi-layer simulation of PBF-LB/M using continuum powder model. Top: Snapshot during process (left) and final part after processing six layers of ten tracks each, colored by temperature during cool down (right). Bottom: Thermal histories of one location during the first four layers of the process for three different scanning strategies (bottom), time normalized by printing time of one layer, $t_{layer}$}\label{fig:appl:pbfScanStrategies}
\end{figure}

\begin{figure}[h]
\centering
\includegraphics[width=0.5\textwidth]{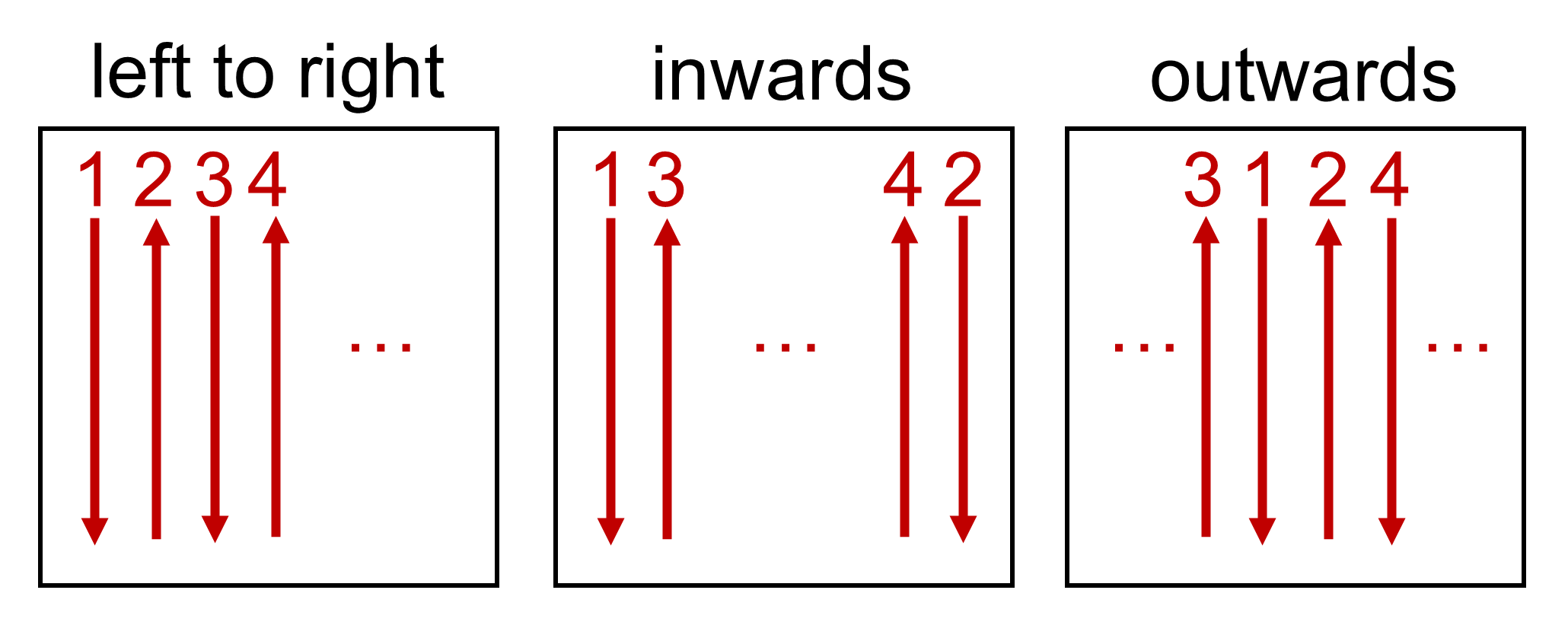}
\caption{Illustration of different exemplary layer scan strategies as referred to in Figure~\ref{fig:appl:pbfScanStrategies}}\label{fig:appl:pbfStrategyExplanation}
\end{figure}

\begin{figure}[h]
\centering
\includegraphics[width=0.99\textwidth]{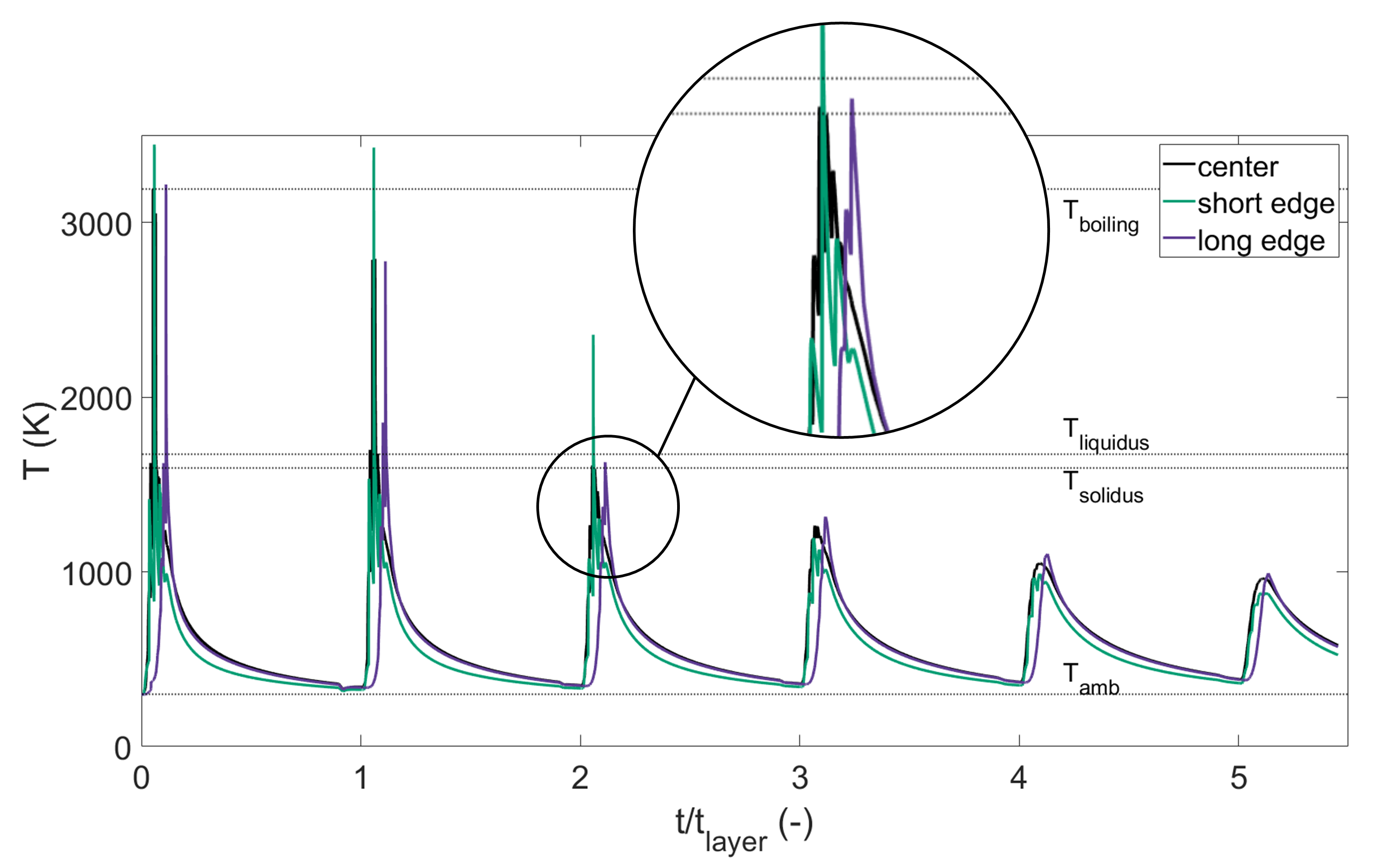}
\caption{Multi-layer simulation of PBF-LB/M using continuum powder model. Thermal histories of three different locations for ``left to right" scan strategy, time normalized by printing time of one layer, $t_{layer}$}\label{fig:appl:pbfPositions}
\end{figure}

\FloatBarrier
\footnotetext{{The experimental images presented in Figure~\ref{fig:appl:pbfDualBeams} are courtesy of Mario Martínez Ceniceros and Luis Portolés Griñán (AIDIMME, Spain) and were obtained with funding from the European Union’s Horizon 2020 research and innovation programme within the project CUSTODIAN, under grant agreement number 825103.}}
In Figure~\ref{fig:appl:pbfScanStrategies}, the model scale is further increased, by simulating multiple layers of the entire PBF-LB/M printing process, while still resolving all relevant fluid mechanical phenomena, such as evaporation, gas dynamics, and surface tension. The shown simulation results feature the thermal history of a point at the center of the rectangular part that is being built, over the course of four printing layers, comparing three different scanning strategies, which are illustrated in Figure~\ref{fig:appl:pbfStrategyExplanation}. Interestingly, depending on the strategy, the point of interest will be fully melted either two or three times, and the cooling rates at which the point solidifies in each layer differ strongly between strategies. Moreover, intuition would tell us that scanning "inwards" will lead to more heat accumulation in the center (where the thermal history is plotted), which should lead to lower cooling rates (favorable in terms of crack susceptibility). However, the cooling rates after each melting cycle are much higher for this strategy. While these simulations are certainly not at the scale of an entire built part, they represent a unit cell of a part (i.e., one simple geometrical element, many of which make up a part, such as a thin wall), and encompass the scales at which accurate representation of phase change and fluid flow are relevant. In Figure~\ref{fig:appl:pbfPositions}, the thermal histories of three different positions in the sample are shown, all for the ``left to right" strategy: at the center of the rectangle forming the layer geometry in the plane normal to build direction, and respectively at the middle of the long and short edge of the rectangle, each at the height of the first layer. It becomes obvious that, for a given part geometry and scan strategy, thermal histories drastically differ from location to location, calling for geometry-specific parameter optimization.

\FloatBarrier
\section{Outlook}
Laser-based manufacturing continues to evolve as a highly versatile and precise tool across a wide range of industries with the potential for mass-customization and high degree of automation.  While laser-material interaction and the subsequent transient physical phenomena are per se highly complex, the advent of novel technologies enabling, for example, dynamic freeform beam shaping, the usage of optical vortices, or hybrid wavelength processing -- to name a few -- leads to infinitely large parameter spaces to be used for process design and optimization. Optimization via experimental campaigns is unrealistic in such high-dimensional parameter spaces and a thorough understanding of the underlying physical phenomena is crucial. Hence physics-based simulations with high predictive capabilities will remain an important field of active research and development. Furthermore, developments in the fields of artificial intelligence, such as physics-informed machine learning and inverse modeling, are expected to result in fast predictive modeling to be used in in-line process control and digital twin applications, in order to quickly find the optimal parameter combinations for a desired process outcome and eventually enable first-time-right manufacturing of highly customized parts.

\FloatBarrier
\bmhead{Funding}
T.F. acknowledges funding through the FWF within the project ``Formidable" (I 6401-N). C.Z. acknowledges funding through the TU Wien Doctoral School within the project ``DigiPhot". M.B. acknowledges funding through the FFG within the project ``GREEN". Part of the results presented in Section~\ref{sec:appl:pbf} have been obtained with funding from the European Union’s Horizon 2020 research and innovation programme within the project CUSTODIAN, under grant agreement number 825103. CUSTODIAN project is an initiative of the Photonics Public Private Partnership.





\FloatBarrier
\bibliography{sn-bibliography}

\end{document}